\documentclass{aa}  
\usepackage{graphicx}
\usepackage{txfonts}
\usepackage{natbib,twoopt}
\usepackage{caption}
\usepackage[]{overpic}  

\usepackage{gensymb} 
\usepackage{subeqnarray}

\usepackage[breaklinks=true]{hyperref} %
\bibpunct{(}{)}{;}{a}{}{,}             %
\makeatletter
  \newcommandtwoopt{\citeads}[3][][]{\href{http://adsabs.harvard.edu/abs/#3}%
    {\def\hyper@linkstart##1##2{}%
     \let\hyper@linkend\@empty\citealp[#1][#2]{#3}}}
  \newcommandtwoopt{\citepads}[3][][]{\href{http://adsabs.harvard.edu/abs/#3}%
    {\def\hyper@linkstart##1##2{}%
     \let\hyper@linkend\@empty\citep[#1][#2]{#3}}}
  \newcommandtwoopt{\citetads}[3][][]{\href{http://adsabs.harvard.edu/abs/#3}%
    {\def\hyper@linkstart##1##2{}%
     \let\hyper@linkend\@empty\citet[#1][#2]{#3}}}
  \newcommandtwoopt{\citeyearads}[3][][]%
    {\href{http://adsabs.harvard.edu/abs/#3}
    {\def\hyper@linkstart##1##2{}%
     \let\hyper@linkend\@empty\citeyear[#1][#2]{#3}}}
\makeatother

\newcommand{\rbh}{\mathbf{R}_\bullet}
\newcommand{\vbh}{\mathbf{v}_\bullet}
\newcommand{\Mbh}{M_\bullet}

\newcommand{\fv}[1]{f_{\mathbf{x}}(\mathbf{v}_{#1})}
\newcommand{\Dirac}[1]{{\cal D}(#1)}
\newcommand{\K}{{\cal K}}

\newcommand{\lstar}{l_\star}
\usepackage{amsmath} 
\usepackage{amssymb}
\usepackage{bm}
\usepackage{tikz}
\usetikzlibrary{calc}
\usetikzlibrary{shapes}
\tikzset{>=latex} 
\usepackage{xcolor}
\usepackage[]{circledsteps}
\pgfkeys{/csteps/inner color=white}
\pgfkeys{/csteps/outer color=black}
\pgfkeys{/csteps/fill color=black}
\colorlet{veccol}{orange!90!black}
\colorlet{myblue}{blue!60!black}
\tikzstyle{vector}=[->, very thick, veccol]

\def\N{9}
\usepackage[utf8]{inputenc} 
\usepackage{mathrsfs}  

\newsavebox{\maboiteA}
\savebox{\maboiteA}
    { 
     \begin{minipage}{0.5cm} \noindent\textit{\normalsize{A)}\\ \vspace{9.95cm} } \end{minipage}
     }  
\newsavebox{\maboiteB}
\savebox{\maboiteB}
    { 
     \begin{minipage}{0.5cm} \noindent\textit{\normalsize{B)}\\ \vspace{9.95cm} } \end{minipage}
     }

\newcommand{\myfm}[1]{\mbox{$#1$}}
\def\spose#1{\hbox to 0pt{#1\hss}}	
\def\ltabout{\mathrel{\spose{\lower 3pt\hbox{$\mathchar"218$}} 
     \raise 2.0pt\hbox{$\mathchar"13C$}}}
\def\gtabout{\mathrel{\spose{\lower 3pt\hbox{$\mathchar"218$}}
     \raise 2.0pt\hbox{$\mathchar"13E$}}}

\newcommand{\unit}[1]{\ifmmode \:\mbox{\rm #1}\else \mbox{#1}\fi}
\newcommand{\ze}{\ifmmode \mbox{z=0}\else \mbox{$z=0$ }\fi }

\newcommand{\clump}[1]{clump \textsc{#1}}
\newcommand{\amuse}{Am$\mu$se}
\newcommand{\lbh}{l_\bullet}
\newcommand{\boldv}[1]{\ifmmode \mbox{\boldmath $ #1$} \else 
 \mbox{\boldmath $#1$} \fi}
\newcommand{\mone}{\myfm{^{-1}}}
\newcommand{\half}{\myfm{\frac{1}{2}}}

\def\nb1{{\sf NBODY1} }

\newcommand{\rmd}{\ifmmode \:\mbox{{\rm d}}\else \mbox{ d}\fi }
\newcommand{\rmD}{\ifmmode \:\mbox{{\rm D}}\else \mbox{ D}\fi }

\newcommand{\rhs}{right-hand side }
\newcommand{\wrt}{with respect to }

\newcommand{\eg}{{\em e.g.\/ }}

\newcommand{\ie}{{\em i.e.\/ }}

\newcommand{\kms}{\unit{km~s\mone}}

\newcommand{\pc}{\unit{pc}}
\newcommand{\kpc}{\unit{kpc}}

\newcommand{\cm}{\unit{cm}}

\newcommand{\solar}{\myfm{_\odot}}
\newcommand{\solarm}{\unit{M\solar}}

\newcommand{\rate}[2]{\myfm{\langle\Delta\mathbf{#1}_{#2}\rangle} }
\newcommand{\rates}[2]{\myfm{\langle(\Delta\mathbf{#1}_{#2})^2\rangle} }

\newcommand{\dgr}{\myfm{^\circ} }

\nolinenumbers

\begin{document}

   \title{Dynamical traction and black hole orbital migration\thanks{Dedicated to the memory of our colleague T. Padmanabhan.}}
   \titlerunning{Dynamical traction and black hole orbital migration}
   \authorrunning{Boily et al.}
   
   \subtitle{I. Angular momentum transfer and a fragmentation-driven instability}

   \author{C.M. Boily\fnmsep\thanks{Corresponding author}
          \inst{1},  T.L. François\inst{1,2},  J. Freundlich\inst{1}, F. Combes\inst{3,4},
         A.-L. Melchior\inst{3}, Y. Hénin\inst{1,5}
          }
   \institute{Observatoire astronomique de Strasbourg, Université de Strasbourg, CNRS UMR 7550, 10 rue de l'Université, Strasbourg F-67000     \email{christian.boily@astro.unistra.fr}
         \and
         School of Mathematics and Physics, University of Surrey, Guildford, Surrey GU2 7XH, U.K.
         \and
          LERMA, Sorbonne Université, Observatoire de Paris, Université PSL, CNRS, F-75014, Paris, France
          \and
          Collège de France, 11 Place Marcelin Berthelot, 75005 Paris, France
          \and
          Télécom Physique Strasbourg, Pôle API, 300 Bd Sébastien Brant, F 67412 Illkirch Graffenstaden Cedex, France
             }

   \date{Received November 15, 2024; accepted July 7, 2025}

  \abstract
   {Observations with the {James Webb Space Telescope} of massive galaxies at red-shift $z$ up to $ \sim 14$, many with quasar activity, require 
   a careful accounting of the orbital migration of seed black holes to the heart of the host galaxy on time-scales of $300 \unit{Myr}$ or less. }
  {We investigate the circumstances which allow a black hole to remain put  at the system barycentre when the equilibrium galactic stellar core is  anisotropic.} 
   {A  Fokker-Planck treatment is developed to analyse  the migration of a massive black hole which focuses on exchanges of orbital angular momentum with the stars. 
   We further use a set of  N-body calculations to study the response of stellar orbits 
   drawn from  a Miyamoto-Nagai disc embedded in a larger, isotropic isochrone (H\'{e}non) background potential.  }
   {When the black hole has little angular momentum initially, but orbits in a sea of stars drawn from an odd $f[E,L_z]$ velocity distribution function, a wake in the stellar density  sets in which pulls on the black hole and transfers angular momentum to it. 
   We call this  {dynamical traction} in contrast with the more familiar 
   Chandrasekhar dynamical friction. We argue that 											  
   this phenomenon takes place whenever the kinetic energy drawn from $f[E,L_z]$ has an excess of streaming motion over its (isotropic) velocity dispersion.  We illustrate this process for a black hole orbiting in a dynamically warm disc with no sub-structures. 
    We then show for a dynamically  cold disc that the outcome depends on both the orbit of the black hole and that of the stellar sub-structures stemming  from a Jeans instability. 
       When the stellar 
       clumps have much binding energy, a black hole may scatter off 
   of them after they formed. In the process the black hole may be dislodged from the centre and migrate outward
    due to dynamical traction.  When 
   the stellar clumps  are less bound, they may still 
   migrate   to the centre where they either dissolve or  merge with the  black hole. The final configuration is similar to a nuclear star cluster which may yet be moving at $\sim 10 \kms $ \wrt\ the barycentre of the system.  }
{ 
The angular momentum transferred to a black hole by dynamical traction delays the migration to the galactic centre by several hundred million years.
The efficiency of angular momentum transfer is a strong function of the fragmented (cold) state of the stellar space density.
In a dynamically cold environment, a  black hole is removed from the central region through  a two-stage orbital migration instability. A criterion against this instability  is proposed  in the form of a threshold in isotropic velocity dispersion compared to streaming motion (i.e., angular momentum). 
For a black hole to settle at the heart of a galaxy on time-scales of $\sim 300 \unit{Myr}$ or less requires that a large fraction of the system angular momentum be dissipated, or, alternatively, that the black hole grows {in situ} in an isotropic environment devoid of massive sub-structures.}

\maketitle
\nolinenumbers 
%
\section{Introduction}
This article investigates the orbital evolution of a black hole moving freely at the heart of a galaxy in the later stages of a major merger. The motivation 
for this work stems from recent observations by the {James Webb Space Telescope} (JWST) of galaxies in the mass 
range $10^9$ to $10^{10}$ solar-masses at cosmological red-shifts up to $z \simeq 12$ (\citealp{naidu2022, labbe2023, CEERS2023}). The short $\lesssim 500$ Myr 
time-scales implied  challenge standard models of the formation of galaxies and super-massive black holes (SMBH's: see e.g. \citealp{galaxyformation2010, boylan2023, matthee2024, barro2024} ) 
because they point to a high level of efficiency of the formation processes. For example, the rapid growth of SMBH's on a time-scale of a few  hundred million years 
by accretion or coalescence suggest that seed BH's of $10^4\, M_\odot$ or more form already at the epoch of recombination \citep{seedBH2024}. In these scenarios, the growth of a single BH by gas accretion at the Eddington limit, or by the stochastic  shredding and loss-cone capture of main-sequence stars,  is made more likely 
by placing the BH in the pit of the (local) gravitational well, where gas and stars are accreted on shorter time-scales 
(\citealp[e.g.][]{galaxyformation2010}; \citealp{schawinski2015}; \citealp{bhgrowth2021}; \citealp[see also the review by][]{volonteri2012}).  But this basic picture bypasses the inevitable mass build-up of the  host galaxy through multiple mergers: 
at the same time as a BH accretes 
from its immediate surroundings, it drifts in space in a time-dependent galactic potential: several recent studies have included BH wandering in their dynamical modelling (\citealp[e.g.][]{bellovary2019, pfister2019, lapiner2021}). Bright cluster galaxies in the Illustris-TNG cosmological simulations point to SMBHs wandering over time-scale of order $~ 1 \unit{Gyr}$  (\citealp{chu2023}). 
The expectation that an SMBH will come to settle at the barycentre of the host galaxy rests on the theory of dissipation of orbital energy by dynamical friction (\citealp{chandrasekhar1943, BT08}), and from observations of local-Universe AGN galaxies where it almost invariably sits at the photometric centre of the galactic bulge. The Milky Way's SMBH, of a relatively low mass $(\simeq 4\times 10^6\solarm$; \citealp[see e.g.,][]{chatzopoulos2015}), is a prime example of this configuration.  
Yet the recent detection of an off-centred SMBH at $z \approx 7.15$ by \citet{ubler2024} casts this expectation in the spotlight and begs for a reassessment of the conditions that enable an SMBH to stay at rest at the heart of a young galaxy.  
We ask how an SMBH can remain at the barycentre of a cored galaxy when it is surrounded by a disc of stars (\ie, stars supported by angular momentum). That configuration is designed to  represent the final stage of an in-falling SMBH 
carried by a major (dry) galaxy merger in an advanced stage, with 
much orbital angular momentum still present in the surrounding stars (see for example \citealp{chapon2013}). 
Thus this set-up 
complements earlier work exploring SMBH wandering in an isotropic background of stars \citep{devita2018,merritt2013}.  
Our focus is on the orbital evolution of an SMBH accreted by a young galaxy with  an anisotropic flow of stars with net rotation (a proto-galactic disc). 
We view the SMBH as an external $m = 1 $ perturbation mode acting on a system in near-equilibrium. 
The expectation for an accretion event carrying little angular momentum is that it will 
pick up angular momentum from background stars (the perturber has a slow pattern speed: \citealp{TW1984}):
 this will slow down its migration to the centre.  
We focus on the case of an SMBH orbiting in the same plane as the proto-galactic disc, with the  goal  to quantify its 
orbital  migration under gravity alone (no gas dissipation).{  Since gravitational dynamics is scale-free,  hereafter we 
will use the acronym BH as  a generic term.  }

Our study should be put in the perspective of earlier works. 
In a system with periodic orbits and a flat (cored) stellar mass density, all orbits have  periods in the  ratio $\simeq 1:1$. The 
{break down of resonant motion}  between the perturber and the background stars {near the system barycentre}  leads to {core-stalling}  (\citealp[e.g.][]{read2006,boily2008,cole2012,kaur2018,banik2022}), when dynamical friction shuts off entirely (\citealp[see also][for orbit-based analyses]{petts2015,banik2021,chiba2024}). This leaves the orbital energy and angular momentum of a massive perturber unchanged. 
\cite{kalnajs1972} had shown that dynamical friction in a self-gravitating disc of 
uniform angular speed $\Omega$ vanishes everywhere. His analysis of global modes of perturbation extends to individual 
orbits, when the disc is fully phase-mixed: the perturbation of stellar orbits of any energy by a massive BH leads to positive 
and negative torques acting on the BH as the stars respond to the perturbation (the BH was dubbed a {catalyst} of energy exchange between stars in \citealp{boily2008}). Recently, \citet{banik2022} and \citet{kaur2022} showed that the exact cancellation of gravitational torques depends on details of the phase-space distribution function, $f$.  The general conclusion is that dynamical friction  will be much reduced in any realistic  near-constant density environment with the BH on a circular orbit. 

The case of a rotating system with disc-like ordered motion is different because of the strongly anisotropic 
velocity field. The response of the stars to an in-falling  BH leads to the growth of density patterns, as shown first 
by \citet{JT1966} where a first-order perturbation analysis developed  exponential growth in the sheared-box approximation. 
In a galactic 
disc in equilibrium, spiral density waves cause mechanical torques which redistribute angular momentum throughout the disc
\citep[\eg][]{LBK1972,TW1984,hopkins2011, alcazar2015}. \cite{TW1984} performed a general  analysis based on resonant orbits and argued that {friction} originates from global resonant modes exerting a net torque on a massive body.  The response of the stars 
remains correlated in phase-space to the perturber's orbit, which is then  said to be {dressed} by these response modes 
(see \citealt{weinberg1998}; \citealt{fouvry2015}; and especially \citealt{lau2021} for a detailed analysis). A strong response by 
the stars leads to polarised orbits: the self-gravity of the perturbed stars triggers  non-linear effects (modes couple to each 
other) which are difficult to track analytically. Indeed we will show with numerical examples how a fragmented (dynamically cool) 
stellar disc has a much different impact on the BH than a warm one. 

We review the basics of dynamical friction in \S\ref{sec:dynamicalfriction}, 
before developing our perturbation analysis based on a Fokker-Planck treatment in \S\ref{sec:theproblem}. Having got some analytical compass, we then 
move on to present a series of restricted N-body models which detail the time evolution of the BH's perturbed orbit (\S\ref{sec:nbody}). 
We show in \S\ref{sec:results} that the orbital migration of the BH triggers the growth of $m = 1$ and even $m = 2$ (bar) modes in the stars. 
For such modes, further evolution of the orbit through gravitational torques becomes possible. {A brief quantitative comparison of the 
numerical results with theoretical expectations is presented in \S\ref{sec:theory}. }
We discuss the implication of the perturbed configurations for the short-term accretion history of the BH in \S\ref{sec:discussion}. 

   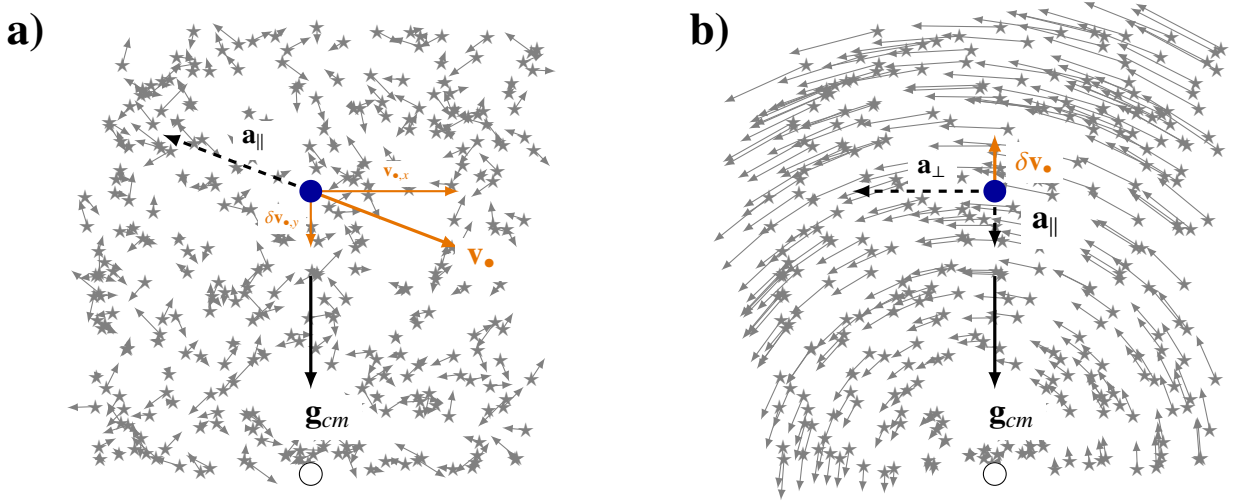
\begin{figure*}
   \hspace{10mm} 
   \begin{tikzpicture}[scale=1.5]
  \def\N{300}
  \foreach \i in {1,...,\N}{
  \def\xa{2*rand} 
  \def\ya{2*rand+2}
  \coordinate(A\i) at (\xa, \ya);
  }
  \foreach \i in {1,...,\N} {
  \draw[->,domain=-2.2:2.2,color=gray]  (A\i) node {$\star$}  -- ++(0.25*rand,0.25*rand) ;
  }
  \draw[->, color=veccol, very thick, style = solid, domain=-2.2:2.2 ] (0,2.5) -- (1.30,2.0) ;
  \draw [domain=-2.2:2.2, color=veccol ] (1.50, +1.90)  node[scale=1.25,fill=white] {$\vbh$ } ;
  \draw[->, color=veccol, thick, style = solid, domain=-2.2:2.2 ] (0,2.5) -- (0.,2.0) ;
  \draw [domain=-2.2:2.2, color=veccol ] (-0.25, +2.25)  node[scale=0.70, fill=white] {$\delta\mathbf{v}_{\bullet,y}$ } ;
    \draw[->, color=veccol, thick, style = solid, domain=-2.2:2.2 ] (0,2.5) -- (1.3,2.5) ;
  \draw [domain=-2.2:2.2, color=veccol ] (+0.75, +2.65)  node[scale=0.70, fill=white] {$\mathbf{v}_{\bullet,x}$ } ;
  \draw[->, very thick, style=dashed,  domain=-2.2:2.2 ] (0,2.5) -- (-1.30,3.0) ; 
  \draw [domain=-2.2:2.2] (-0.50, +2.95)  node[scale=1.05, fill=white] {$\mathbf{a}_\parallel$ } ;
  \draw[->, very thick, style=solid,  domain=-2.2:2.2 ] (0,1.75) -- (0,0.75) ; 
  \draw [domain=-2.2:2.2] (0.15, 0.50)  node[scale=1.25, fill=white] {$\mathbf{g}_{cm}$ } ;
  \fill[myblue] (0,2.5) circle (0.100);
  \fill[white] (0.0,0) circle (0.105);  
   \draw[black] (0.0,0) circle (0.100);  
  \node[font=\bfseries, scale=1.7] at (-2.5,3.9) {a)};
\hspace{90mm} 
\newdimen \Rx ; \newdimen \Ry ;
 \foreach \i in {1,...,\N} {
 \pgfextractx\Rx{\pgfpointanchor{A\i}{center}} 
 \pgfextracty\Ry{\pgfpointanchor{A\i}{center}} 
  \draw[->,domain=-2.2:2.2,color=gray]  (A\i)  node {$\star$}  --  (\Rx - 0.20*\Ry, \Ry + 0.20*\Rx ) ;
  }
   \draw[->, color=black, very thick, style = dashed , domain=-2.2:2.2 ] (0,2.5) -- (0.,2.0) ;
   \draw [domain=-2.2:2.2, color=black] (0.45, +2.2)  node[scale=1.25, fill=white] {$ \mathbf{a}_\parallel $ } ;
   \draw[->, very thick, style=dashed,  domain=-2.2:2.2 ] (0,2.5) -- (-1.25,2.5) ; 
   \draw [domain=-2.2:2.2] (-0.55, +2.70)  node[scale=1.05, fill=white] {$\mathbf{a}_\bot$ } ; 
       \draw[->, color=veccol, very thick, style =solid , fill=white, domain=-2.2:2.2 ] (0,2.5) -- (0.,3.0) ;
       \draw [domain=-2.2:2.2, color=veccol  ] (0.35, +2.75)  node[scale=1.05, fill=white] {$\delta\vbh $ } ;
   \draw[->, very thick, style=solid,  domain=-2.2:2.2 ] (0,1.75) -- (0,0.75) ; 
   \draw [domain=-2.2:2.2] (0.15, 0.50)  node[scale=1.25, fill=white] {$\mathbf{g}_{cm}$ } ;
   
   \fill[myblue] (0, 2.5) circle (0.100);
  
   \fill[white] (0.0,0) circle (0.105);  
   \draw[black] (0.0,0) circle (0.100);  
  \node[scale=1.7, font=\bfseries] at (-2.5,3.9) {b)};
\end{tikzpicture}

\caption{Sketch of gravitational accelerations for two different situations. The panels show a massive body (blue dot) orbiting among stars, where the system's barycentre 
at the bottom is the focus of  the global gravitational acceleration $\mathbf{g}_{cm}$ (open circle).  The mean 
field acceleration gives rise to a velocity component $\delta v_\bullet$ along the vertical y-axis. 
On panel (a), a massive perturber moves at some velocity $\vbh$ in an isotropic field of stars; the result of the polarisation of the stellar orbits 
is a net dynamical {friction} force  $\mathbf{a}_\parallel$, parallel but in the opposite sense to $\vbh$.  On panel (b), the  
massive body is ejected from the origin at a velocity 
$\delta\vbh$ pointing upwards. It orbits  in a strongly anisotropic (rotating) stellar field and slows down due to the steady negative acceleration  $\mathbf{g}_{cm}$. However, dynamical friction now contributes two terms, one  $\mathbf{a}_\parallel$ as before, and a second one,  $\mathbf{a}_\bot$, orthogonal to $\delta\vbh$. That non-zero component due to the streaming stars 
causes a steady transfer  of angular momentum to the perturber.  We refer to this second component as dynamical {traction} acting on the 
perturber. } \label{fig:Cartoon}
\end{figure*}

%
\section{Dynamical friction}\label{sec:dynamicalfriction}
The main driver of BH orbital evolution is the dynamical 
friction exerted by low-mass field stars (and gas) on the BH; the density wake forming behind 
a BH of mass $M_\bullet$ moving at $\mathbf{v}_\bullet$  causes a frictional acceleration of the BH
(\citealt{chandrasekhar1943} ; \citealt[\S 8.1]{BT08}) 

\begin{equation}
\frac{\mathrm{d} \vbh}{\mathrm{d} t } = \langle\Delta\vbh\rangle_\parallel = -  \frac{\Gamma}{\pi} \,\frac{\Mbh}{m_\star}\,
\int \mathrm{d}^3\mathbf{v}\, f(\mathbf{v}) \frac{\vbh - \mathbf{v}}{| \vbh - \mathbf{v}|^3}\, , \label{eq:chandrasekhar}
\end{equation}
where $M_\bullet, \vbh$ are the black hole's mass and velocity; $\mathbf{v}$ the background stars velocity of phase-space density $f(\mathbf{v})$, and the 
integral is performed over all bound stars (orbital energy $E < 0$). The Coulomb term reads $\Gamma = 4\pi\, G^2 m_\star^2\, \ln \Lambda $   with 

\[ \Lambda = \frac{p_{max}}{p_{90^\circ}} \simeq 0.3 N \, ,\]
where $p$ is the impact parameter evaluated for a hyperbolic stellar orbit in the star-BH barycentre, and $\Lambda$ is the ratio of the maximum value to the 
value leading to a $90^\circ$ deflection angle \citep[e.g.][]{spitzer1987, heggie03}. The last relation to the number of stars $N$ applies for a system of bounding volume $V$ and uniform mean mass density, $\rho_\star = N m_\star / V$.  

As a consequence of Eq.~\ref{eq:chandrasekhar}, when the field stars are distributed isotropically in velocity space (\ie, when $f[\mathbf{v}] = f[v]$ is a function of the norm $v$ alone) the net deceleration runs parallel to $\vbh$ as depicted on 
Fig.~\ref{fig:Cartoon}(a). The steady loss of kinetic energy brings the massive body at  the barycentre of the galaxy, 
the only stable point in equilibrium for a body at rest.  The  time interval for this to take place is estimated from 

\begin{equation}
\Delta t_{fric} = \frac{6}{5\ln\Lambda} \frac{M(<r)}{\Mbh} \times \sqrt[2]{\frac{3\pi}{G\rho_\star}} \equiv  \frac{6}{5\ln\Lambda} \frac{M(<r)}{\Mbh} \, t_{dyn} 
\label{eq:tdyn} 
\end{equation}
with the background stars of density $\rho_\star$ adding to a mass inside the orbital radius $M(<r)$ much larger than the 
massive body $\Mbh$ and we have introduced a definition of the dynamical time $t_{dyn}$. (In numerical applications the ratio $\Delta t_{fric} / t_{dyn} \simeq 3$ ; see \S\ref{sec:nbody}.)

The type of kinetic energy diffusion that results from (\ref{eq:chandrasekhar}) is drawn from 
summing over non-resonant  star-BH hyperbolic encounters. The unperturbed stellar trajectories are straight lines and vanish 
to infinity after the encounter with the perturber. In reality the host galaxy spawns a finite volume and stars with similar  
orbital energy visit each other and the perturber BH repeatedly: this will cause resonant coupling between 
the orbits when their period ratio is commensurate \citep{BT08}. Such resonant response leads to efficient energy exchanges and 
the notion of {buoyancy} and core-stalling \citep{read2006,banik2022,kaur2018} ; this will be revisited briefly in \S\ref{sec:discussion}.
In the next 
paragraph \S\ref{sec:theproblem} , we outline the coupling between the BH orbit and the background stars based on the non-resonant Fokker-Planck treatment. 

%
\section{Position of the problem}\label{sec:theproblem}
To achieve our goal, several simplifying assumptions are required.  
The first is that a well-defined galactic rotation pattern is in place. We also presume that the BH  has already bound to the host 
 system (no dissipative terms, no gas) and remains embedded in the rotating flow of stars. 
 Also, the BH is a small fraction of the total mass (less than 1\% of the stellar disc, but $\approx 0.1 \%$ of the system, in line with observational trends ; \citealp{kormendy2013}) and much more 
 massive than the background stars. The problem then is to identify a small set of initial conditions for the BH's orbit 
 that are idealised yet useful to our analysis. 

 \subsection{Orders of magnitude calculation}
Let a BH sit at rest at the origin of coordinates surrounded by a spheroid of stars drawn from a distribution 
function $f(E,L_z)$, where $E$ is the energy and $L_z$ the z-component of the angular momentum. 
For simplicity, we lump together the {undressed} BH mass and the stars strongly bound 
to it  (when stellar orbits evolve adiabatically in the BH potential, \citealp[\eg,][\S2.3]{BT08, spitzer1987}).  
The stars give rise to an axi-symmetric potential $\Phi(R,z)$ and the whole system is in equilibrium (à la \citealt{tremaine1994}).  In the context of a young galaxy in an advanced stage of formation, some sub-structures of significant mass may still induce time-dependant  perturbations in the global potential.  
We set the BH in motion through an $m = 1$ perturbation mode, which we model as a hyperbolic (shock) encounter 
with a body of mass $M_p$, impact parameter $p$, passing by the BH at velocity $V_p$ parallel to the z-axis  \citep[\S5.2]{spitzer1987}.  The velocity imparted to the BH is then 

\begin{equation}
	 \left\| \delta\vbh \right\| \simeq \frac{2GM_p}{p V_p}   \label{eq:perturbation}
 \end{equation}
with a velocity vector $\delta\vbh$ which we align with the y-axis of a Cartesian frame of reference (cf. Fig.~\ref{fig:Cartoon}[b]).  
When the (dynamically hot) spheroid of stars spawns a uniform density $\rho_\star$, orbital motion is flagged with a reference angular frequency $\omega = \sqrt{4\pi G\rho_\star / 3} $ and the amplitude of the BH orbit  can be estimated from the solution for a harmonic oscillator, 

\newcommand{\xbh}{\mathbf{x}_\bullet}
\begin{equation}
\delta \xbh \simeq  \frac{\delta\vbh}{\omega} \, . 
\label{eq:dxBH}
\end{equation}
The impact parameter $p$ and  relative velocity $V_p$ in Eq.~(\ref{eq:perturbation}) are free to choose.  The  mass $M_p$ however can be constrained by requiring that the 
perturbing  body survives the encounter. For an extended body (such as a clump of gas or a star cluster), this will be 
achieved if $M_p$ exceeds the local Jeans mass \citep[\S3]{BT08}, 

\begin{equation}
M_p \ge \frac{4\pi}{3} \rho_\star \left(\frac{\lambda_J}{2}\right)^3 =  \frac{\pi^\frac{5}{2}}{6} \, \left( \frac{ \sigma_\star}{\sqrt{G\rho_\star}} \right)^3 \rho_\star   \equiv M_J\, .  \label{eq:Jeansmass}
\end{equation}
In this equation the length $\lambda_{J}$ is the Jeans wavelength and $\sigma_\star$ the stars' one-dimensional velocity dispersion. 
A massive BH will generate a strong tidal field in a volume where it contributes much of the mass; furthermore,  the impact parameter $p$ should be greater than the size of the perturbing 
body $M_p$, otherwise it could overlap with the BH at closest approach. For these reasons, in numerical applications we will set

\begin{subequations} 
\begin{equation} p \gtrsim \max\left\lbrace \frac{\lambda_J}{2}, \lbh\right\rbrace\, ;\end{equation} \vspace{-5mm}
\begin{equation}   \lbh \equiv \left[ \frac{3}{4\pi}\, \frac{\Mbh}{\rho_\star} \right]^{\,\displaystyle{\frac{1}{3}}}\, ,\label{eq:lbh} \end{equation}
\end{subequations} 
 for self-consistency. The length $\lbh$ is sometimes referred to as the BH radius of influence (as in \citealp{merritt2004} up to a factor 2; $\lbh$ should not to be confused with the event horizon of relativistic gravitation). 

We consider  a naked BH mass but 
include all stars with binding energy (to the BH) exceeding the energy perturbation applied 
to the BH itself. For stars in a quasi-Keplerian orbit around the BH, 
this rough approximation implies that the BH mass includes 
stars confined to a \citet[]{bahcallwolf} cusp where $\rho_\star \propto r^{-7/4} $.  We will discuss the implications of this 
simplification in the closing section (\S\ref{sec:conclusions}). Below we derive expectations from a Fokker-Planck treatment of the diffusion of kinetic energy. Mathematical details can be found in Appendix~\ref{sec:FP}. 

\subsection{Sample distribution function}
A Fokker-Planck treatment requires a choice of phase-space distribution function $f(\mathbf{x}, \mathbf{v})$.  
{We apply a Dirac $\delta$-function to select loop orbits otherwise  drawn randomly  from  $f$. 
 The distribution function (d.f.) takes the form }

\begin{equation}
 \left. f( \mathbf{x}, \mathbf{v}) \right|_\mathbf{x} \equiv  \fv{\star} = f_\bot( \mathbf{v}_\bot) \,\, \Dirac{ v_\phi - R\Omega} \label{eq:df}
\end{equation}
where $R, \phi, z$ are cylindrical coordinates; $\Omega(R,z=0)$ the local angular frequency;  $\Dirac{v}$ is the one-dimensional Dirac operator \citep[with units of $v^{-1}$; see \eg][for further applications of the Dirac operator]{fukushige2000, renaud2011}. The $R, z$  components of the velocity field are made to be irrotational and isotropic by fixing 

\begin{equation}
f_\bot (  \mathbf{v}_\bot) = \frac{\rho(R,z)/ m_\star}{2\pi\,\sigma^2_\star} \, \exp\left( - \displaystyle{ \frac{v^2_\bot}{2\sigma_\star^2} } \right) \, .\label{eq:df_ortho}
\end{equation}
We  set the azimuthal velocity component of the BH, $v_\phi$, equal to the parallel component $v_\|$ in Eqs.~(\ref{eq:diffrates}a-c), so 
\[ v_\phi \simeq v_\| \]
throughout.  This will prove sufficiently accurate to recover the evolution  with the restricted N-body numerical models of \S\ref{sec:nbody}. 

With the definition 
\begin{equation}
        \K^2 \equiv  ( R\Omega - v_\phi)^2 + v_\bot^2 
\end{equation}
and the help of Eqs.~(\ref{eq:diffrates}),  (\ref{eq:dKdv})  and the relations (\ref{eq:df}) and (\ref{eq:df_ortho}), 
we find 
\begin{equation}
\langle \Delta v_\phi \rangle = \Gamma \frac{\Mbh}{m_\star}\frac{v_\phi}{v}\, f_\bot (0) \left[ \mp 2\pi + x e^{x^2} \, \mathrm{erfc}(x) \right] \times \frac{\mathrm{d}\K}{\mathrm{d}v} \label{eq:rate_vpar}
\end{equation}
where $ x \equiv \K \sqrt{\pi/2\sigma_\star^2} $ and erfc($x$) is the complementary error function.  The key feature of (\ref{eq:rate_vpar}) is that the contribution to the variation in kinetic energy $v_\phi \, \langle \Delta v_\phi \rangle$ changes sign 
with $R\Omega - v_\phi$  (see [\ref{eq:Ediffusion2}]). It is the exact same feature of gain or loss of kinetic 
energy depending whether the orbit trails or leads a resonant density pattern as found in earlier studies (\citealt{TW1984}; \citealt{rauch1996}; \citealt{weinberg1989}; \citealt{kocsis2015}). We have not treated the growth of resonant features here, only the response of the background stars and its impact on the BH's orbit. 
%

\begin{figure*}[h]
\usebox{\maboiteA{}} 
   \centering 
   \includegraphics[width=0.45\textwidth]{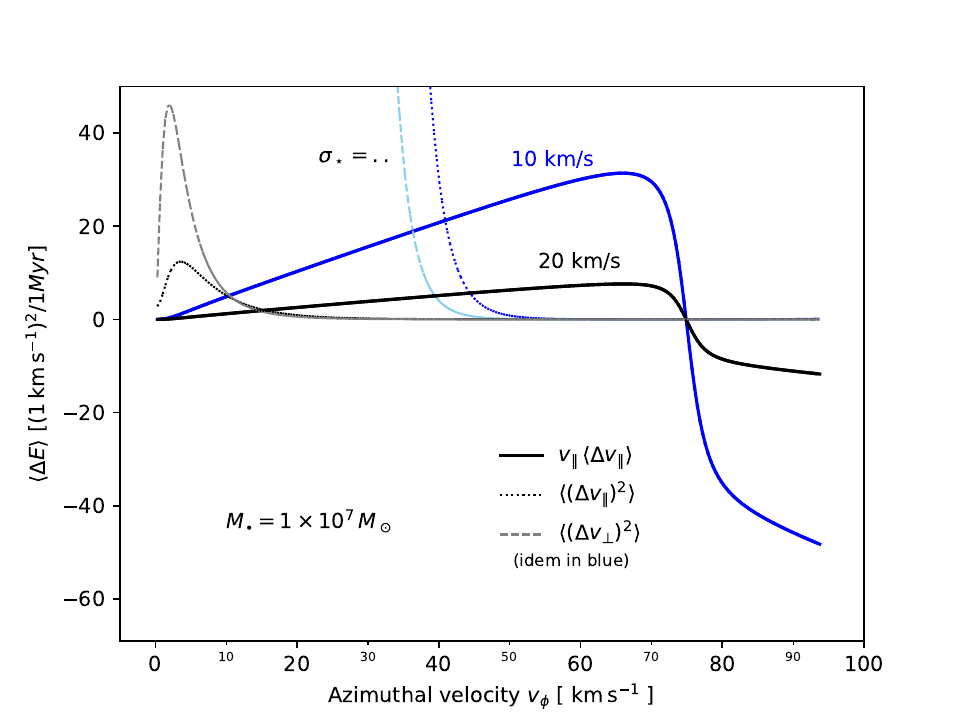}
   \usebox{\maboiteB{}} 
     \includegraphics[width=0.45\textwidth]{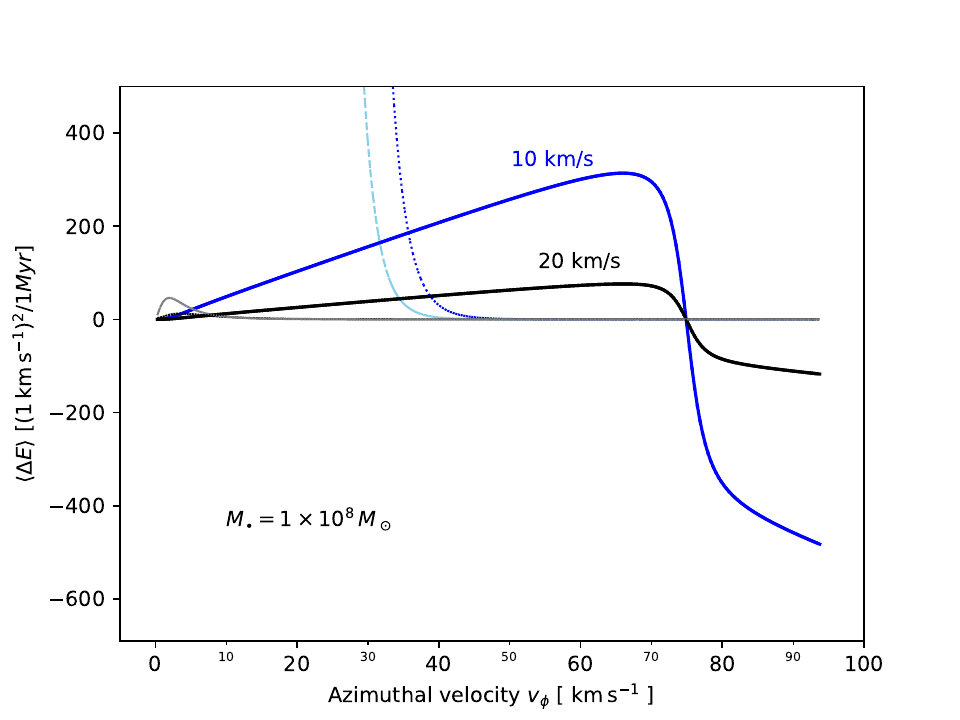}
     \vspace{-4.95cm}
     \caption{Rates of specific kinetic energy diffusion, $\langle\Delta E \rangle$ as a function of the  azimuthal velocity $v_\phi$. The rates are evaluated from Eqs.~(\ref{eq:Ediffusion2}) and (\ref{eq:diffrates})  for two values of the stellar velocity dispersion, $\sigma_\star: 20 \kms $ (in black or grey); and $10\kms$ (in blue or skyblue).  A)  Setting $\Mbh = 1 \times 10^7 \solarm$,  with $\Mbh / m_\star \simeq 10^3$, the figure graphs each quadratic component and the first-order parallel one ( $v_\|\,\langle\Delta v_\|\rangle$, dark solid   lines;   note that it  has the same  sign as $R\Omega - v_\phi$). B) The same but for an BH with $\Mbh= 1\times 10^8 \solarm$ ($\Mbh/m_star \simeq 10^4$).  The much 
     higher inertia of the BH leads to a ten-fold increase in the parallel term but only subtle differences in the diffusion (quadratic) ones. 
     The light- and dotted-blue curves peak off scale in either case. 
      The (central) angular velocity $\Omega = 0.05 / \unit{Myr} $ is such that  $R\Omega \approx 70 \kms $ at $R = 1.5 \kpc$, a good match to the circular velocity at that radius.}
        \label{fig:Ekrates}        \label{fig:Ekrates100M}
\end{figure*}

\subsection{The transition to dynamical traction}\label{sec:traction}
Figure~\ref{fig:Ekrates} plots the  rates of specific kinetic energy variation entering the \rhs of (\ref{eq:Ediffusion2})  as function of the azimuthal velocity component, $v_\phi$. For the reference calculations displayed on Fig.~\ref{fig:Ekrates}(a), the Coulomb potential was set to 
  $\ln \Lambda = 21\, (N = 10^{5}) $ and  $m_\star = 1 .6 \times 10^4 \,\solarm,\ \Mbh = 1\times 10^7\,\solarm  $ ; the rotational velocity    $R\Omega \simeq  70 \kms$ at cylindrical radius $R = 1 .5\kpc$ fixes the angular frequency.    Two sets of results are shown, obtained for a stellar  velocity dispersion $ \sigma_\star = 10 \kms$ (in blue) and $20 \kms$ (in black). 
   
The high sensitivity to $\sigma_\star$ is clearly visible for the quadratic terms $\rates{v}{\bot}$ and $\rates{v}{\|}$. 
With $\sigma_\star = 10\kms$,  both curves rise steeply at low $v_\phi$, and reach a maximum value well off-scale on the figure.  Thus at low $v_\phi$, 
the gain or loss of kinetic energy is completely dominated by random scattering. The expectation, therefore, is that unless 
the BH 
is on an orbit with a significant azimuthal velocity, it is unlikely  to acquire further streaming motion. Instead, a gain in radial- and vertical  velocity would set it on  a new orbit of increased eccentricity (no gain in angular momentum) at a different energy, $E$. The turnover from a gain in radial- to one in azimuthal 
kinetic energy occurs at $v_\phi \approx 45 \kms$, when the rate $v_\|\,\rate{v}{\|} \simeq 20 (\kms)^2 / \unit{Myr} $ 
reaches a much larger amplitude than the quadratic terms. 
By contrast, when the velocity dispersion $\sigma_\star = 20 \kms$, the rates $\rates{v}{\bot}$ and $\rates{v}{\|}$ peak at a much lower 
value of $v_\phi \approx 6 \kms$, and decrease rapidly thereafter. The gain in azimuthal kinetic energy becomes  dominant at $v_\phi \approx 20 \kms$ and larger values, which translates in a net gain in azimuthal motion from that point on-ward. This behaviour is expected because a warmer stellar population will be harder to polarise, and so the effect of random scattering off of the BH orbit is much reduced (less effective gravitational focusing). 
The streaming flow however remains the same, so that the response in azimuth is much less affected by variations in $\sigma_\star$.  We say that the gain in angular momentum is triggered by a {\it dynamical traction transition}\footnote{We have not shown that this transition takes place through a topological critical point in solution space, hence we will not refer to this dynamical process as a bifurcation. We thank the referee for bringing this to our attention.} because 
    for some critical value of $v_\phi$ the BH's orbit gains angular momentum systematically, transforming its orbit from  a 
    low-$L_z$ box orbit, to a high-$L_z$ loop. The on-set on this seemingly irreversible trend is a sensitive function of the velocity field of the  background stars. We cover this point more fully with applications later in \S\ref{sec:critical}. 

Another important parameter is the choice of BH mass. On Fig.~\ref{fig:Ekrates}(b), we graph the results for the same 1D velocity dispersions $\sigma_\star$, for an BH of mass $ \Mbh = 1\times 10^8\solarm$. The ten-fold increase in $\Mbh$ means that the rate $v_\|\,\rate{v}{\|}$ also increases ten times (see the multiplicative factor in [\ref{eq:Ediffusion2}]). 
The most noticeable effect is a slight shift to the left of the curves for $\rates{v}{\bot}$ and $\rates{v}{\|}$, which both cross 
the run of  $v_\|\,\rate{v}{\|}$ at lower values of $v_\phi$. For example, for the case with  $\sigma_\star = 20 \kms$ 
 the turnover to a regime where  $v_\|\,\rate{v}{\|}$ is dominant  takes place at $v_\phi \approx 12 \kms$ (it was $\approx 20 \kms$ on Fig.~\ref{fig:Ekrates}[a] for the lower-mass BH). This suggests that the more massive BH may yet gain angular momentum  from a more slowly-rotating  stellar  flow (all other quantities being unchanged). To back up this intuition, 
  we have checked that reducing the angular frequency $\Omega$ from $\approx 7.0 \times 10^{-1} \unit{Myr}^{-1}$ to $ 3.5 \times 10^{-1} \unit{Myr}^{-1}$ brings the 
 crossover value to $v_\phi \approx 20 \kms$, when $\sigma_\star = 10 \kms  $ ; and to $v_\phi  \lesssim 5 \kms$, when
  $\sigma_\star = 20 \kms$\footnote{These results were obtained with $\Mbh = 1\times 10^7 \solarm$.}. 
  The crossover value goes down by a factor of more than two from the reference calculation, highlighting the strong link with the rotation curve. The calculations with the reduced   angular speed $\Omega$ still had a circular velocity $\simeq 35 \kms$ at $R = 3/2 \kpc$, 
  significantly larger than the two values for the velocity dispersion $\sigma_\star $  that we tested. A reduced $\Omega$ implies  longer time-scales for the on-set of the dynamical traction transition (\ie, when the net gain in angular momentum will   exceeds the loss to dynamical friction, and transform the orbit of the BH to a loop orbit). 
\newline 

There is no gain or loss in azimuthal energy from dynamical traction 
 when $v_\phi = R\Omega$ (co-moving BH) because $v_\|\,\rate{v}{\|}  = 0$ there. In the event of the BH running ahead of the stars, 
the rate $v_\| \rate{v}{\|} < 0  $ changes sign. If and when that is the case, the BH {loses} kinetic energy and angular 
momentum, when it was gaining before. This is the signature of an 1:1 resonant trapping at  co-rotation. That result was obtained 
by an argument of symmetry, rather than a proper linear perturbation treatment of the stellar response. No such analysis was 
needed here because we focused on the perturber (the BH) and not the system as a whole. 

\subsection{Time evolution}
A self-consistent treatment of the orbital evolution of the BH with diffusion coefficients would require a full  2D Fokker-Planck integration. 
The configuration we are treating is far for equilibrium and will evolve on a dynamical time-scale. For that reason, we will make more 
progress with a set of restricted N-body integrations (see \S\ref{sec:nbody}). Here we focus on the linear- and diffusion 
terms of Eq.~(\ref{eq:Ediffusion2}) to compute the increase in kinetic energy of each parallel- and orthogonal degrees of freedom by treating their  independent evolution. With this rough treatment we can estimate quantitatively the time-scale over which the 
dynamical traction becomes dominant and the BH acquires significant angular momentum, $L_z$. 

The specific kinetic energy $E_k$ of the BH is a sum over quadratic components of the velocity $v^2 = v^2_\| + v^2_\bot $. We split $E_k = E_\| + E_\bot$ such that 

\begin{eqnarray}
 E_\| & = &   \frac{1}{2} v^2_\| + \frac{1}{2} \sigma^2_\| \, , \\ 
 E_\bot & = & \frac{1}{2} \sigma^2_\bot \, . 
\end{eqnarray}
We compute \eg $E_\bot$ from Eq.~(\ref{eq:Ediffusion2}) with a simple summation over time ; likewise for the two components of $E_\|$. For instance, 

\[ \sigma_\bot = \sqrt{{2} E_\bot} = \left( {2} \int_0^t \rates{v}{\bot}\, \mathrm{d} t \right)^{\frac{1}{2}}\, .  \] 
By assuming that the diffusion terms always have the same (positive) sign, we optimise the growth of kinetic energy 
orthogonally to the rotation flow. 
Since the flow of rotating stars singles out the torque acting on the BH, we argue that this approach slows down the 
transfer of angular momentum to the BH because the BH would take longer to develop a velocity vector that runs 
parallel, or nearly parallel, to the flow of stars. 

\begin{figure}
   \centering
   \includegraphics[width=0.5\textwidth]{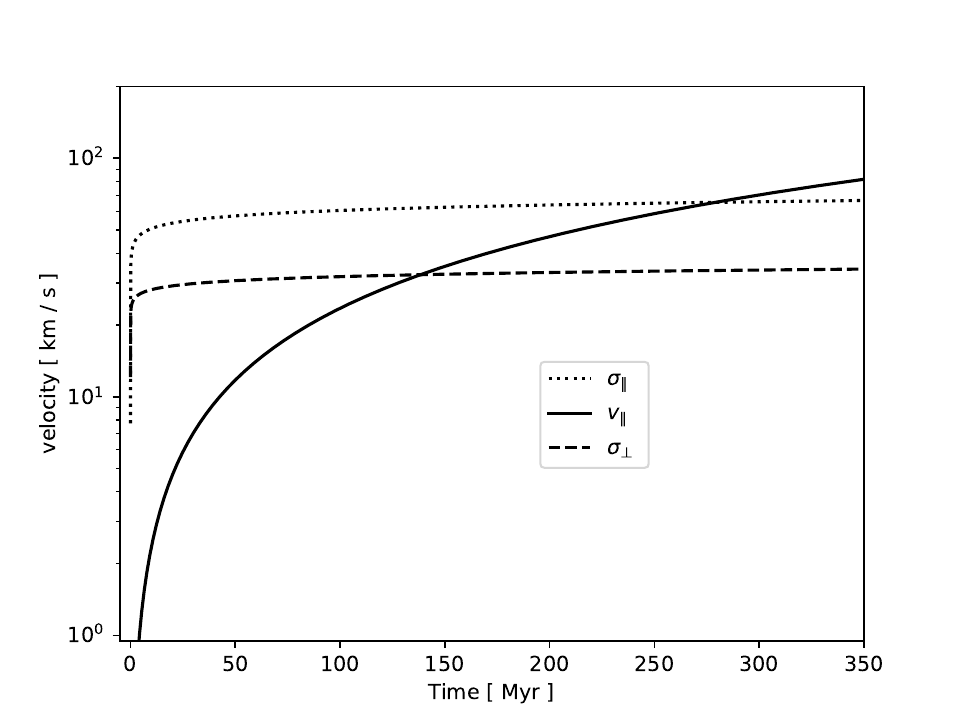}
     \caption{Amplitude of the BH velocity components as a function of time. 
     The figure graphs the velocity dispersion for each parallel- and orthogonal component (denoted as $\sigma_{\|\, \mathrm{or}\, \bot}$ in the legend)    and the (first-order integrated) streaming velocity ($v_\|$, black solid line). The period of a circular orbit at radius $R = 1 \kpc \ (1.5 \kpc) $ is $\simeq 61\,\unit{Myr}\  (95 \unit{Myr})$.}
        \label{fig:VvsT}
\end{figure}

To avoid a singularity in the diffusion coefficients, we set up the BH with a small velocity vector such that $v_\| = v_\bot \simeq 3 \kms$ initially at a cylindrical radius $R = 1.5 \kpc$ ; we also set $\sigma_\star = 15 \kms $ in Eq.~(\ref{eq:df_ortho}) to boost the growth of velocity components (dispersion and stream: cf. Fig.~\ref{fig:Ekrates}). 
The results of the integrals over time are displayed on Fig.~\ref{fig:VvsT} for an BH of mass $\Mbh = 1 \times 10^7 \solarm$, with a ratio $\Mbh/m_\star = 10^3$. The initial 
rapid growth of the velocity dispersion components $\sigma_{\|\,\mathrm{or}\, \bot}$ is clearly visible, but note how it quickly 
levels off as the BH velocity increases (large relative velocities diminish the efficiency of 
energy diffusion). On the other hand, the component parallel to the stellar flow, \ie $v_\| = v_\phi$,  reaches a larger amplitude than either of the two dispersions $\sigma_{\|\,\mathrm{or}\, \bot}$ after $\approx 270$  Myr of evolution. 
     The square streaming velocity $v^2_{\|}$ of the BH  soon exceeds the sum of the two squared dispersion components $\sigma^2_\bot +\sigma^2_{\|}$ at times 
     $ t \simeq 300 $ Myr (see \S\ref{sec:critical} below). In other words, there is more specific kinetic energy in the streaming motion (carrying angular momentum)  after $300$ Myr of evolution. This is close to three full  periods of $\simeq 95$ Myr at a distance $R = 1.5 \kpc$ and with a rotation flow $R\Omega = 100 \kms$. Not surprisingly, perhaps, this coincides exactly with the estimated time to dissipate
     kinetic energy in isotropic systems given by Eq.~(\ref{eq:tdyn}). But where one effect drains out orbital energy, the traction mode opposes this trend. 

\begin{figure}
   \centering
   \includegraphics[width=0.45\textwidth]{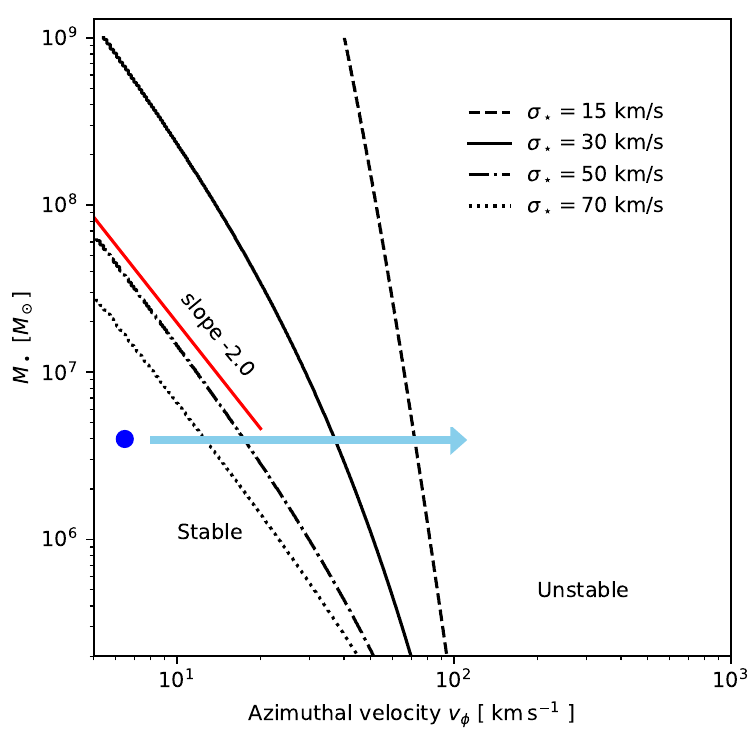}
     \caption{Critical line of stability for various configurations. The vertical axis is  the BH mass $\Mbh$ plotted against azimuthal velocity $v_\phi$. 
     At fixed $\Mbh$, the minimum azimuthal velocity $v_\phi$ to reach such that the dynamical traction sets in is found by crossing one of the black curves, from left, to right.
      Four curves are plotted, each corresponding to a different value of the stellar velocity dispersion $\sigma_\star$ as indicated in the legend. The blue dot marks the estimated coordinates of the Milky Way's BH. The red curve highlights an asymptotic trend of slop  $\simeq -2$ found in the limit of low $v_\phi$. }
        \label{fig:VpBH}
\end{figure}

\subsection{The on-set of  the transition to dynamical  traction} \label{sec:critical} 
As seen from Fig.~\ref{fig:VvsT}, at some point in time the BH acquires a velocity vector such that its main component is $v_\phi \approx v_\|$ so that it moves almost (though not exactly) in the direction of the flow of stars. Its z-component angular momentum $L_z = R v_\phi$ also increases with time. We stress that the time integral that we performed 
did not take into account any motion in  $R$ or $z$. 
Despite this caveat, it is interesting to predict  when the growth of angular momentum through the transition to dynamical traction  
must set in and 
transform the BH orbit into a loop orbit. We define this turning point as the value of $v_\phi$ such that the rate $v_\|\,\rate{v}{\|} $ 
exceeds the sum of the rates due to the diffusion terms (see Fig.~\ref{fig:Ekrates}). We find that turning point  for $\Mbh$ in the bracket $[10^5, 10^9]\, \solarm$ and plot the results for four values of the stellar velocity dispersion $\sigma_\star$. 
We look at a few cases more closely below. 

\subsubsection{Example for an $\Mbh = 10^7\solarm $ BH}
The outcome for that case is shown on Fig.~\ref{fig:VpBH}. Consider the case of a BH with $\Mbh = 10^7 \solarm$. 
At low values of $v_\phi$, the 
BH gains and looses velocity components isotropically from the background stars. The gain and loss of such increments acquired 
stochastically mean that we do not expect a net increase in angular momentum, and the configuration is deemed stable. If and when 
the BH has acquired a significant azimuthal component, however, the situation changes: when $\sigma_\star \simeq 70 \kms$ and $v_\phi \geqslant 7 \kms$, the positive transfer rate $v_\|\,\rate{v}{\|} $ much exceeds the other terms, and hence $v_\phi$ will continue to grow. Hence as the $\Mbh = 10^7\solarm$ BH   crosses the dotted line on Fig.~\ref{fig:VpBH}, the system switches from a stable to an unstable configuration. 
The physical reason for this is that the BH continues to induce a response in the stellar density as described originally by \citet{chandrasekhar1943}, whereas the weak response coming from the quadratic diffusion coefficients only becomes weaker as the 
motion of the BH aligns with that of the flow of stars. (To first order one has $\rate{v}{\bot} = 0 $ by symmetry, which is  
exact if we set $v_\phi$ strictly equal to $ v_\|$.) 

When the stellar velocity dispersion $\sigma_\star$ is reduced, an BH  triggers a response from a wider range of stellar orbits. 
Consequently the stochastic evolution of the BH's orbit persists until a larger azimuthal velocity component is 
reached and the transition to a systematic increase in $L_z$ takes place.  As an example, when $\sigma_\star$ is reduced from 70 
to $30 \kms$, we find that the transition occurs at $v_\phi \approx 35 \kms$ (solid line on Fig.~\ref{fig:VpBH}).  If we fix $\sigma_\star = 50 \kms$, the transition point shifts back to $v_\phi \approx 15 \kms$. 
A similar interpretation follows from increasing the mass $\Mbh$ at constant $\sigma_\star$. A more massive BH will trigger a strong 
linear response from the flow of stars, and so will acquire angular momentum steadily even if it is surrounded by stars of a 
relatively high velocity dispersion. On the contrary, if the stellar velocity distribution function is very cold (low $\sigma_\star$), an BH would trigger a strong response out to a large radius, in all directions. In that situation, it would 
not undergo a steady growth of its  angular momentum unless it is given a large azimuthal velocity, for example, from 
an external perturbation. The shift over to a critical value of $v_\phi$ is sketched with an arrow on  Fig.~\ref{fig:VpBH}.

\subsubsection{The case of the Milky Way BH}
We have indicated with a large blue dot on Fig.~\ref{fig:VpBH}  the position of the Milky Way's BH at $\Mbh = 4 \times 10^6\,\solarm$ and with a  velocity in azimuth $\approx 2 \kms $, an 
upper limit for the velocity of Sgt A$^*$ in the Galactic frame of reference (\citealp{reid2004}). The situation is complicated by the presence of a nuclear cluster of stars (NSC) around the BH (\citealp{neumayer2020} for a review). \citet{chatzopoulos2015} estimate the half-light radius of the NSC at $\simeq 7 \pc$, where the velocity dispersion is on the order of $35 \kms$. 
Their modelling of the NSC suggests significant rotation, with an isotropic rotator model \citep{BT08} of 
angular frequency 
\footnote{Notice how $\Omega_{nsc} \simeq 10^4 \kms/\kpc \simeq$ 
20 times  the angular rotation  frequency at galactic scales: Gaia data suggests $\Omega \simeq 50 \kms/\kpc$ for the innermost 
100 pc of the Milky Way (\citealp{gaia2023}; see also \citealp{arguelles2023}).} $\Omega_{nsc} \approx 30 \kms/ 3.9 \pc$. 
 Despite the large NSC angular velocity, the mean azimuthal  velocity $v_\phi \simeq 0.8\,\sigma_\star \simeq 28 \kms$ for the stars inside the half-light radius \citep{neumayer2020}. 
The solid line on Fig.~\ref{fig:VpBH} gives a 
critical azimuthal velocity $v_\phi \approx 40 \kms$ for a stellar velocity dispersion $\sigma_\star \simeq 35 \kms$. This is 
significantly larger than the measured rotation for the stars. In conclusion, the BH would not suffer a dynamical traction 
transition even if it had acquired a $v_\phi$ of the same amplitude as that of the stars. The ratio $v_\phi/\sigma_\star \simeq 0.8$ implies that the NSC is nearly isotropic in velocity space, 
so the BH would sink back to the origin quickly \citep{chatzopoulos2015}.

That line of  reasoning follows from experience with fully isotropic velocity distribution functions, $f(E)$. For that case a massive body always loses momentum \citep{BT08}.  Consider the estimated mass of the NSC's stellar population at $\approx 9 \times 10^6 \solarm$, which  exceeds that of the BH itself only by a factor $\times 2$.  If we apply an $m = 1$ perturbation to the BH, its motion  would trigger a strong response from the NSC stars and bring back the BH to the barycentre on a short mass-segregation time-scale given by 

\begin{equation} 
t_{ms} \propto t_r  \simeq \frac{m_\star}{\Mbh} \times\frac{ N_\star\, t_{dyn}}{ \ln N_\star} \sim  \frac{2}{ \ln N_\star}\,  
 t_{dyn} 
\end{equation} 
  with $t_r$ denoting the two-body relaxation time, and  $ \Mbh \simeq N_\star\, m_\star /2$. The NSC dynamical time \[t_{dyn}\sim \frac{ r_h}{ \sqrt{\overline{v^2_\star}} } \simeq 0.2 \unit{Myr}\]  with $r_h$ the half-light radius of $\approx 7 \pc$ and a root-mean   square velocity $\approx 30 \kms$.  See \eg, \citet{merritt2013}, \citet{subr2014} or \citet{panamarev2022} for further details and discussion on the collisional dynamics of the  NSC. 
  
On the other hand, the impact of the  nuclear stellar disc (NSD) on a scale of $ \sim 150 \pc$, of an estimated mass $\sim 10^9 \solarm$,  is much more likely to trigger a sustained transition to dynamical traction  (\citealt{launhardt2002}; \citealt{fritzetal2016}; \citealt{nogueras2020}; \citealt{nieuwmunster2024}; \citealp[but see][]{zoccali2024} for a critical assessment). The dynamical time-scale $t_{dyn} \sim 1 / \sqrt{G\rho} \sim 1 $ Myr. Reading the rotation velocity of $\simeq 50 \kms$ at radius $r =  100\,\pc$ from the Gaia survey (\citealp{gaia2023}), we compute a period of $\simeq 12 $ Myr for a circular orbit at that radius. The growth rate of azimuthal velocity displayed on Fig.~\ref{fig:Ekrates} over several $\times 10 $ Myr would be significant. We note that the NSD has been the 
site of recent star formation events  over times of $30 $ to 100 Myr \citep{nogueras2020}, while some low-metallicity stars may be as young as 
100 Myr within the central-most NSC (scale of $10 \pc$; \citealp[see][]{fellenberg2022,chen2023}). These are indirect indications of past 
perturbation on the right time-scale for the full growth of the transition to dynamical traction. We will not attempt to model the central region of the Milky Way in this contribution. Instead we present a first numerical exploration of the on-set of the transition in \S\S\ref{sec:nbody} and \ref{sec:results}.

%

\section{Restricted N-body calculations}\label{sec:nbody}
Recently, \cite{kaur2018} and  \cite{banik2022} studied the exchange of mechanical energy between stars and an orbiting BH analytically and with a set of numerical integrations \citep[see also][]{boily2008}). These studies show how the response of stars either drains energy from, or invests energy in, the BH's orbit depending on 
the character of the stellar orbits and their parameters such as their phase, inclination angle, and so on. 
We view these analyses as a strong hint that different families of stellar orbits couple to a different degree with an 
BH, and in particular that anisotropy in the stellar velocity field is a key ingredient to the BH's orbital evolution. 
Here we borrow from \cite{boily2008}, who noted that co-planar stellar orbits couple much more effectively to the BH than those at high inclination angle (see their \S5.3 and Fig.~14b). This allows a simplification of the numerical treatment, which we  lay out below. 

\subsection{Initial conditions} \label{sec:ICs}
Firstly, we divide the stellar population in a {spherical}- and a {disc} components, whereby the dynamically cooler disc component is 
made up of stellar orbits with an angular momentum orthogonal to the orbital plane of the  BH. We freeze the 
other (dynamically hot) stellar orbits completely. The system to evolve is made of two potentials: for the frozen one, we picked the spherical isochrone model of Hénon, 

\begin{equation}
\Phi_{Is} ( r ) = - \frac{GM_{Is}}{a + \sqrt{a^2+r^2} }\, , \label{eq:Henon}
\end{equation}
where $a$ is a radial scale length, and $M_{Is}$ the total mass. For the second, live potential,  we opted for an axi-symmetric 
 Miyamoto-Nagai disc in cylindrical coordinates $R,\theta, z$ \citep[see \S2 of][]{BT08}, 

\newcommand{\Z}{{\cal Z}}
\begin{equation}
\Phi_{D} (R,z) = - \frac{GM_D}{\sqrt{  (a + {\Z } )^2  + R^2} }\, , \label{eq:MiyamotoNagai}
\end{equation}
where we have defined $\Z^2 = z^2 + b^2$, $M_D$ is the total disc mass, and $a,b$ are radial- and vertical scale lengths, 
respectively. We have opted to fix $a = 1.5 \kpc$ in both potentials, and a ratio $b/a = 1/6$ for the disc potential. 
 
These potentials are combined and their joint stellar phase-space distribution function solved in action space with the \href{https://github.com/GalacticDynamics-Oxford/Agama}{\textcolor{blue}{Agama}} package \citep{vasiliev2019}. A restricted N-body realisation of the stellar disc is obtained from a Monte-Carlo sampling of the joint distribution function. 
We noted how the response of centrophobic loop orbits subjected to the motion of a massive perturber is the strongest   
when they are co-planar.  
The root-mean square variation  in mechanical energy for loop orbits reaches $\approx 80\%$ of the peak co-planar 
value whenever  their orbital angular momentum remains within $\pm\approx 18\dgr $ of the 
polar vector of the perturber's orbital plane \citep[see Fig. 14b of][; it becomes progressively weaker for larger inclination angles]{boily2008}. The solid angle spawned around the pole  is  close to $5\%$ of  the whole sphere.  
An isotropic distribution function would yield 
orbits with angular momentum vectors covering the full sphere uniformly.  The disc potential (\ref{eq:MiyamotoNagai}) of  
aspect ratio $b/a = 1/6$ gives rise to near-circular motion within a wedge of opening angle 
$ =  \arctan{1/6} \simeq 9.5\dgr$ for  close to $85\%$ of the orbits (the distribution function becomes progressively  more isotropic as we approach the centre, within a cylindrical radius $R < a $ ; it is very nearly isotropic inside a small region bounded by  
$R \simeq b = 250 \pc$). This sits comfortably within the  solid angle for which we have estimated the orbits will respond
the strongest to the perturbation. With these parameters we therefore focus on {discy}  (nearly co-planar)  orbits. We could have extracted co-planar orbits directly from the spherical isochrone distribution. 
Instead, we recall our motivation for more anisotropic systems in the context of a galaxy merger, or the end-state of the 
formation of a disc galaxy. For that reason we chose to boost the sample of disc orbits by fixing the mass parameters in (\ref{eq:Henon}) and (\ref{eq:MiyamotoNagai}) in the ratio

\[ \frac{M_D}{M_{Is}} = \frac{1}{6}\, \simeq 0.167 . \] 
This increases the fraction of disc orbits by about 3 times what we would have drawn from an isotropic d.f. in spherical symmetry.

A physical mass of $M_{Is} = 1 \times 10^{10} \solarm $ was chosen to match the mass of the central kiloparsec of a typical 
 galaxy. The central  atomic hydrogen density of the isochrone sphere $\rho_{Is} \simeq 7.16\, \cm^{-3}$ is only slightly lower than that for the  disc, for which $\rho_D \simeq 8.6\,\cm^{-3}$ ; these values are roughly twice the estimated value  of 3.9 $\cm^{-3}$ for the Milky Way at the solar radius \citep{blandhawthorn2016}.

%
	\begin{table*}
	     \caption[]{Default parameters of the potentials and \amuse\ numerical integrators.}\label{tab:parameters}
	     \vspace{-5mm}
	     \begin{center}
	\begin{tabular}{cccccccccc}
	\multicolumn{5}{c}{Isochrone (Hénon) potential} & \multicolumn{5}{c}{Miyamoto-Nagai disc   } \\[2pt] 
	\multicolumn{5}{c}{\rule{90mm}{0.5pt}}  & \multicolumn{5}{c}{\rule{90mm}{0.5pt}} \\[2pt] 
	$M_{Is}$  &  $a$  &   $\sigma_r$ & \multicolumn{2}{c}{comment}  &  $M_{D}$  &  $a$  & $b$ & $\{\sigma_R, \sigma_z\}$ & comment \\[-2mm]
	\multicolumn{5}{c}{\rule{90mm}{0.5pt}} & \multicolumn{5}{c}{\rule{90mm}{0.5pt}}\\[2pt] 
	 $ \solarm $ & $\kpc $ &  $\kms$ &  & & $ \solarm $ & $\kpc $ & $\kpc $  &  $\kms $ &   {\small  warm : } \\
	 $1\times 10^{10} $ & 3/2 &  22.7 & \multicolumn{2}{c}{frozen, infinite extent} & $1.6\times 10^{9} $ & 3/2 & 1/4 &  $\{4.2,  4.6\}$  & {$\sigma_{R\ \textrm{or}\ z} \simeq 12\,\kms$ }\\[10pt]

	      		\multicolumn{5}{c}{Barnes-Hut integrator   } & 	      \multicolumn{5}{c}{Hermite (direct)  integrator   } \\[2pt] 
	\multicolumn{5}{c}{\rule{90mm}{0.5pt}} & \multicolumn{5}{c}{\rule{90mm}{0.5pt}}\\[2pt]  
	Name & $N_{2b}$ & $\theta_c$ & $l_\star [\pc] $ & accuracy & Name & $N_{2b}$ &  $l_\star [\pc] $ & mode & accuracy \\[-2mm]
		\multicolumn{5}{c}{\rule{90mm}{0.5pt}} & \multicolumn{5}{c}{\rule{90mm}{0.5pt}}\\[2pt]  
	\textsc{BHTree} & 16 or 32 & 0.50 & 125 &  quadrupolar  & \textsc{Ph4} &  2 &  10 &  openmp & 4th order  \\[5pt]
\end{tabular} 
     \end{center}
\end{table*}

\subsection{Numerical integration} \label{sec:numerical}
We performed numerical experiments using the \href{https://www.amusecode.org/}{\textcolor{blue}{Am$\mu$se}} suite of community codes to integrate the equations of motion \citep{pelupessy2013,portegieszwart2018}. 
The python package  \amuse\ v.5.2 (November 2023) offers an interface to treat each component of a system as a set endowed with its own methods and attributes. We opted to couple a live set of stars to a subset of BH's and test stars using the bridge methodology \citep{fujii2007,portegieszwart2013}. A bridge easily allows to  
couple a  Barnes-Hut Tree integrator \citep{BH1986} and the 
 direct-summation 4th-order accurate Hermite integrator \textsc{Ph4} \citep{portegieszwart2018} to integrate the disc stars 
and BH orbits, respectively. Both integrators are `bridged' to a frozen background potential standing in for the extended galactic centre. 
The \textsc{BHTree} code has a set-up with critical opening angle $\Theta_c$ which we sat to $ 0.5 $ (down from 
the default $0.75$), and a critical number of near-neighbours for multipolar expansion  which we boosted, from the default 12, to 16 and in some cases, 32. 
This tactic meant that 
more stars were integrated with a 2nd-order accurate leap-frog symplectic scheme, whereas distant stars were integrated with a
 quadrupolar expansion of the stars' potential. 
 A large softening length $l = 125\,\pc$ was used for the two-body star-star interaction. We found this to 
 be the largest value that still gave a sensible resolution of the vertical structure of the disc (scale height b = $250 \pc$) while allowing also to resolve any Jeans-type mode of instability. The stars-BH interactions were treated with a much smaller softening kernel where $ l = 10\,\pc$. The more compact kernel allowed for larger deflection angles of the stars and a accurate rendering of the BH potential (the effective size of the stellar mass elements $ \sim 40\,\pc$, see below). The parameters used for the models and integrators are listed in Table~\ref{tab:parameters}. 
 
Calculations were done by  sampling the disc d.f. with $N = 100\, 000 $ stars, so each star has a mass  $m_\star =  0.16 \times 10^5 \solarm $. The BH has a reference mass $\Mbh = 1.25 \times 10^7 \solarm $ for a mass ratio  $ \Mbh / m_\star  = 781$. This takes us clearly in the galactic dynamics domain with each 'star' the mass equivalent of a  rich open cluster of (real) stars. The disc density $\rho _D \simeq 8.6 \unit{H/cc}$ implies a mass inside the volume spawned by $l = 125 \pc $ of $ 1.67 \times 10^6 \solarm$, or about 100 of our model stars. We argue this makes all galactic sub-structures well resolved on a scale $\sim 100 \pc $ and above. The physical mass of the BH much exceeds (by a factor  $\approx 7.5$) the stellar 
mass in a spherical volume spawned by $l$, hence 
the influence of the BH is correctly resolved on scales exceeding $\sim 100 \pc$. The dynamics of the stars is also well recovered 
on such scales, because their individual mass is recovered from a spherical volume of radius $ \approx 45 \pc $. 
Finally, we designed a global units converter to ensure that the 
\amuse\ integrators and the Agama module work hand in hand. 

A series of test cases are presented in Appendix~\ref{sec:validation}. We 
checked that the parametric set-up of the numerical integrators allowed to recover circular orbits in the frozen potentials to high accuracy. For a live 
disc component, we note that the total system angular momentum is conserved to about 0.4\% accuracy for a runtime of $t = 1.5 \unit{Gyr}$, sufficient 
to monitor the transfer of the orbital angular momentum from the stars, to the BH, with three significant digits. In Appendix~\ref{sec:stability} we address the question of the stability of the configurations with a frozen background component. 

\begin{center}
\begin{figure*}[h]
\begin{minipage}{\textwidth}
\centering 
\begin{overpic}[abs, unit=1mm, scale=0.65]{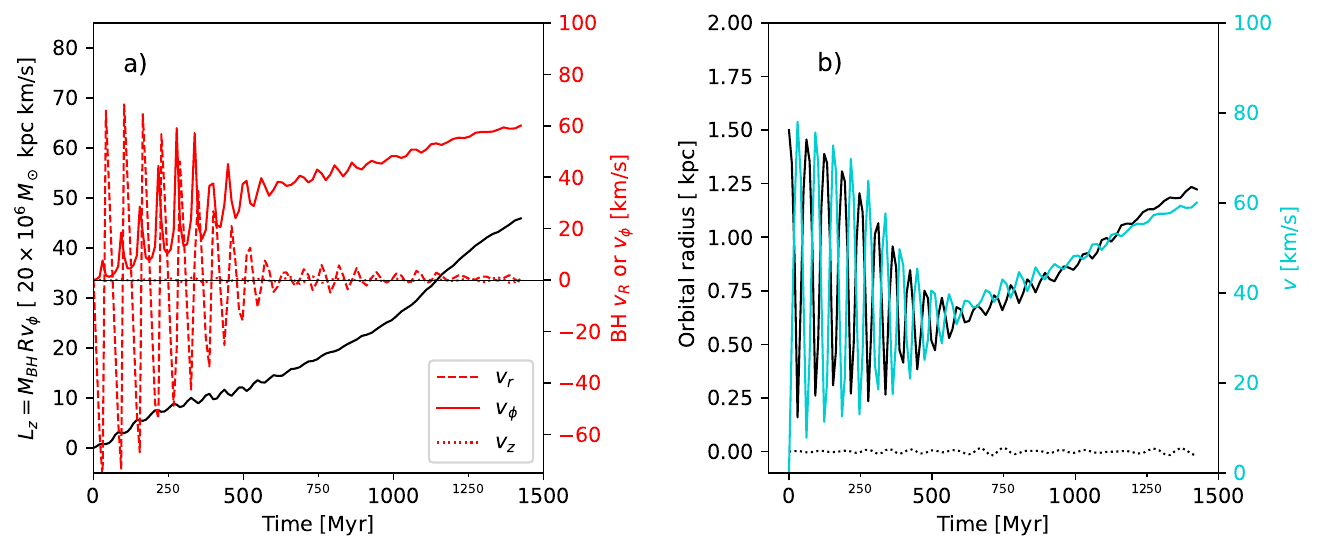}
\put(25,51){{\small \textit{infall, warm disc} } } 
\end{overpic} 
\caption{The case when an BH is let go from $R = 1.5 \kpc$ in a (dynamically) warm stellar disc (Table~\ref{tab:parameters}). 
On panel (a) we graph the run of $L_z$ as the black solid curve, with radial- and azimuthal velocity components 
 shown in red. On panel (b), the cylindrical radius $R$ and the amplitude of $\vbh$ are displayed, each with their own color-coded 
vertical axis. In the early stage,  dynamical friction 
makes the orbit shrink, however the perturber BH acquires angular momentum throughout (panel [a], black curve). Both $R$ and $\vbh$ increase systematically from $t \approx 500 \unit{Myr}$ onward. 
The dotted line at the bottom is the $z$-coordinate bounded to $\pm 15 \pc$. Compare with the test case displayed on Fig.~\ref{fig:BHTimeEvol}. }
\label{fig:BH_LRhot}
\end{minipage}
\begin{minipage}{\textwidth}
\centering 
\begin{overpic}[abs, unit=1mm, scale=0.65]{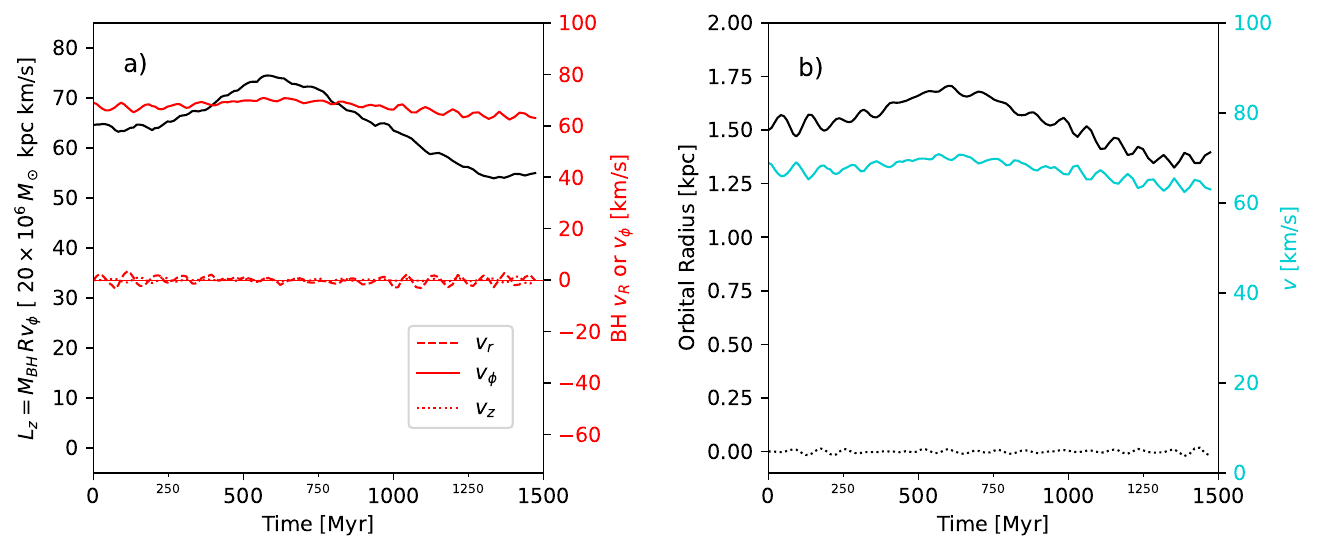}
\put(22,35){ {\small \textit{circular, warm disc} } } 
\end{overpic} 

\caption{As for Fig.~\ref{fig:BH_LRhot}, but now for the BH started at  $R = 1.5 \kpc$ on a circular orbit. Note the sharp response of the 
axi-symmetric disc to the BH which triggers radial motion. The disc adjusts to the black hole perturbation and settles progressively
over several orbital times to a new equilibrium. The BH's orbit however does not migrate significantly over the full integration time. 
Panel (a):  run of $L_z$ as the black solid curve, with radial- and azimuthal velocity components 
 shown in red. Panel (b): the cylindrical radius $R$ (in black) and amplitude of $\vbh$ (in turquoise). 
The dotted line at the bottom is  the $z$-coordinate bounded to $\pm 17 \pc$. }
\label{fig:BH_LRhotcircular}
\end{minipage}
\end{figure*}
\end{center}

\begin{center}
\begin{figure*}[h]
\begin{minipage}{\textwidth}
\includegraphics[scale=0.6012]{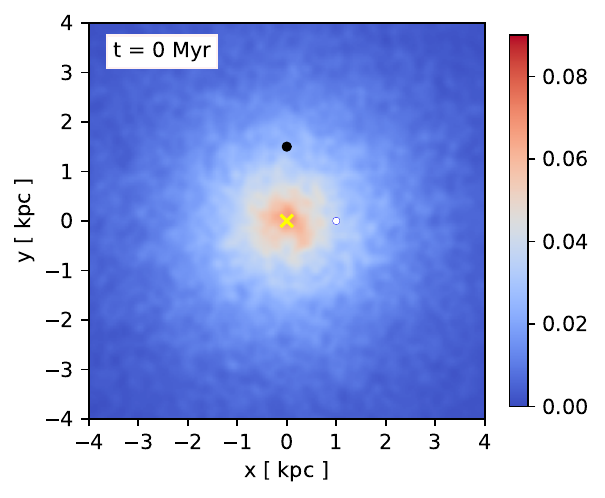} 
\includegraphics[scale=0.6012]{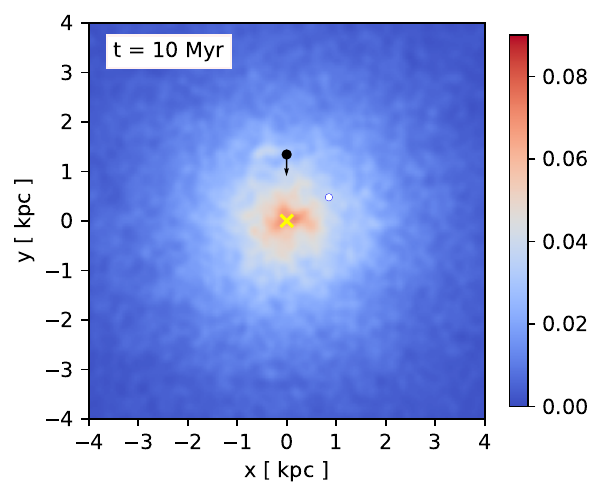} 
\includegraphics[scale=0.6012]{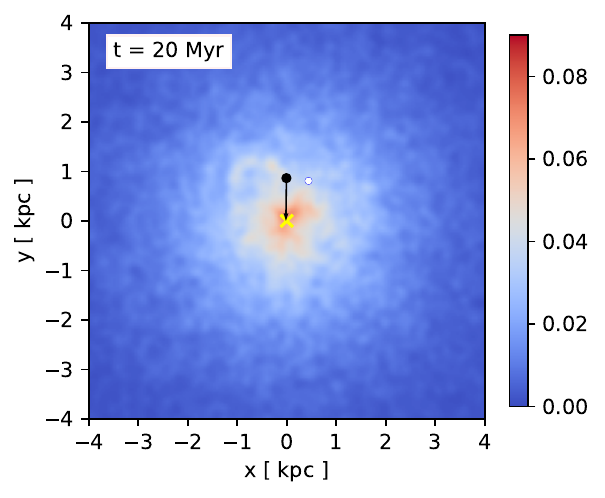} 
\includegraphics[scale=0.6012]{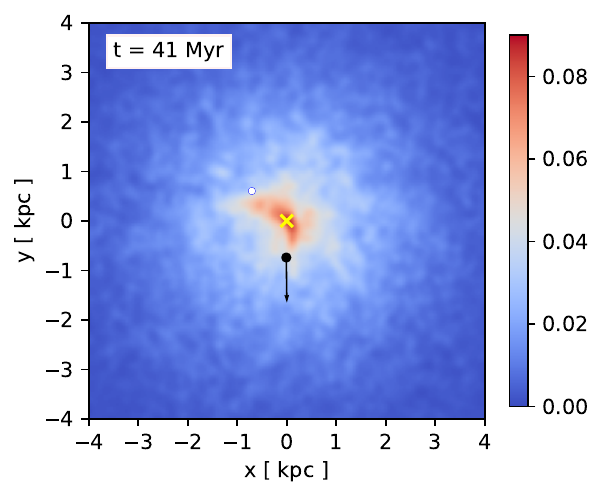} 
\includegraphics[scale=0.6012]{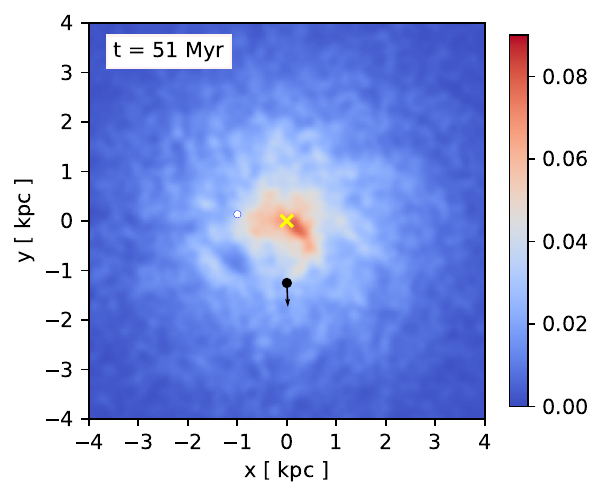} 
\includegraphics[scale=0.6012]{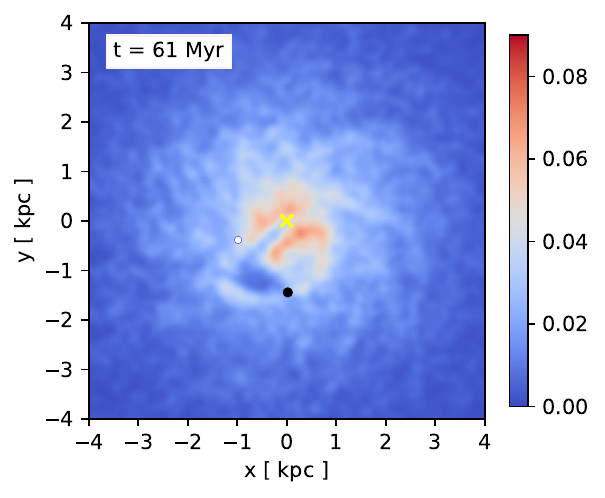} 
\includegraphics[scale=0.6012]{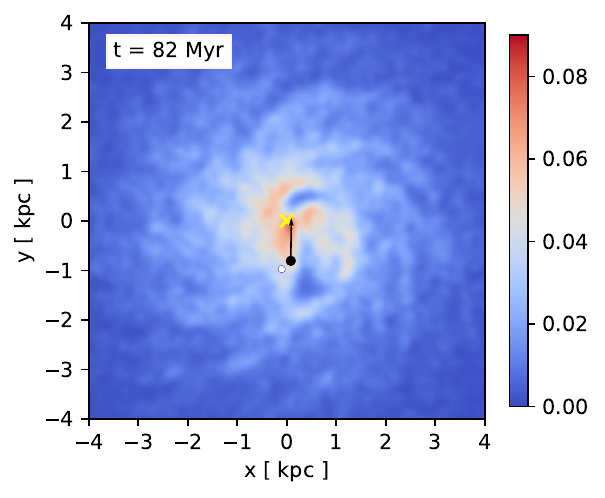}
\includegraphics[scale=0.6012]{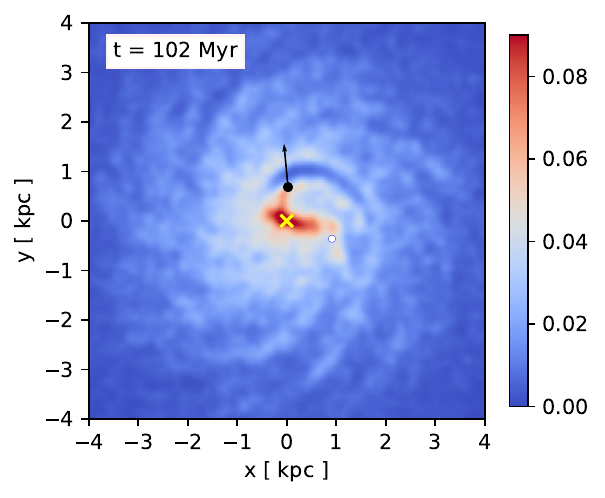}
\includegraphics[scale=0.6012]{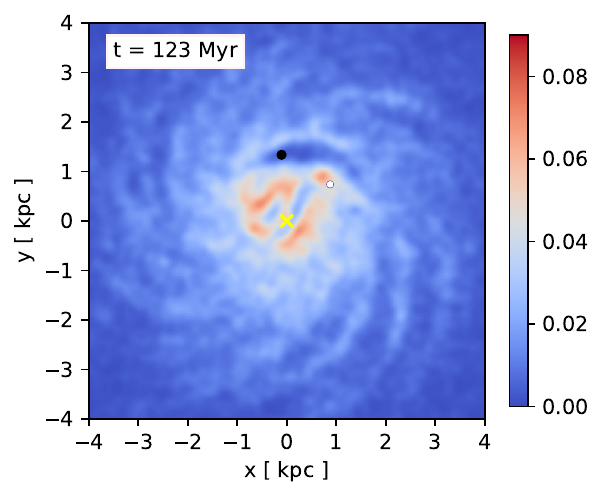}
\caption{Top left to bottom right: evolution of the system when the BH is released from rest at at radius $ r = 1.5 \kpc$. The black arrow indicates the velocity vector. Large density fluctuations soon develop in the inner $\sim 1 \kpc$ where the BH potential becomes dominant. After $t \approx 100 $ Myr of evolution strong density fluctuations appear, including 
large-scale voids (deep blue). The BH used in the calculation had a mass $\Mbh = 1.25\times 10^7 \solarm$.} 
\label{fig:BHRadialOrbit}
\end{minipage}
\end{figure*}
\end{center}  

\begin{center}
\begin{figure*}[h]
\begin{minipage}{\textwidth}
\centering
\includegraphics[scale=0.6125]{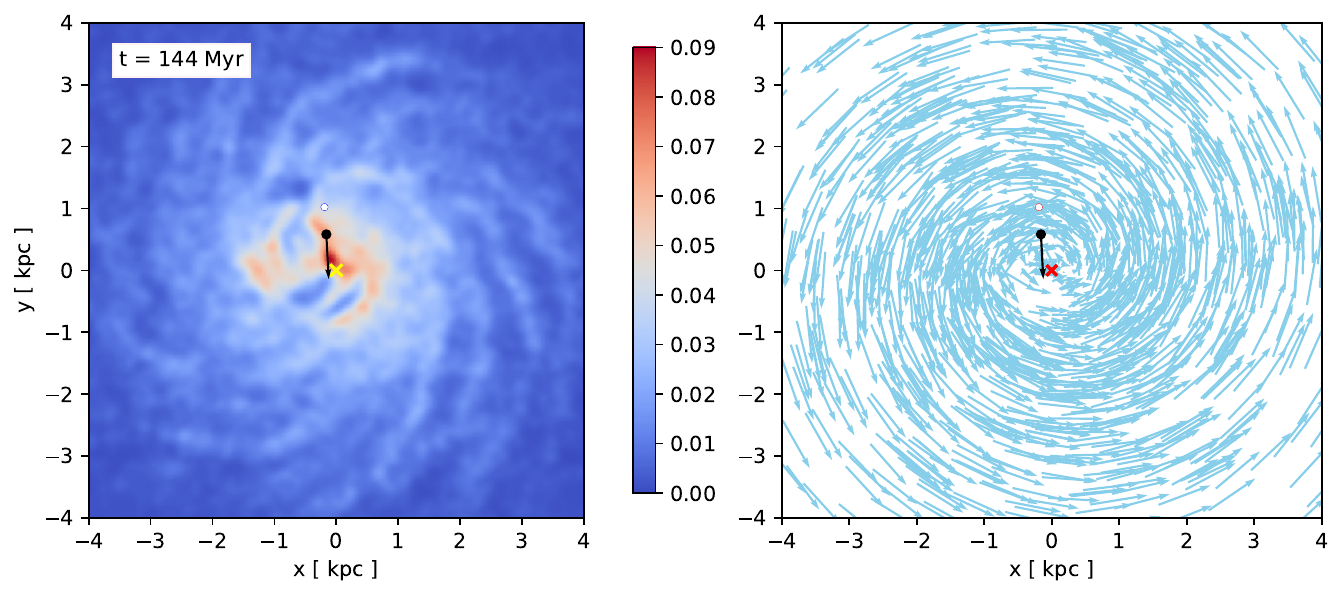}
\includegraphics[scale=0.6125]{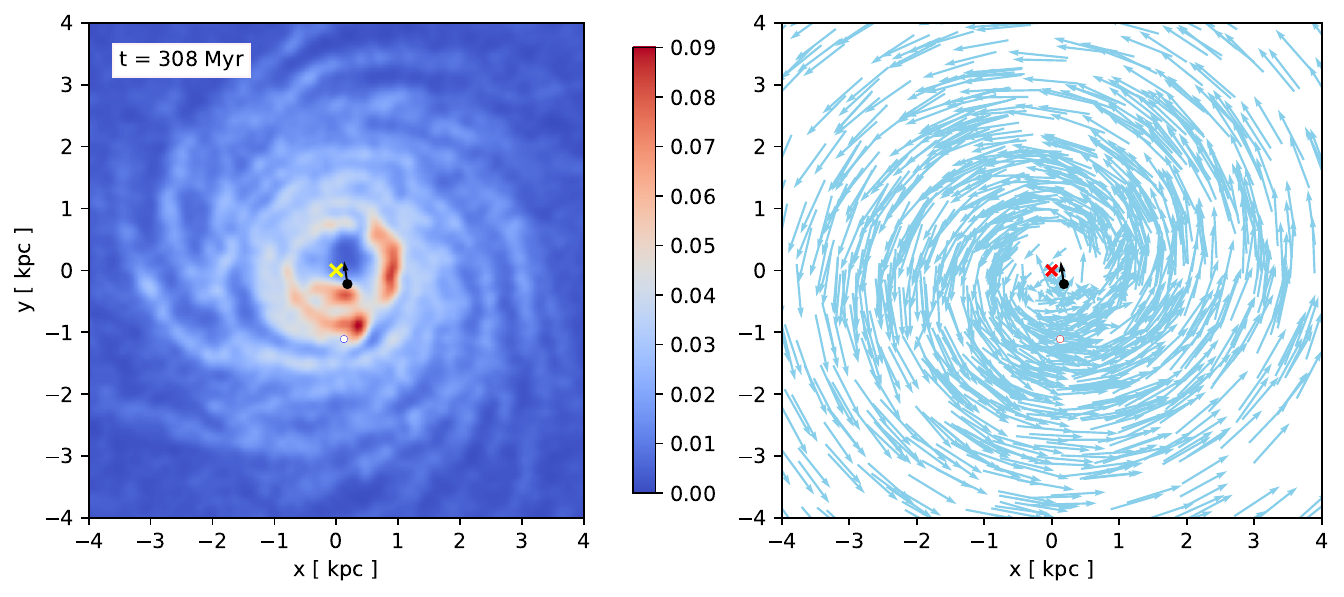}
\includegraphics[scale=0.6125]{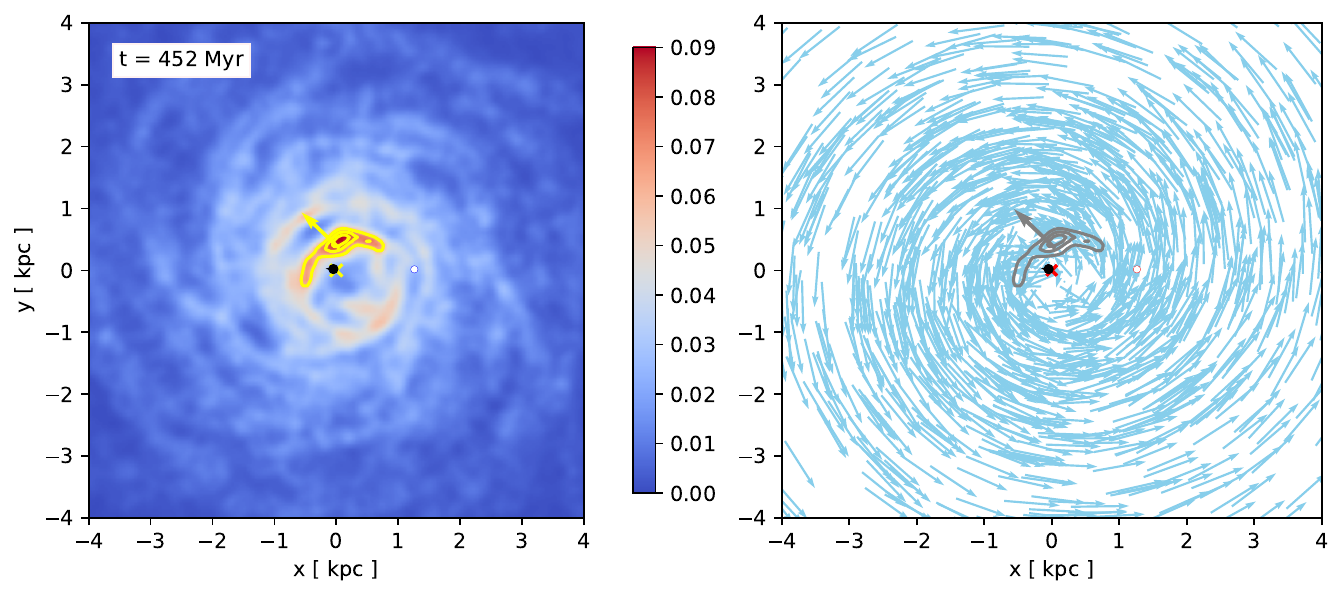}
\caption{Transition to a fragmented morphology  for a galactic disc perturbed by a massive black hole. Left-hand panels: density map of stars with peaks indicated in red colour. Right-hand panels: the velocity field. A black dot indicates the BH of $1.25 \times 10^7 M_\odot$ (or, $\simeq 1\%$ the mass of the stars), with its velocity vector also in black. 
The configurations are displayed at three times of 144 (top), 308 (middle) and 452 Myr (bottom row) of evolution. 
Contour levels in yellow (left-hand panel) and grey (right-hand panel)  identify the member stars of the clump formed at $t = 452 $ Myr. The BH sits at $\simeq 52 \pc$ of the coordinates centre. The relative velocity between the clump's centre of mass motion and the BH is indicated with an arrow on both panels, pointing in a north, north-west direction (west is to the left). 
} 
\label{fig:ModelGalaxy}
\label{fig:Clump452}
\end{minipage}
\end{figure*}
\end{center}

\begin{center}
\begin{figure*}[h]
\begin{minipage}{\textwidth}
\centering
\begin{overpic}[abs, unit=1mm, scale=0.65]{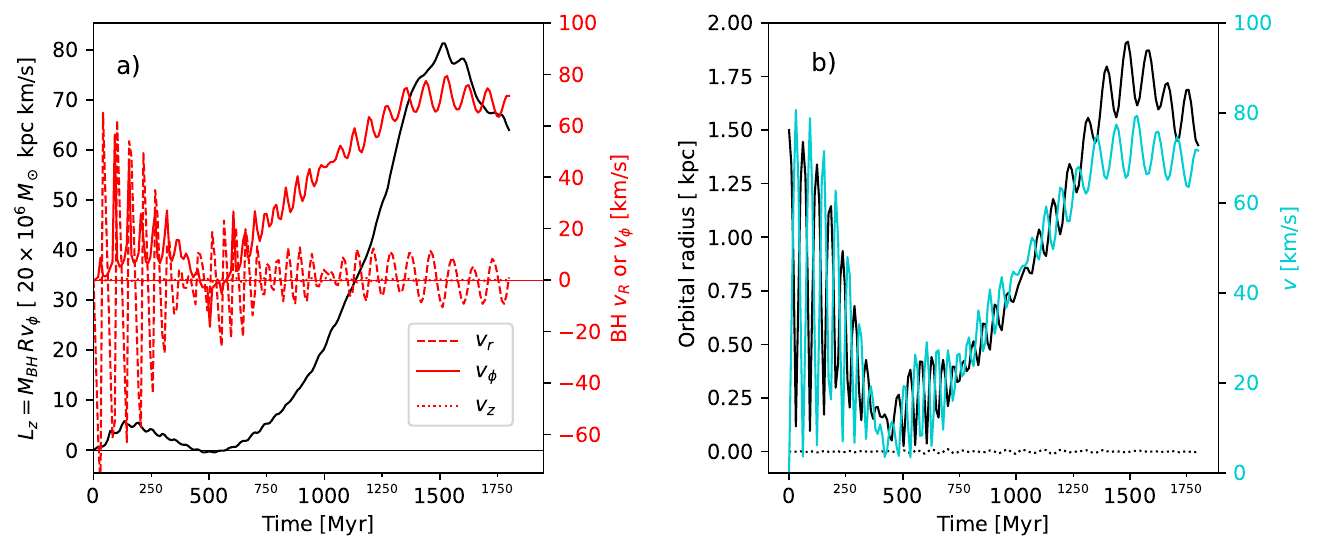}
\put(18,50){{\small \textit{infall, cool disc} } } 
\end{overpic} 

\caption{The case of an BH falling from rest on a radial orbit. a) The solid black curve graphs the angular momentum accrued over time (left-hand axis with black labels) ; the right-hand axis is the scale for each velocity component (red curves) ; b) The cylindrical radius (in black, left-hand axis) and velocity (right-hand axis, in turquoise) as function of time. The dotted line at the bottom is the $z$ coordinate bounded to $\pm 12 \pc$. }
\label{fig:BH_LR}
\end{minipage}
\begin{minipage}{\textwidth}
\centering 
\begin{overpic}[abs, unit=1mm, scale=0.65]{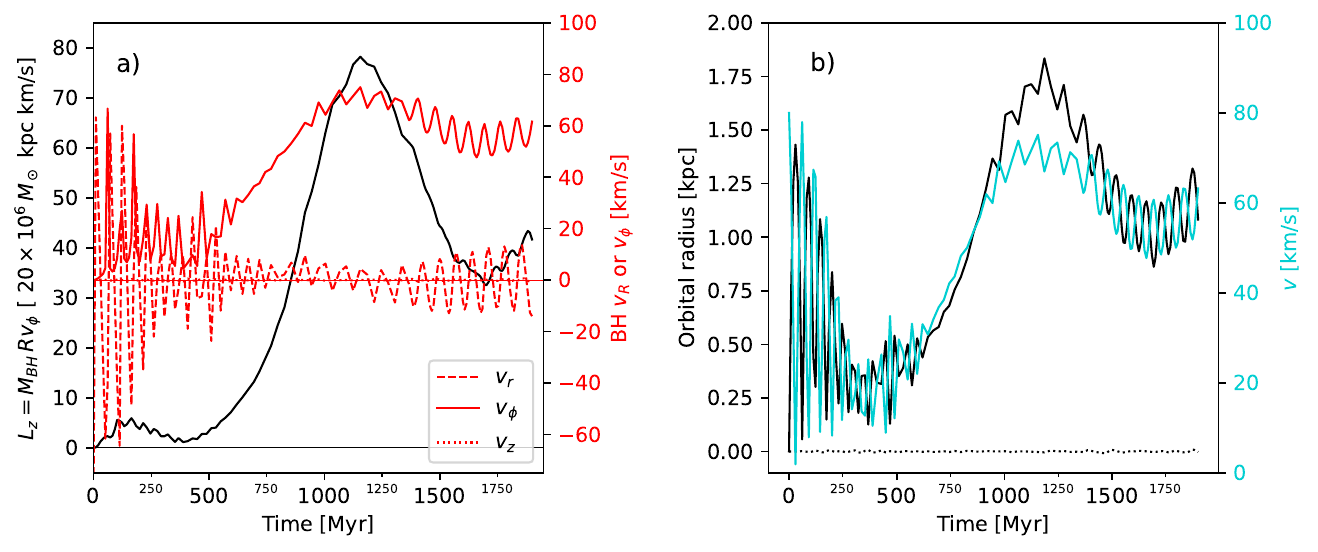}  
\put(17,52){ { \begin{minipage}{1.5cm}  {\small \textit{ejected, \\ cool disc} } \end{minipage} } } 
\end{overpic} 
\caption{As for Fig.~\ref{fig:BH_LR}, but now the BH is launched from the origin at $v_y = 80\,\kms $, and comes to rest at 
a radius $R \simeq 1.35\,\kpc$. The run of all quantities are remarkably similar but for a shift in time of $\approx 200 \unit{Myr} (\approx$ two orbital times) when $L_z$ peaks. Notice also that the phase of dynamical friction does not bring the BH close to the origin as 
its apocentre distance always exceeds $250 \,\pc$. The gain in angular momentum seen on panel (a; black curve) begins before $500 \unit{Myr}$. }
\label{fig:BH_LRfromOO}
\end{minipage}
\end{figure*}
\end{center}

%

\section{Results} \label{sec:results}
We  proceed with the problem of an in-coming BH as  outlined in \S\ref{sec:theproblem}.  
To keep the initial conditions close to the reference configuration of \S\ref{sec:numerical}, we use the same disc/halo 
configuration but shift the BH to a radius $a = 1.5 \kpc$, at the edge of the core of approximately constant density 
(recall that the same radial scale length was used for the disc and isochrone halo). No angular momentum is given to the BH initially. 
The initial configuration is similar to what is depicted on Fig.~\ref{fig:Cartoon}(b), after an $m = 1$ mode dislodged the 
BH from the origin. The analysis leading to Fig.~\ref{fig:VvsT} did not factor in the orbital evolution (especially the radial 
distance, $R$) which must occur on a time-scale $\sim 100 $ Myr for the chosen radial orbit. This zero-momentum BH orbit 
is an idealised case of a scenario where the BH is being accreted by the system. 

\subsection{The case of a warm disc}\label{sec:warmdisc}
When they form in a cold disc, density  waves can carry angular momentum in the same way as the gravitational focusing leading to dynamical friction. To minimise 
the impact of such waves, we discuss first the case of a warm disc (Table~\ref{tab:parameters}) which remains  
axially symmetric throughout the integration time (barring the response to the BH's motion). 

\subsubsection{Radial orbit} \label{sec:warmradial} 
For an BH set on a radial orbit in a warm disc, the 
orbital evolution develops significant  dynamical friction early-on as the BH falls from rest to the system's barycentre. The run over time is 
graphed on Fig.~\ref{fig:BH_LRhot} for the radius, velocity components, and $L_z$. One can clearly see how both the radial velocity $v_r$ and 
cylindrical radius $R$ diminish during the first $ \sim 500 \unit{Myr}$ of evolution. This corresponds to about 4 periods of the circular 
orbit at the initial radius, in good agreement with the time-scale given by Eq.~(\ref{eq:tdyn}). But where the dissipative friction drops 
in amplitude with the velocity $\vbh$ (the Stokes limit), the dynamical traction exerted by the stream of stars remains steady. In fact, on Fig.~\ref{fig:BH_LRhot}(a), 
the black curve mapping the angular momentum $L_z$ increases constantly, as does the azimuthal velocity component (solid curve in 
red). If the effect of dynamical friction is to kill off the radial velocity component $v_r$, one sees that dynamical 
traction, on the contrary, boosts the orthogonal $v_\phi$ component steadily. Hence the BH never reaches the centre, and in  fact begins to migrate outward as it gains angular momentum (Fig.~\ref{fig:BH_LRhot}[a] and [b]). 

\subsubsection{Circular orbit}
To determine whether the BH's orbital evolution is correctly recovered from the Fokker-Planck treatment of \S\ref{sec:FP}, 
we re-started the numerical integration but now with the BH initially set on a circular orbit  (same stellar disc and set-up otherwise). 
The expectation from Eq.~\ref{eq:diffrates} and Fig.~\ref{fig:Ekrates} is that the BH should remain trapped at constant 
radius, since there is in principle no net gain or loss of kinetic energy extracted from locally co-moving stars. 

The results are displayed on Fig.~\ref{fig:BH_LRhotcircular} with the same colour-coding and line styles as on Fig.~\ref{fig:BH_LRhot}. 
The expected circular orbit shows significant evolution over time, for several reasons. Firstly, 
large deflection-angle stars-BH encounters (collisions) lead to small but noticeable time-evolution of the BH's radius and angular 
momentum during the first $\sim 200 \unit{Myr}$. This is close to two orbital periods.  Collisions take place 
 because the epicyclic motion stemming 
from the  stellar velocity ellipsoid brings stars on and out of reach of the BH radius of influence, and this happens over a few 
revolutions about the centre. 
Secondly, the (global) reflex response of the stars to the BH's perturbation also contributes to destabilise  the BH's orbit on the same 
time-scale\footnote{It would have been simple to set up a quiet start to nullify that effect, for example by growing a BH adiabatically, but that would have been contrary to the context of an accretion event viewed here as an external perturbation.}. 
Further orbital evolution takes place over a longer time-scale of $\sim 500 \unit{Myr}$ as the BH 
radius varies with time in the bracket $1.3 - 1.7 \kpc$ (Fig.~\ref{fig:BH_LRhotcircular}[b]).  
We have not explored the reasons for  this behaviour  in any detail. 
 Note however that small radial displacements will be amplified if and when dynamical traction (or, friction) takes place, since it will increase (or, decrease) the orbital angular momentum and so boost the orbit's eccentricity. Also worth noting is that the time-scale of $500 \unit{Myr}$ 
 compares well with the time of $ \sim 300 \unit{Myr}$ needed to  transition to an azimuthally-dominant increase of kinetic energy (see Fig.~\ref{fig:Ekrates100M}).   
Our conclusion is that exact balance between dynamical friction and traction is not 
enough to maintain an BH on a circular orbit due to the response of the stars, first locally through strong scattering; and then globally, by the relaxation to a new equilibrium configuration of the system as a whole. 
 Despite these caveats, 
we find no systematic and continued radial migration for a BH starting with a high-$L_z$ orbit, in contrast with the case of initially strictly radial motion. \newline

The results for these example orbits give confidence that the basic mechanics of friction/traction and transfer of angular momentum is well 
recovered by the numerical treatment. We also find that the amplitude of the motion and time-evolution stay well above the 
noise level graphed on Fig.~\ref{fig:BHTimeEvol}. The higher velocity dispersion of disc stars, of total amplitude $\approx 25 \kms$, 
exceeds the circular velocity in the BH's potential at the linear resolution $ l = 125 \pc$, for which we compute 

\[ v_{circ} = \sqrt{\frac{G\Mbh}{l}} \approx 18 \kms\, .  \] 
The relative velocity between stars and the BH orbiting at the same radius (so sharing the same velocity in azimuth) 
  may yet be high enough to prevent a significant number of stars to bind with the BH. Visual inspection of density maps 
  suggests that only a few stars bind up with the BH. That situation will change when the BH orbits in a dynamically cooler 
  disc, as we argue in the next sub-section.

\subsection{The case of a dynamically cool disc}\label{sec:cooldisc}
We  now turn to the more complex case of a cool disc, when the vertical stellar velocity dispersion is reduced to $\sigma_z \approx 4.5 \kms$ ; the disc effective scale height $\simeq 100 \pc$ is lower by more than a factor of two compared with the 
warm disc (when $h \simeq 240 \pc$).  Self-gravitating 
Jeans fragments and large-scale spirals  are triggered by the  BH's gravity, which we view as an $m = 1$ perturbation mode. In the reminder of this section, the discussion follows the reference configuration for a dynamically cooler disc listed in Table~\ref{tab:parameters} unless stated otherwise.  

\subsubsection{Radial BH orbit} \label{sec:ClumpPart1}
Let  the BH start from  rest at $R = 1.5 \kpc$ initially, as done before for the warm disc  (\S\ref{sec:warmradial}). 
The early evolution is staged on Figs.~\ref{fig:BHRadialOrbit} and \ref{fig:ModelGalaxy}, where we find the BH initially falling 
to the system's barycentre due to the mean gravitational field. A full orbit is completed in $\simeq 123 $ Myr. 
These first few orbital times show the BH triggering large-amplitude density waves, and these last 
for several hundred Myrs, seen both in density maps and in the velocity field  (see Fig.~\ref{fig:ModelGalaxy}). 
In the time interval running up to $ t \simeq 400 $ Myrs, the BH acquires significant angular momentum, as shown on Fig.~\ref{fig:BH_LR}(a), where we plot  the z-component of $\mathbf{L} = \Mbh\,\rbh\times\vbh $ along with the radial- and 
azimuthal components of $\vbh$. At $t = 200 $ Myr, we compute $L_z \approx 120 \times 10^6 \solarm \kpc/\kms $ or $\approx 20 \times $ the peak value obtained in the test calculation with initially $\vbh = \mathbf{0}$ (Figs.~\ref{fig:BHatrest} and \ref{fig:BHTimeEvol}). Therefore this early evolution is not the result of numerical inaccuracies or approximations made. 

Despite the strong coupling and efficient transfer of angular momentum between stars and BH, dynamical friction soon takes over and drives  the rapid decrease in radius, bringing the BH systematically closer to the 
centre (Fig.~\ref{fig:BH_LR}[b]). Specifically on Fig.~\ref{fig:ModelGalaxy}, bottom frame when $t = 452 $ Myr, we find the BH within $r_\bullet \simeq 52 \pc $ of the centre, well below our spatial resolution limit of $125 \pc$.  At that time, it is moving at a  velocity of $\approx 11.3 \kms$ in the direction of the large banana-shaped clump  directly above it to the north. The phase-space coordinates of the BH are  consistent with harmonic motion in the frozen background potential as depicted on Fig.~\ref{fig:BHatrest}. However, the kpc-size over-density breaks the axi-symmetry of the background 
potential and exerts a net pull on the BH.  We wondered what impact that  would have on its  orbital  evolution. 

To investigate this, we first draw an outline of the over-dense structure with 
a surface density threshold of $> 0.06 $, 
or four standard deviations above the 
mean\footnote{In computational units,  the surface density $\Sigma$ falls in the range $[2.68\cdot 10^{-4}, 9.01\cdot 10^{-2}]$, with a mean value $1.25(6)\cdot 10^{-2}$ and standard deviation $\sigma_\Sigma = 1.12(4)\cdot 10^{-2}$\, .} . 
We 
display the selection with three density contour levels on Fig.~\ref{fig:Clump452}. Stars that fall inside that area were picked up
by estimating the density at the position of each star using the same Gaussian kernel  as for the  gridded maps of the disc. The set of 
stars selected together make up a mass of $32 \times 10^6 \solarm$ (2002 member stars), or close to $2.6 \times $ the mass of the BH; we refer to this set of stars as  \clump{S452}.  
To inspect the internal cohesion of \textsc{S452}, we computed its total mechanical energy $E = E_k + W $ and found $ E \approx 8046.6 \, (\unit{km}/\unit{s})^2 \solarm > 0 $ with a ratio $ E_k / |W| \gtabout 1.77$ to $1.88$, depending on 
whether the \clump{} was trimmed of a subset of the most distant stars to its barycentre, or not. The positivity of the mechanical energy is robust against details of the selection procedure, and implies that the clump is not bound. An unbound over-density should be interpreted as a wake caused by dynamical friction triggered  by the 
massive perturber  (\citealp{chandrasekhar1943, BT08}).  To assess the likelihood  that a subset of 
 \clump{S459}'s member stars are not, by themselves, a bound sub-system, 
we re-computed the energy $E$ of each star in the rest frame of the selected clumped stars taken in isolation. 
We identified stars that had a negative energy, and found out that these make up 62\% of the 2002 stars selected. 
We then removed all the stars with a positive energy and recomputed the potential of the new stellar clump with (now) only 
843 stars: that system had a total positive energy, with a  kinetic- to gravitational energy ratio very close to 
$E_k/|W| = 1.40$. The 
conclusion we draw from this analysis is that self-gravity plays only a minor role in the interaction of the clump with 
the BH, because unbound stars contribute a too-important fraction of the clump's gravitational potential. 

 \textsc{S452}  member stars disperse progressively over a time-scale of $\sim 200 \unit{Myr} $, close to two orbital periods, somewhat slowed down by their own gravity.   
We ask whether the initial configuration of \clump{S452} may efficiently drag the BH out of the origin. 
 Switching to the rest-frame of the BH, we find the clump moving away on a north, north-west trajectory at a velocity of $\simeq 9.8 \kms$ (the velocity vector is displayed on Fig.~\ref{fig:Clump452}). This configuration matches very much the  $m = 1 $ mode of perturbation  discussed in \S\ref{sec:theproblem}, here  on a radial scale of $\simeq 1 \kpc$. 
 The distance of the centre of mass of the clump to the BH is $\simeq 340 \,\pc $, 
 which gives rise to a monopole acceleration of $GM_{clump}/|| \rbh - \mathbf{R}_{clump}||^2 \simeq 0.00131 \kpc/\unit{Myr}^2$.  An  acceleration of this amplitude maintained for $\Delta t \approx 40 \unit{Myr}$ (the dynamical time-scale) 
  would displace the BH  by 

\[ \Delta x \approx \frac{ GM_{clump}}{|| \rbh - \mathbf{R}_{clump}||^2} \frac{\Delta t^2 }{2} \approx 1 \kpc\, . \]

That this expectation is not fully  realised is seen on Figs.~\ref{fig:BH_LR}(b), where the BH orbital radius increases to a much lesser extent over $\simeq 50 \unit{Myr}$ of evolution. 
The steady, nearly-linear rise of the BH's  radial distance and  velocity in the interval $t = 452 \unit{Myr} $ to 
$ \approx 1 \unit{Gyr}$ seen 
on Fig.~\ref{fig:BH_LR}(b) instead suggests that the acceleration field applied to the BH does not remain steady, but 
fluctuates in amplitude  over 
time\footnote{We checked that the bend in the angular momentum $L_z = \mathbf{R}\times\mathbf{v} \propto R v_\phi $  grows more slowly than $t^{5/2}$ in the interval $t : [450, 1500 ] \unit{Myr} $, consistent with a quasi-linear relation $\propto t^\gamma, \gamma < 1.25$ for both $v_\phi$ and $R$.}. 
In other words, 
the clump of stars exerting a pull on the BH certainly does not preserve enough of  its mass and morphology to have 
a significant impact on its long-term orbital evolution. Because the size of the clump compares well with the Jeans length $\lambda_J \sim 670 \pc$ of Eq.~(\ref{eq:lJeans}), we initially concluded, wrongly,   that the 
self-gravity of the clump enhanced significantly  the amplitude and duration 
of the traction  on the BH.  The dissolution of \clump{S452} argues against that.

\subsubsection{The core region, fragments}\label{sec:corefragments}
The BH perturbs the core region significantly as soon as its orbital radius $ \simeq 200 \pc$ or lower, since inside that 
volume the mass $\Mbh$ compares with the {total} gravitational mass of stellar- and frozen components. 
We found the velocity dispersion near the centre increasing over time to more than $20 \kms$ (up to a peak value of $25.5 \kms$ around the time $t \approx 452 \unit{Myr}$), distorting both the radial flow  and rotation curve. A high velocity dispersion 
facilitates the coupling between the BH and the streaming stars (e.g., Fig.~\ref{fig:Ekrates}), 
because it reduces the amplitude of Brownian motion by the perturber. After the BH 
has reached deep in the core region, the axi-symmetry of the stellar flow is broken: we  argued in \S\ref{sec:ClumpPart1} that a 
single impulse by a massive clump of stars gives enough linear momentum to shift the BH away from the centre in a 
region where the stellar stream becomes dominant again. The angular momentum 
is transferred to the BH progressively thereafter, driven by dynamical traction from the streaming stars. 
 It is worth to note that the BH follows closely the mean disc rotation, from the time $t = 750 \unit{Myr}$, onward. The BH has essentially the same 
specific angular momentum as the surrounding stars. A closer look at Fig.~\ref{fig:BH_LR}(b) shows an orbital  radius  oscillating with significant amplitude, the more so  as time passes : 
the BH is on an eccentric orbit, with the implication that it rotates, in turn, faster and  slower 
(in angular speed) than the stars as it 
shifts from the perigee to the apogee of its galactic orbit. It is therefore trapped around the radius of co-rotation with the surrounding  stars (see Fig.~\ref{fig:Ekrates} and Appendix~\ref{sec:FP}). 

\begin{center}
\begin{figure*}[h]
\begin{minipage}{\textwidth}
\includegraphics[scale=0.7012]{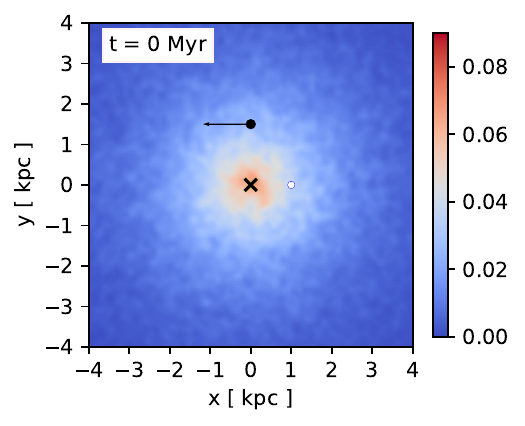} 
\includegraphics[scale=0.7012]{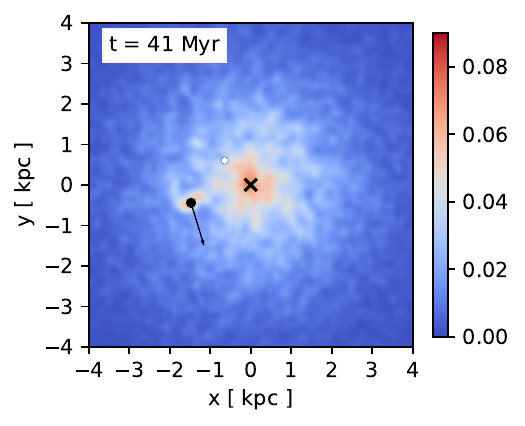} 
\includegraphics[scale=0.7012]{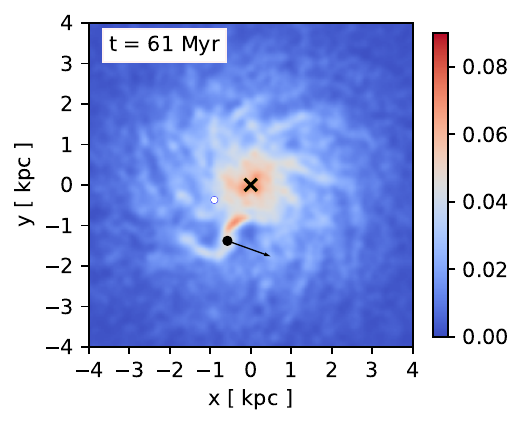} 
\includegraphics[scale=0.7012]{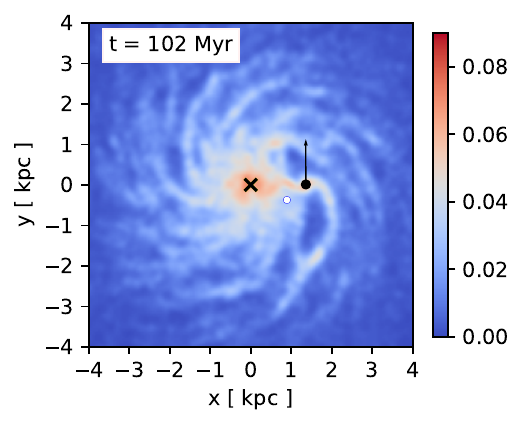} 
\includegraphics[scale=0.7012]{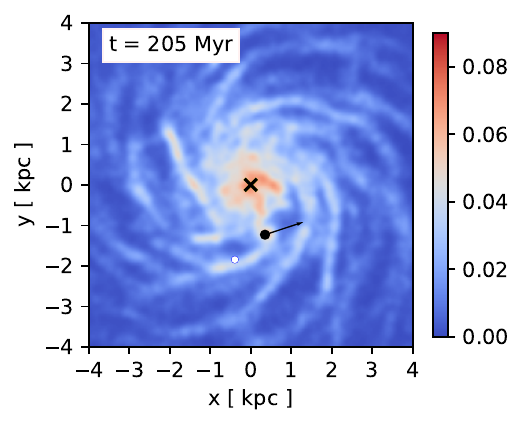} 
\includegraphics[scale=0.7012]{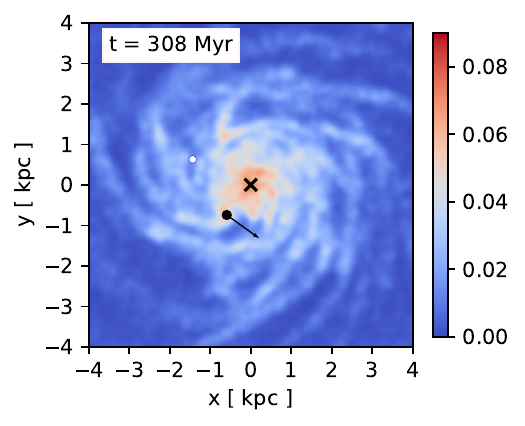} 
\includegraphics[scale=0.7012]{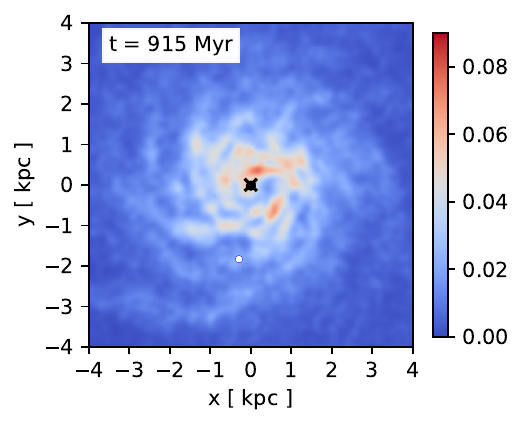}
\includegraphics[scale=0.7012]{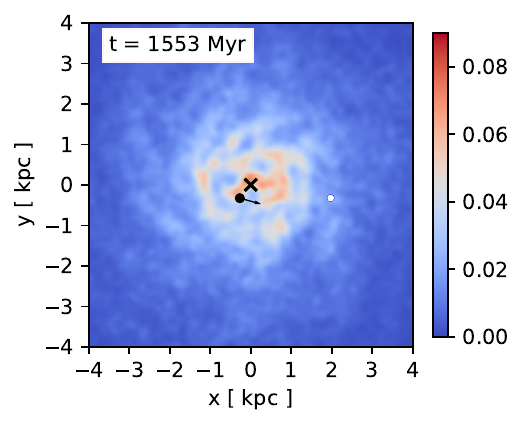}
\includegraphics[scale=0.7012]{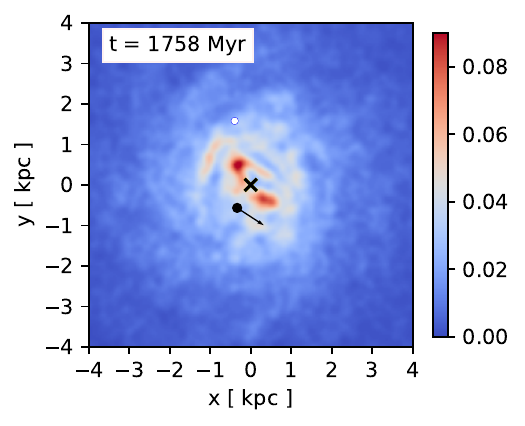 }
\caption{Top left to bottom right: evolution of the  BH set on  a circular orbit initially of radius $ r = 1.5 \kpc$. The black arrow indicates the velocity vector. Large spiral waves and Jeans fragments develop quickly due to the lower 
stellar velocity dispersion. The BH binds with stars and opens a gap in its wake, which leads to a steady loss of angular momentum 
and inward migration. This is most visible in the frames with time running from $t = 41 \unit{Myr}$ to $915 \unit{Myr}$. 
At $t = 915\unit{Myr}$ the BH sits at $r \approx 20 \pc$ with  a rest-frame velocity $\simeq 9 \kms$. Non-axi-symmetric clumps of stars 
dislodge the BH which acquires significant angular momentum by dynamical traction in the later stages of evolution (on a time-scale $\sim 500 \unit{Myr}$). 
  Calculations performed with an BH of mass $\Mbh = 1.25\times 10^7 \solarm$.} 
\label{fig:BHCircularOrbit}
\end{minipage}
\end{figure*}
\end{center}

\begin{center}
\begin{figure*}[h]
\begin{minipage}{\textwidth}
\centering 
\begin{overpic}[abs, unit=1mm, scale=0.65]{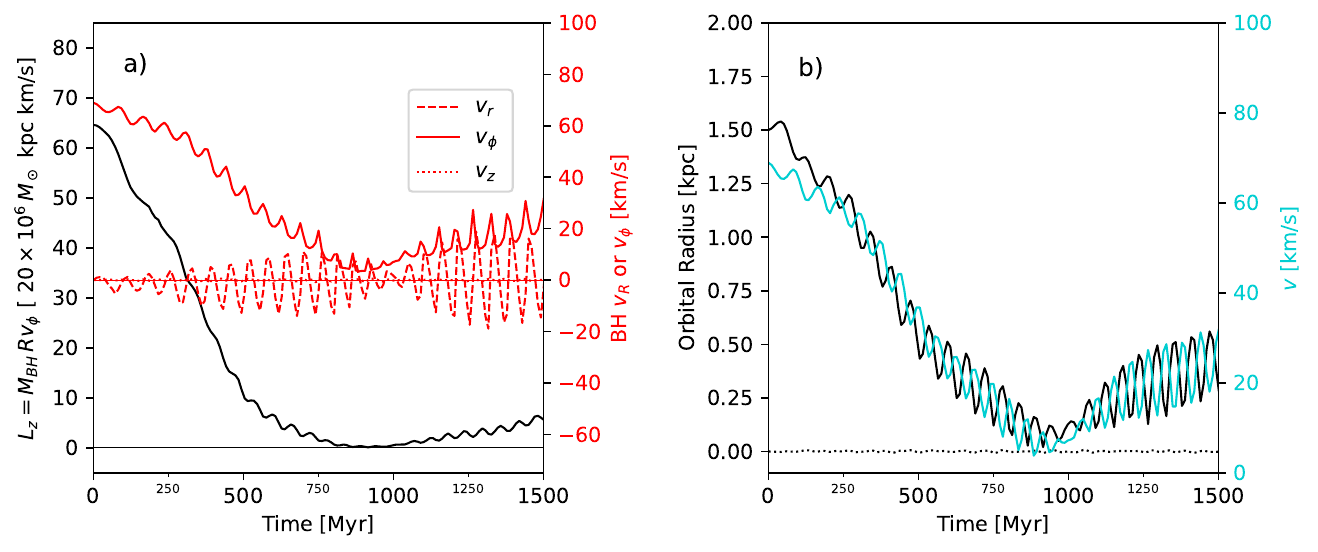}
\put(18,50){ { \begin{minipage}{2.3cm}  {\small \textit{circular, cool disc}}  \end{minipage} } } 
\end{overpic} 
\caption{Runs of angular momentum, radius  and velocity components as function of time for an BH started on a circular orbit. a) The solid black curve graphs the angular momentum (left-hand axis with black labels) ; the right-hand axis is the scale for each velocity component (red curves) ; b) The cylindrical radius (in black, left-hand axis) and norm of the velocity (right-hand axis, in turquoise). The dotted line at the bottom is the $z$ coordinate bounded to $\pm 10 \pc$. The radius and velocity components reach a minimum at $t = 915 \unit{Myr}$.  }
\label{fig:BH_LR_CircularCold}
\end{minipage}
\end{figure*}
\end{center}
\begin{center}
\begin{figure*}[h]
\begin{minipage}{\textwidth}
\centering 
\begin{overpic}[abs, unit=1mm, scale=0.65]{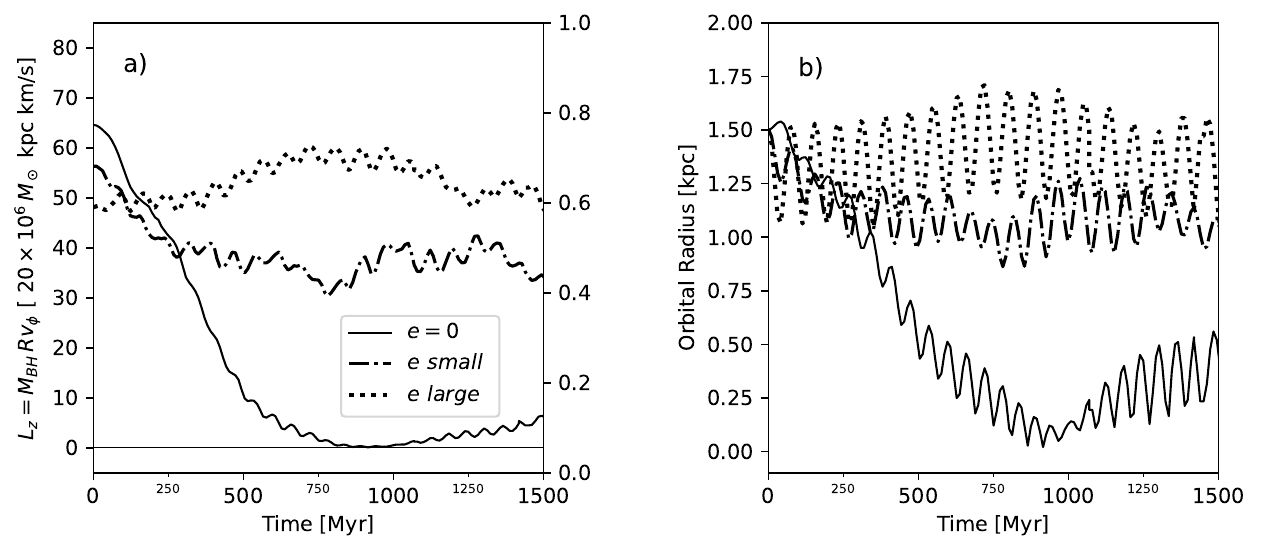}
\put(20,50){ { \begin{minipage}{2.5cm}  {\small \textit{eccentric, cool disc}}  \end{minipage} } } 
\end{overpic} 
\caption{Runs of angular momentum and radius  for three BH orbits: circular; an elliptical orbit with {small} eccentricity $e \approx 0.55$; and a third 
case with a {large} $ e \approx 0.67$. a) The run of angular momentum for each case as indicated in the legend; the solid black curve graphs $L_z$ 
for the circular initial orbit displayed on Fig.~\ref{fig:BH_LR_CircularCold}.  b) The cylindrical radius  as function of time for the same three cases. Note how the BH orbit with the more elliptical orbit actually evolves at a near-constant mean radius. }
\label{fig:BH_LR_MultipleCold}
\end{minipage}
\end{figure*}
\end{center}

\subsection{Dynamically cold discs and BH orbits with angular momentum}\label{sec:BHwithmomentum}
We turn to a more likely situation where the BH starts off on an orbit that carries angular momentum. The case of the circular 
orbit is the most illustrative. The radial- and azimuthal velocity dispersion of the stars mean that the BH falls victim of the 
asymmetric drift \citep[][\S4.4, p. 326]{BT08}, implying a faster BH velocity in azimuth than the mean for the surrounding stars. 
 The faster BH motion causes a net negative drag (from the stars) and a loss of angular momentum. In practice the stellar velocity dispersion 
 is so low that the asymmetric drift causes a difference of only $ \sim 1 \kms$ with circular motion, much lower still than the 1D dispersion of $\approx 4 \kms $ (see Tab.~\ref{tab:parameters}). We expect the asymmetric drift to have a negligible impact here. 
 
 On the other hand, a reduced stellar velocity dispersion  boosts the local gravitational focusing of the stars by the BH. 
 This phenomenon enhances the number of stars that bind with the BH. At the spatial resolution $r = l = 125 \pc$, 
 the circular velocity    in the BH's gravitational field 
 
 \[ v_c = \left( \frac{G\Mbh}{l} \right)^{\half} \simeq 18 \kms\] 
is much higher than the 3D velocity dispersion $\approx 7 \kms$ of the unperturbed stellar disc component. A significant 
number of stars in the annulus $R = 1.5 \pm 0.125 \kpc$ will therefore interact strongly with the BH (the annulus covers close to 2.8 \% of the disc's surface but contains about $5\%$ of the disc's mass ; the BH weighs $\simeq 0.8 \%$ of the disc). 

\subsubsection{Evolution from a circular orbit}
The time evolution is mapped out on Fig.~\ref{fig:BHCircularOrbit}. It is noticeable on the frames running from $t = 40 \unit{Myr}$ to $308 \unit{Myr}$ that the BH attracts stars and "cleans up" its orbit so that a trail of low-density (a gap) forms. The bridges of higher density that form on either side of its orbit 
lead to the  efficient angular momentum transport and orbital evolution, as in proto-planetary circumstellar disc dynamics \citep[so-called Type-II migration; \eg,][]{kley2012,Papaloizou2021} . 
An importance difference is that no ram pressure exists in the disc material, and the spiral features  are never close 
to axial symmetry. Despite these caveats, it is clear that gravitational {torques} acting on the BH's orbit, rather than the 
net effect of dynamical traction / friction, drive this rapid inward migration : a wake of stars  slowing down the  BH would show  up as a trailing over-density on Fig.~\ref{fig:BHCircularOrbit}, in the frames $t = 61 \unit{Myr}$ to $308 \unit{Myr}$. Instead we 
find an area devoid of stars in front and behind the BH. Compare this situation with the case of the radial orbit (Fig.~\ref{fig:BHRadialOrbit}; see also Fig.~\ref{fig:Clump452}): clearly for that case, an over-dense  wake trails the motion of the BH. 
The rapid evolution of the disc response over a few periods lead the clumps of disc material to interact and merge with each other, eventually forming 
dense substructures of roundish shape. That situation is depicted on frame $t = 915 \unit{Myr}$ when the BH is within $20 \pc$ of the centre, virtually 
at rest since $\vbh \simeq 9 \kms$ measured at that time lies within the numerical uncertainty for a BH starting at rest (Fig.\ref{fig:BHTimeEvol}). 

It is important to note that the fragmented mass density at $t = 915 \unit{Myr}$ breaks the axi-symmetry of the system. A direct consequence of this is that the BH does 
not remain still, but is pulled out of the origin by one large clump, or a few  small ones. 
The energy and angular momentum gained by the BH means that by 
$t = 1553 \unit{Myr}$, it is now orbiting at a radius $ R \approx 300 \pc$; it has acquired significant angular momentum which persists until at least 
$t = 1750 \unit{Myr}$, when the disc material defines a  bar-like potential on  a scale of $\sim 500 \pc$ (the BH radial orbit is then at a radius $\simeq 710 \pc$). 
The run of radius, velocity components and angular momentum is plotted on Fig.~\ref{fig:BH_LR_CircularCold}. It is striking how the angular momentum drops to essentially zero at $t \approx 915 \unit{Myr}$, and rises systematically thereafter. The same comment applies to its orbital radius, seen on 
Fig.~\ref{fig:BH_LR_CircularCold}(b). The fact that the BH's orbit does not circularise exactly (small but persistent oscillations in radius $R$) may be one reason why symmetry breaking is enhanced  near the origin, \ie, on a scale matching the BH's 
radius of influence, or $ r \approx 300 \pc$.

\subsubsection{Eccentric orbits}
To underscore the singular character of the (initially) circular BH motion, we ran two more cases with the BH starting on  mildly eccentric orbits. 
On Fig.~\ref{fig:BH_LR_MultipleCold}, we display the runs of orbital angular momentum $L_z$ (panel [a]) and radius (panel [b]) for the initially circular 
orbit discussed above ($e = 0$), along with two orbits that begin with lower $L_z$. The first one has $L_z \rightarrow 5 L_z / 6$ leading to  an orbit of small eccentricity $(e \approx 0.55$) ; the second one has $L_z$ reduced by a second factor $5/6$ so $L_z \rightarrow 0.7 L_z$ compared to circular motion, and a  larger orbital eccentricity  ($e \approx 
0.67$). We estimated the eccentricities from the ratio of radii at peri- and apo-centres. 

To understand the trends 
in $L_z$ and radii over time, it is useful to think in terms of epicyclic motion. In this regime, the angular momentum is conserved as the object moves in and out radially. We may assume that this is true of the unperturbed stellar orbits. 
Here, the BH may dominate  the gravitational potential locally and polarises stellar orbits of lower and higher $L_z$, inside and outside of its orbit, 
respectively. When the BH starts off on  a nearly circular orbit, its own epicycle does not generate significant radial motion 
compared with the radial stellar velocity dispersion $\sigma_{R} \simeq 4.3 \kms$. Thus in  the limit  of small $e$, 
the BH act as a gravitational focal point. It  is able to open a gap along its orbit, when it becomes a catalyst for 
angular momentum transfer, from the stars 
inside its orbit, to the stars outside it. This is a very similar situation to the instability studied by \citep{JT1966} for a thin, 
cold fluid of stars. 
On the contrary, when the BH is launched from a low-angular momentum orbit, of higher 
eccentricity $e$, it is unable to clear a gap in the stellar population because its own significant radial motion 
effectively boosts its velocity relative to the stars. This in turn makes the BH a less effective gravitational attractor. The BH orbit 
labelled $large\ e \approx 0.67$ on Fig.~\ref{fig:BH_LR_MultipleCold}(b) gives a mean-amplitude (rms) radial velocity of $v_R 
\simeq 6.8 \kms $ 
which is in fact larger than the stellar radial velocity dispersion, at least initially. 
Hence the BH stops being a point of gravitational focusing, and 
fewer stars respond strongly to its gravitational pull. A gap does not form, and the BH orbit preserves its angular momentum 
more efficiently. 

\subsection{Transition from gravitational torques to dynamical friction} \label{sec:transition}
Our study of BH orbital evolution points to a more complex picture than for warm discs. The time evolution 
of a BH orbit initially with zero angular momentum is first driven by dynamical friction in competition with 
dynamical traction (see \S\ref{sec:warmradial}). This happens both in a dynamically warm- or 
cold stellar disc, and can lead to a transition, from a contracting orbit losing angular momentum, to an expanding orbit 
gaining angular momentum. When the disc is cold and the BH starts on  a near-circular orbit, the mechanics at play is 
not that of dynamical friction because a massive body will suck in stars that are close in phase-space: the formation of a 
gap of under-dense regions stretching in front and behind the massive body is the signature of this process (Fig.~\ref{fig:BHCircularOrbit}).  We have argued, based on two cases of BH orbits that start on mildly eccentric 
orbits, that gap formation shuts off if the BH is on a sufficiently eccentric orbit, and the interplay of dynamical friction / traction 
kicks in. This transition to an evolution of the orbital momentum driven by gravitational focusing (friction or traction) 
occurs at a critical orbital angular moment for fixed values of $\Mbh$ and stellar velocity dispersion $\sigma_\star$ (the 
radial component mostly controlling the Toomre stability parameter $Q$, cf. Eq.~\ref{eq:ToomreQ}). 
We have not identified a relation between $Q, L_z, $ and $\Mbh$ that quantifies this transition. We anticipate that this point is worthy of a study all by itself, to which we hope to return in a future contribution. In the next section, we seek to justify in more 
quantitative details the application of the Fokker-Planck treatment to cold systems. 

\begin{figure*}[h]
\centering
\includegraphics[scale=0.600]{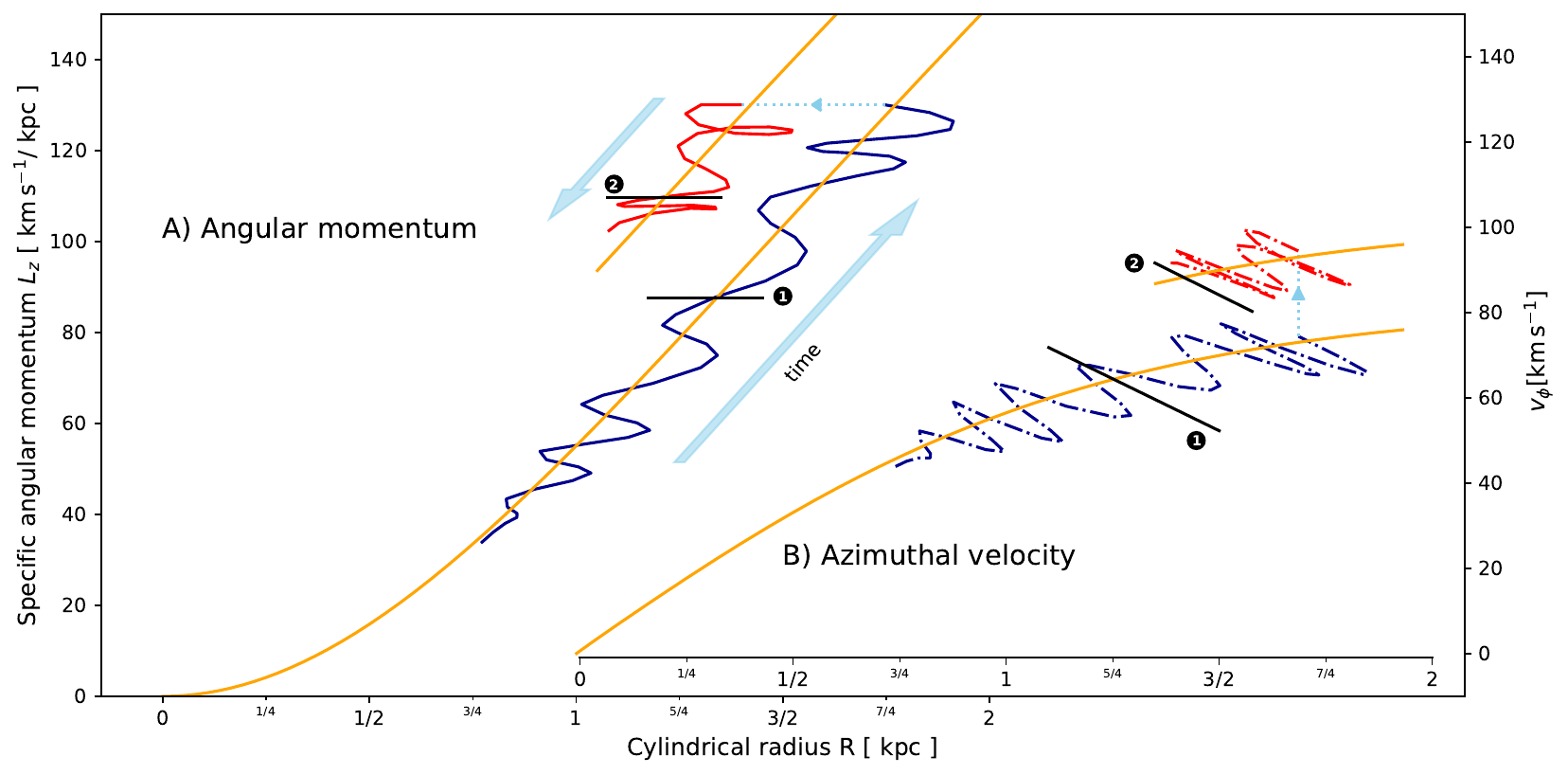} 
\caption{A) The figure graphs the specific angular momentum $L_z$ as a function of the cylindrical radius $R$ during the 
orbital migration of the BH ; the sky-blue  large arrows indicate the run of time $t$. The orange solid lines trace the expectation  
values for the model galaxy. The curve in blue (red) shows the 
coordinates before (after) $L_z$ has reached a maximum. The black lines indicate evolution at constant $L_z$. The billiard balls 
numbered 1 and 2 single out specific phases of evolution. B) Run of azimuthal velocity $v_\phi$ against radius $R$. The axes 
are shifted for clarity. The colour-coding for the other curves and symbols  is as for (A).}
\label{fig:LandV_vsR}
\end{figure*}

%
\section{Comparison with  theory} \label{sec:theory}
Our analysis  of kinetic energy diffusion of Appendix \ref{sec:FP}  was developed with a 
smooth homogeneous background in equilibrium in mind. The surface density maps of our numerical cold discs models, indeed images of  high-z galaxies (\eg, the  GOODS survey, \citet{elmegreen2007}), on the contrary, generally point to a complex, fragmented mass 
distribution in the core region visited by the BH. We limit our exploration to the case when the BH starts on a radial orbit 
in a dynamically cold disc (Table~\ref{tab:parameters}).  
We want to assess whether the basic picture of repeated gain and loss 
of azimuthal velocity is indeed at work in the later stages of evolution, when the BH has acquired angular momentum from the  streaming of stars (cf. Fig.~\ref{fig:Ekrates} and Appendix \ref{sec:FP}). 

To that end, we graph on Fig.~\ref{fig:LandV_vsR} the runs of specific angular momentum $L_z$  and azimuthal velocity $v_\phi$ as a function of the cylindrical radius $R$ for the BH orbit displayed on Figs.~\ref{fig:BHRadialOrbit} and 
 \ref{fig:Clump452} . The rotation curve for the 
model galaxy is displayed in solid orange, and two phases of evolution are displayed with different colours: prior to $L_z$ reaching 
a maximum (in dark blue); and at later stages when $L_z$ decreases episodically (in red). The two phases are shifted for clarity (the light-blue dotted lines connect the end-point of one phase, to the start of the next), with large light-blue arrows indicating the flow of time. 
The output data was sampled at intervals of 10 Myr, from $t = 1008 \unit{Myr}$ to $t = 1800 \unit{Myr}$, with the transition through a peak $L_z$ value at $t = 1522 \unit{Myr}$. 

We discuss first the early phase, from $1008$ to $1522 \unit{Myr}$ (dark blue curves). 
 Fig.~\ref{fig:LandV_vsR}(a) graphs  continued increase in $L_z$ with time, despite significant fluctuations (positive or negative) in  radial migration. Let us focus on a time slice around the coordinates marked with the circled number \Circled{\tiny\textbf{1}}.  
Near the symbol \Circled{\tiny\textbf{1}}, the steady increase in $L_z$ is evident as the radius increases from $\approx 1.2$ to $1.5 \kpc$; notice how $L_z$ (in blue) sits {above} the expected value (orange curve) at $R \approx 1.2 \kpc$. 
Consider now the evolution of $v_\phi$ over the same interval, where we find again that the BH has excess azimuthal velocity at $R = 1.2 \kpc$, sitting above the 
orange curve. Its larger azimuthal velocity means that it jumps ahead of the stars, which exert a negative acceleration on it. Therefore $v_\phi$ decreases in absolute 
terms as we move forward in time, to a larger radius $R$. The evolution at constant $L_z$ would make $v_\phi$ drop and fall on a path parallel to the black straight line, 
which is clearly not the case (Fig.~\ref{fig:LandV_vsR}[b]). In fact, the azimuthal velocity still decreases such that it crosses the orange curve, and soon the BH has a significant deficit of 
azimuthal velocity (dot-dash blue line, near \Circled{\tiny\textbf{1}} on the figure). The situation is now reversed, with the BH lagging behind the stellar flow at that 
radius: the dynamical {traction} exerted by the stars now pulls it forward and it soon catches up and overtakes the stellar flow. This see-saw pattern repeats itself 
up to $ R \approx 1.8 \kpc$ which is near the point where the rotation curves flattens off; we find for our model galaxy  that  $d v_\phi / d R < 0 $ when $ R \gtrsim 2 \kpc$,  off-scale on Fig.~\ref{fig:LandV_vsR}(b).

The second time slice of interest is marked with the circled number \Circled{\tiny\textbf{2}}. In this   later stage, the orbit  
evolves differently, with episodic loses of angular momentum, followed by time intervals of near-constant $L_z$. This is illustrated most  
clearly near the points \Circled{\tiny\textbf{2}} on  Fig.~\ref{fig:LandV_vsR}(a), when the horizontal black line (of constant angular momentum) traces well the orbit of the BH ; in (b), the case of the azimuthal velocity, the black curve as a slope matching that of the data, in red. It is interesting to observe that the 
time intervals of near-constant $L_z$ are close to one full orbital time, or $\approx 100 \unit{Myr}$.
We suspected that this could be an indication of a bar-like structure or some other non-axisymmetric density fluctuations 
developing inside the BH's orbit. But the much increased radial stellar velocity dispersion, now exceeding $\sigma_R \approx 18 \kms$ close to $R = 0 \kpc$  (it started at $4.2 \kms$) works against bar- or any substructure formation.  
 A casual inspection of density maps, not shown here, instead suggests that at the BH's orbit of $R \approx 1.7 \kpc$ where 
 $\sigma_R \approx 8 \kms$ is much lower, the BH is still able to open a gap (albeit of low contrast) in the stellar distribution. 
 In other words, the drop in angular momentum is most likely  caused by gravitational torques, which here again work
  to lower its orbital 
 angular momentum. We have not investigated the duration of such torques or their net impact quantitatively 
 but suspect that they account for episodes of rapid angular momentum loss. 
By contrast, a steady decline in $L_z$ would be attributable to dynamical friction. 
This interpretation implies that the time-evolution of the BH's orbit is coupled to both the inner- and outer disc stellar \
populations through their velocity dispersion, \ie their local stability to Jeans modes of fragmentations. A future study with 
spectral analysis should provide a more definitive diagnostic of angular momentum transport in these systems.

A related question which sheds light on the analysis is whether the time derivative  $\rate{v}{\|}$ obtained numerically 
matches the values computed from (\ref{eq:diffrates}) and (\ref{eq:Rosenbluth}). We tailor these equations 
 to the numerical set-up by fixing $N_s  = 1\times 10^5$ and $ m_\star = 1.6 \times 10^4\solarm $.  We chose to round up the 
 ratio  
$\Mbh / m_\star$ from 731,  as for the curves displayed on Fig.~\ref{fig:Ekrates}, to $ \simeq 10^3$. This simplifies the algebra and 
is of little consequence to the outcome.  We also set $\sigma_\bot = \sqrt{2} \sigma_\star$ for the two-dimensional velocity dispersion. All quantities can be expressed in units of $(\kms)^2 / \unit{Myr}$ with a 
single multiplicative constant. In the case of $\sigma_\star = 10 \kms$, we read on Fig.~\ref{fig:Ekrates} a value of 

\[ v_\| \rate{v}{\|} \simeq 40\ (\kms)^2/\unit{Myr} \]
when $v_\| \simeq 60 \kms$. This velocity component can also be read off Fig.~\ref{fig:BH_LR}(b) in the time interval  $> 1000 \unit{Myr}$. 
If we isolate for  $\rate{v}{\|}$ from the above relation, we find that $\rate{v}{\|} \approx 0.67 \kms/\unit{Myr}$. A naive time-integration over 
the sampling interval $\mathrm{d}t = 10 \unit{Myr}$ leads to $\Delta v_\| \approx 6.6 \kms$. Inspection of the output logs from the numerical integration
gives us typical time-variations $\Delta v_\phi \approx 4$ to $5 \kms$, over the same time interval. In the case when $\sigma_\star = 20 \kms$, the same 
exercise leads to $\rate{v}{\|} \approx 1/6 \kms/\unit{Myr}$ and $\Delta v_\| \approx 1.66 \kms$ (same sampling rate). Thus we conclude that the
analytical rates in (\ref{eq:diffrates}) are of the right amplitude, but fall short when the dispersion $\sigma_\star$ increases much above $10 \kms$. 
Inspecting the runs for  times $t \ge 329 \unit{Myr}$ we 
found a steady radial decrease in $\sigma_\star$ in the interval $R = 0$ to $2 \kpc$, when $\sigma_\star$ falls from  $\ge 20 \kms$, to $\approx 10 \kms$. 
In other words, the variations in $\Delta v_\phi$ become larger as we move outwards since locally $\sigma_\star$ is reduced. This comforts the conclusion that dynamical traction accounts quantitatively for much of the outward orbital evolution of the BH.

%

\begin{center}
\begin{figure*}[h]
\usebox{\maboiteA{}} 
\centering 
\includegraphics[scale=0.65]{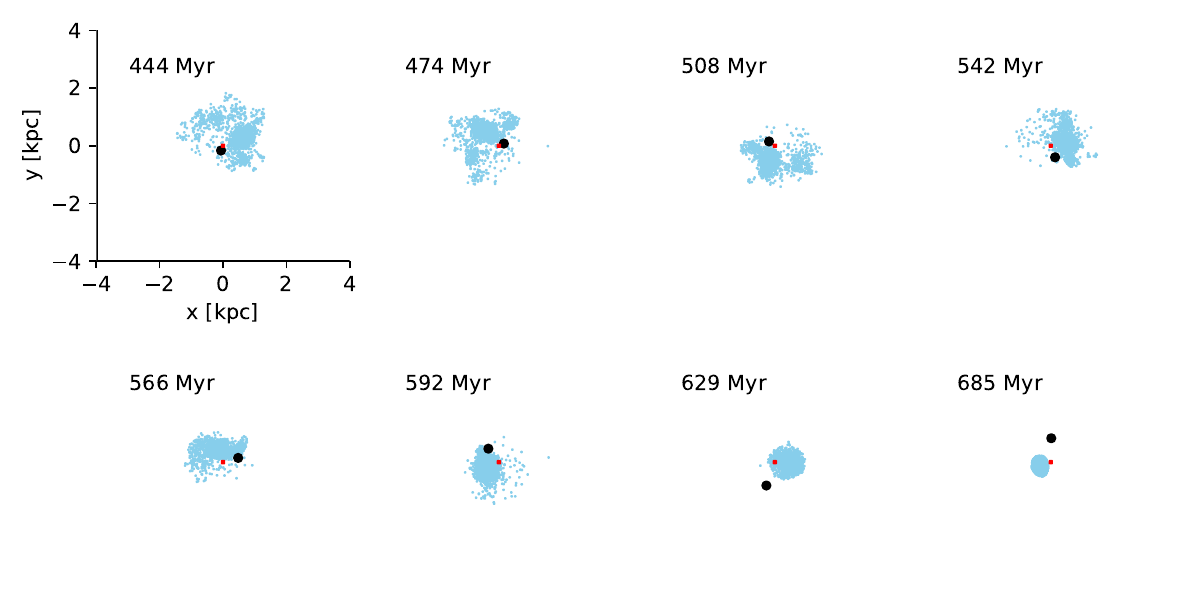}\\
\vspace{-5.15cm}
\usebox{\maboiteB{}} 
\includegraphics[scale=0.65]{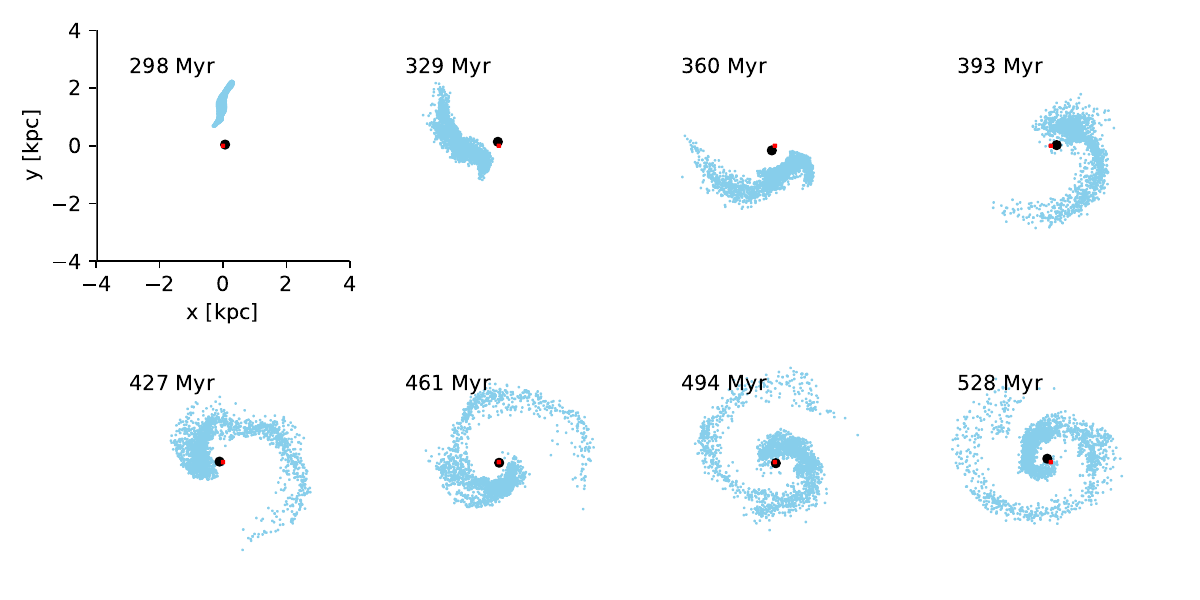}
\vspace{-5.05cm}
\caption{Two examples of the interaction between 
the central BH and  stellar sub-structures. In (A, top row), when the background spherical Isochrone 
 potential is partially frozen, the disc stars form clumps of higher mass and binding energy: such structures may dislodge 
the BH. This two-stage instability then grows as the BH extracts angular momentum from the flow of rotating stars around the centre. 
In (B, bottom row) the situation for a self-gravitating disc + Isochrone potential means that the clumps are much less massive and of a lower 
binding energy. In that case, a clump will dissolve in the strong tidal field of the BH. The centre of coordinates is indicated with a red dot. See text for more details. 
}
\label{fig:SelfGravityRuns}
\end{figure*}
\end{center}

\section{Discussion}\label{sec:discussion}
The problem of a massive perturber evolving in cored galactic potentials was revisited here with a focus on a background of stellar orbits with non-zero streaming motion drawn from a simple distribution function. 
A local treatment of gravitational focusing derived from Fokker-Planck diffusion coefficients was performed. The effects of resonant motion was left out of the analysis. A consequence of anisotropic gravitational focusing,  
we argued, is that the steady streaming of stars leads to an effective {traction} on the BH after a few orbital periods (see Appendix \ref{sec:FP} and Fig.~\ref{fig:VvsT}).  Dynamical traction competes with the more familiar Chandrasekhar {friction} 
and we have illustrated the transition from one regime to the other for an BH on  a radial orbit in a hot disc (Fig.~\ref{fig:BH_LRhotcircular}). This is the expected evolution for low-angular momentum BH orbits.

Theoretical  expectations are largely fulfilled in 
configurations where the streaming stars are stable against Jeans fragmentation modes (dynamically hot discs). 
This is not so with cold discs, where we found that stellar sub-structures are key players in the orbital evolution of the BH. 
Indeed stellar clumps shut out dynamical traction almost entirely, so that an BH may reach the galactic centre on a time-scale 
of $\sim 300 \unit{Myr}$ (compare Figs.~\ref{fig:BH_LRhotcircular} and \ref{fig:BH_LR}). In this idealised framework, 
the formation of stellar clumps is triggered by the response of the stellar orbits to the migrating  
BH (no clumps form when the BH is at rest at the origin: Fig.~\ref{fig:BHatrest}). Thus when the BH reaches the centre, the stellar mass distribution is far from axial symmetry. The configuration will be unstable if and when sizeable clumps 
orbit near the centre. We spoke of a two-stage instability whereby an $m = 1$ mode pulls the BH out of the centre, at which 
point dynamical traction boosts its angular momentum and outward migration ensues. 
Hence the degree of binding energy of these clumps (their internal cohesion, not resolved properly in our numerical integrations)
is crucial to the stability of the configuration. We address this specific point below.  

\subsection{Fragmentation and self-gravity}
Fragments that develop through Toomre-unstable 'cold' stellar orbits in the disc triggered by the BH perturbation  mimic observations of young galaxies undergoing active star formation. The extended and spherical (isochrone) component made up of 'hot' stellar orbits couples more weakly with the BH;  we froze the spherical component for that reason in the reference numerical set-up. 
 A drawback is that the stellar clumps themselves  do not exchange 
 mechanical energy with the Isochrone potential (by polarising orbits from that component, for instance). In that way the binding energy of the 
 clumps is enhanced artificially, and we expect that relaxing the constraint of a frozen component will change the outcome of the 
 numerical integration significantly. From here onward we use the name 'bulge' as short-hand for spherical isochrone stellar component. 
 
To inspect the impact of a live bulge, we performed two calculations, the first with a fraction of live bulge orbits such 
that they match in number (and mass) those of the live disc, or $100 k$ mass elements each ; and a second run with a fully self-gravitating bulge and disc. The second run was performed with 
the Am$\mu$se GPU-enhanced tree code \textsc{Bonsai} with a total of $8.4 \times 10^6 $ mass elements. The model parameters and bulge/disc mass ratios are those of the standard model of Table~\ref{tab:parameters}. The kernel sizes for numerical integration was reduced to $ l = 32 \pc$ to account better for the dynamics of the clumps. 

Fig.~\ref{fig:SelfGravityRuns} graphs for each case the distribution in space of a subset of stars that form a bound sub-structure at a time when the BH has reached the centre. 
The run shown on the top row (a) for a partially self-gravitating bulge yielded stellar clumps of strong binding energy. The run shows how the clumps merge together and form a structure which far outweighs the BH (by $4 \times$ its mass). The large structure of $\simeq 3318 $ mass elements  seen at the end of the sequence hovers close to the centre of coordinates indicated with a red dot. At the end of the time sequence, we find that $90.3\%$ of the mass elements identified as 
belonging to the clump are bound together. This is a prime example of the important r\^ole that the net clump binding energy can have on 
the stability of the BH's position in the galaxy. 

Compare these results with the second run displayed on Fig.~\ref{fig:SelfGravityRuns}(b), where now the whole system is 
self-gravitating. Now some $2887$ mass elements add up to $\simeq 3.8 \times 10^6 \solarm$ ( 1/3 the mass of the BH). 
The important difference is that now only about $62\%$ of the mass of the clump has negative binding energy. As soon as 
it approaches the centre the tidal field of the BH dissolves the structure which forms a loose spiral pattern at the end of the 
run. The BH remains at the heart of the system, and may accrete stars and (possibly) form a nuclear star cluster. 
No instability will develop that can remove the BH from the centre unless more massive sub-structures form in the course of 
time: nothing of the sort happened over the course of $1.5 \unit{Gyr}$ of evolution for that model. 

 These calculations tell  us that if the anisotropic stellar stream is embedded in  a large, much more massive hot stellar component of 
 nearly isotropic velocity field (like the current bulge), then fragmentation modes that are triggered by a migrating BH are 
 unlikely to develop strong biding energy and create a situation whereby the massive compact object can be ejected from the 
 centre. The BH migrates to the centre on a time-scale $\sim 300 \unit{Myr}$ as anticipated from dynamical friction. 
 We have not determined the minimum mass ratio of isotropic-  to anisotropic (disc-like) components that will lead to this 
 configuration: the calculations that we performed suggest that this ratio has to be at least 2 : 1. It would be interesting to extract the degree 
 of anisotropy (essentially, rotation) from calculations of galaxy formation simulations that resolve the central kpc adequately 
 to provide a better set of constraints on the initial configuration to use for a refined study of BH migration. 

\subsection{Cool and warm discs}
During the early stages of evolution of a low-$L_z$ orbit, the BH undergoes dynamical friction. 
Two cases were considered, one where the disc is dynamically warm, the other, cool. 
No substructures 
form in  a warm disc,  and the evolution is very smooth (Fig.~\ref{fig:BH_LRhot}).  

The situation is much different in cooler environments, when the disc fragments and density waves carry angular momentum outward. 
The BH  may well sink to the centre, but be ejected by a symmetry-breaking configuration of stellar clumps which it triggered on 
the way in. When that happens. the BH acquires angular momentum by dynamical traction from the surrounding
 streaming stars. We have checked that the 
BH's azimuthal velocity $v_\phi$ increases at a rate well matched with Fokker-Planck energy diffusion coefficients in that situation (Fig.~\ref{fig:LandV_vsR}). 
This is important because these coefficients are obtained from {local measurements} of rotation velocity,  dispersion and surface brightness. 
Therefore the dynamical state of an off-centred BH in an external galaxy could possibly be gauged directly from a well-resolved young galaxy with a rotating bulge (or highly anisotropic bulge). 
Several such galaxies have long been observed at red-shifts $z \sim 1 $ to 3  \citep[\eg,][]{elmegreen2007}.  The recent 
JWST detection of an off-centre AGN at $z \sim 7.3$, when the Universe is $\approx 700 \unit{Myr}$ old \citep{ubler2024}, offers a great opportunity to tests and understand better BH dynamics in the early stages of the formation 
of a  galaxy. The environment 
of a BH in a young galaxy should include gas dynamics and the formation of stars for a realistic diagnostic of the internal 
dynamics at the core of young galaxies.  We stressed already how the binding energy of stellar sub-structures, surely determined 
by dissipation processes at formation, will impact on the stability of the BH at the centre of galaxies. 
Analyses of the impact of gas on the orbit of a massive perturber include, \eg,  
\citet[][]{ostriker1999, chapon2013} for radiative gas;  and the recent work  by \citet{lescaudron2023} which includes turbulence. These last authors find a direct link between 
the amplitude of the turbulent power-spectrum and the orbital evolution of the BH.  The density fluctuations brought by turbulent modes parallel the role played here by stellar clumps. There is a bridge to be built between such studies, because a star-forming region will 
proceed from gas cooling in turbulent configurations, and form large stellar aggregates \citep{renaud2021}. 
It is intriguing that the 
BH discovered in the GA-NIFS survey \citep{ubler2024} of an estimated mass $\simeq 5 \times 10^7 \solarm$ is off-set from the host galaxy's 
photometric centre by $R \simeq 1 \kpc$ in projection (see their Fig.~2; the broad-line region that is clearly mis-aligned). Should the physical 
scales be confirmed, the BH displacement would be on the same scale as the one we found in our models. More detailed modelling 
is required to assess the likelihood of an event like this, yet such observations point to a physically plausible configuration. 

%
\section{Conclusions}\label{sec:conclusions}
We have presented analytical arguments and showed with numerical N-body calculations that 
the mechanics of dynamical traction (see \S\S\ref{sec:traction} and \ref{sec:transition})  works hand-in-hand with the classic picture of orbital decay of a massive BH 
derived from dynamical friction ; they are the same phenomena of gravitational  focusing, but work with 
opposite effect to each other: an anisotropic velocity field drawn from a d.f of the form $f(E,L_z)$ may 
lead to a net {gain} of angular momentum by a massive BH and outward migration. This is so 
for a dynamically warm stellar velocity field. The competition between dynamical friction vs traction causes  
a continued transition  in solution space, when one effect comes to dominate over time.

A fragmented, dynamically cold configuration helps  bring the perturber BH to the barycentre
faster, but these very fragments make it difficult to keep it there: we have identified a two-stage 
instability that results in outward migration of a massive BH by dynamical traction. 
No dissipative terms were included in our analysis: the scale-free nature of gravitational 
dynamics mean that all quantities of time, mass and length can be rescaled to fit observational constraints. 
We have focused on a galactic system of total mass = $1.16 \times 10^{10} \solarm$ in equilibrium with a 
characteristic length $a = 1.5 \,\kpc$. The orbital {outward} migration of an $\Mbh = 1.25 \times 10^7 \solarm$ BH in this system may reach $ \sim 1 \,\kpc$ 
over a time-scale of $\sim 750 \unit{Myr}$ to $1 \unit{Gyr}$. It is noteworthy that the recent AGN 
identified by the JWST  at red-shift $ z \simeq 7.3$ shows an off-set from its photometric centre by the same 
length in projection \citep{ubler2024}. On-going and future work with fully self-gravitating calculations should shed light 
on the dynamics of young galaxies by resolving better the core region of these fascinating systems. 

\begin{acknowledgements}
We thank the anonymous referee for a thoughtful report that led to a much improved manuscript. 
We acknowledge  G. Monari, B. Famaey and F. Renaud for their comments and suggestions about this work. K. Kraljic also pointed out several recent papers bearing on black hole dynamics in galaxies. 
The research reported here made extensive use of the community projects Agama and \amuse. We thank the developers for 
making their software available (and documented). 
    Part of this work was supported by the French-Ukrainian DNIPRO research grant
     number Ts~17/2--1 awarded in 2022. Carbon footprints: this project made use of 1680 hours of organically-grown intelligence (OGI), 
     equivalent to $\sim 30 \unit{kWh}$ of energy \citep{brainpower}, together with 6000 CPUh of  computational resources expended at the HPC centre of the
University of Strasbourg (grants g2023a97-c and -g) during the academic year 2023-24. We thank M. Ringenbach and D. Brusson for support. 
\end{acknowledgements}

\bibliographystyle{aa}  
\bibliography{thispaper,biblio_local}

\begin{thebibliography}{80}
\expandafter\ifx\csname natexlab\endcsname\relax\def\natexlab#1{#1}\fi

\bibitem[{{Angl{\'e}s-Alc{\'a}zar} {et~al.}(2015){Angl{\'e}s-Alc{\'a}zar},
  {{\"O}zel}, {Dav{\'e}}, {Katz}, {Kollmeier}, \& {Oppenheimer}}]{alcazar2015}
{Angl{\'e}s-Alc{\'a}zar}, D., {{\"O}zel}, F., {Dav{\'e}}, R., {et~al.} 2015,
  \apj, 800, 127

\bibitem[{{Arg{\"u}elles} \& {Collazo}(2023)}]{arguelles2023}
{Arg{\"u}elles}, C.~R. \& {Collazo}, S. 2023, Universe, 9, 372

\bibitem[{{Bahcall} \& {Wolf}(1976)}]{bahcallwolf}
{Bahcall}, J.~N. \& {Wolf}, R.~A. 1976, \apj, 209, 214

\bibitem[{{Balasubramanian}(2021)}]{brainpower}
{Balasubramanian}, V. 2021, Proc. Natl. Acad. Sci., USA National Library of
  Medicine, 32, 118

\bibitem[{{Banik} \& {van den Bosch}(2021)}]{banik2021}
{Banik}, U. \& {van den Bosch}, F.~C. 2021, \apj, 912, 43

\bibitem[{{Banik} \& {van den Bosch}(2022)}]{banik2022}
{Banik}, U. \& {van den Bosch}, F.~C. 2022, \apj, 926, 215

\bibitem[{{Barnes} \& {Hut}(1986)}]{BH1986}
{Barnes}, J. \& {Hut}, P. 1986, \nat, 324, 446

\bibitem[{{Barro} {et~al.}(2024){Barro}, {P{\'e}rez-Gonz{\'a}lez}, {Kocevski},
  {McGrath}, {Trump}, {Simons}, {Somerville}, {Yung}, {Arrabal Haro}, {Akins},
  {Bagley}, {Cleri}, {Costantin}, {Davis}, {Dickinson}, {Finkelstein},
  {Giavalisco}, {G{\'o}mez-Guijarro}, {Hathi}, {Hirschmann}, {Holwerda},
  {Huertas-Company}, {Kartaltepe}, {Koekemoer}, {Lucas}, {Papovich}, {Pirzkal},
  {Seill{\'e}}, {Tacchella}, {Wuyts}, {Wilkins}, {de la Vega}, {Yang}, \&
  {Zavala}}]{barro2024}
{Barro}, G., {P{\'e}rez-Gonz{\'a}lez}, P.~G., {Kocevski}, D.~D., {et~al.} 2024,
  \apj, 963, 128

\bibitem[{{Bellovary} {et~al.}(2019){Bellovary}, {Cleary}, {Munshi}, {Tremmel},
  {Christensen}, {Brooks}, \& {Quinn}}]{bellovary2019}
{Bellovary}, J.~M., {Cleary}, C.~E., {Munshi}, F., {et~al.} 2019, \mnras, 482,
  2913

\bibitem[{{Binney} \& {Tremaine}(2008)}]{BT08}
{Binney}, J. \& {Tremaine}, S. 2008, {Galactic Dynamics}, $2^{\mathrm{nd}}$
  edn. (Princeton: The University Press)

\bibitem[{{Bland-Hawthorn} \& {Gerhard}(2016)}]{blandhawthorn2016}
{Bland-Hawthorn}, J. \& {Gerhard}, O. 2016, \araa, 54, 529

\bibitem[{{Bogd{\'a}n} {et~al.}(2024){Bogd{\'a}n}, {Goulding}, {Natarajan},
  {Kov{\'a}cs}, {Tremblay}, {Chadayammuri}, {Volonteri}, {Kraft}, {Forman},
  {Jones}, {Churazov}, \& {Zhuravleva}}]{seedBH2024}
{Bogd{\'a}n}, {\'A}., {Goulding}, A.~D., {Natarajan}, P., {et~al.} 2024, Nature
  Astronomy, 8, 126

\bibitem[{{Boily} {et~al.}(2008){Boily}, {Padmanabhan}, \&
  {Paiement}}]{boily2008}
{Boily}, C.~M., {Padmanabhan}, T., \& {Paiement}, A. 2008, \mnras, 383, 1619

\bibitem[{{Boylan-Kolchin}(2023)}]{boylan2023}
{Boylan-Kolchin}, M. 2023, Nature Astronomy, 7, 731

\bibitem[{{Chandrasekhar}(1943)}]{chandrasekhar1943}
{Chandrasekhar}, S. 1943, \apj, 97, 255

\bibitem[{{Chapon} {et~al.}(2013){Chapon}, {Mayer}, \& {Teyssier}}]{chapon2013}
{Chapon}, D., {Mayer}, L., \& {Teyssier}, R. 2013, \mnras, 429, 3114

\bibitem[{{Chatzopoulos} {et~al.}(2015){Chatzopoulos}, {Fritz}, {Gerhard},
  {Gillessen}, {Wegg}, {Genzel}, \& {Pfuhl}}]{chatzopoulos2015}
{Chatzopoulos}, S., {Fritz}, T.~K., {Gerhard}, O., {et~al.} 2015, \mnras, 447,
  948

\bibitem[{{Chen} {et~al.}(2023){Chen}, {Do}, {Ghez}, {Hosek},
  {Feldmeier-Krause}, {Chu}, {Bentley}, {Lu}, \& {Morris}}]{chen2023}
{Chen}, Z., {Do}, T., {Ghez}, A.~M., {et~al.} 2023, \apj, 944, 79

\bibitem[{{Chiba} \& {Kataria}(2024)}]{chiba2024}
{Chiba}, R. \& {Kataria}, S.~K. 2024, \mnras, 528, 4115

\bibitem[{{Chu} {et~al.}(2023){Chu}, {Boldrini}, \& {Silk}}]{chu2023}
{Chu}, A., {Boldrini}, P., \& {Silk}, J. 2023, \mnras, 522, 948

\bibitem[{{Cole} {et~al.}(2012){Cole}, {Dehnen}, {Read}, \&
  {Wilkinson}}]{cole2012}
{Cole}, D.~R., {Dehnen}, W., {Read}, J.~I., \& {Wilkinson}, M.~I. 2012, \mnras,
  426, 601

\bibitem[{{de Vita} {et~al.}(2018){de Vita}, {Trenti}, \&
  {MacLeod}}]{devita2018}
{de Vita}, R., {Trenti}, M., \& {MacLeod}, M. 2018, \mnras, 475, 1574

\bibitem[{{Elmegreen} {et~al.}(2007){Elmegreen}, {Elmegreen}, {Ravindranath},
  \& {Coe}}]{elmegreen2007}
{Elmegreen}, D.~M., {Elmegreen}, B.~G., {Ravindranath}, S., \& {Coe}, D.~A.
  2007, \apj, 658, 763

\bibitem[{{Finkelstein} {et~al.}(2023){Finkelstein}, {Bagley}, {Ferguson},
  {Wilkins}, {Kartaltepe}, {Papovich}, {Yung}, {Arrabal Haro}, {Behroozi},
  {Dickinson}, {Kocevski}, {Koekemoer}, {Larson}, {Le Bail}, {Morales},
  {P{\'e}rez-Gonz{\'a}lez}, {Burgarella}, {Dav{\'e}}, {Hirschmann},
  {Somerville}, {Wuyts}, {Bromm}, {Casey}, {Fontana}, {Fujimoto}, {Gardner},
  {Giavalisco}, {Grazian}, {Grogin}, {Hathi}, {Hutchison}, {Jha}, {Jogee},
  {Kewley}, {Kirkpatrick}, {Long}, {Lotz}, {Pentericci}, {Pierel}, {Pirzkal},
  {Ravindranath}, {Ryan}, {Trump}, {Yang}, {Bhatawdekar}, {Bisigello}, {Buat},
  {Calabr{\`o}}, {Castellano}, {Cleri}, {Cooper}, {Croton}, {Daddi}, {Dekel},
  {Elbaz}, {Franco}, {Gawiser}, {Holwerda}, {Huertas-Company}, {Jaskot},
  {Leung}, {Lucas}, {Mobasher}, {Pandya}, {Tacchella}, {Weiner}, \&
  {Zavala}}]{CEERS2023}
{Finkelstein}, S.~L., {Bagley}, M.~B., {Ferguson}, H.~C., {et~al.} 2023, \apjl,
  946, L13

\bibitem[{{Fouvry} {et~al.}(2015){Fouvry}, {Binney}, \& {Pichon}}]{fouvry2015}
{Fouvry}, J.-B., {Binney}, J., \& {Pichon}, C. 2015, \apj, 806, 117

\bibitem[{{Fritz} {et~al.}(2016){Fritz}, {Chatzopoulos}, {Gerhard},
  {Gillessen}, {Genzel}, {Pfuhl}, {Tacchella}, {Eisenhauer}, \&
  {Ott}}]{fritzetal2016}
{Fritz}, T.~K., {Chatzopoulos}, S., {Gerhard}, O., {et~al.} 2016, ApJ, 821, 44

\bibitem[{{Fujii} {et~al.}(2007){Fujii}, {Iwasawa}, {Funato}, \&
  {Makino}}]{fujii2007}
{Fujii}, M., {Iwasawa}, M., {Funato}, Y., \& {Makino}, J. 2007, \pasj, 59, 1095

\bibitem[{{Fukushige} \& {Heggie}(2000)}]{fukushige2000}
{Fukushige}, T. \& {Heggie}, D.~C. 2000, \mnras, 318, 753

\bibitem[{{Gaia Collaboration} {et~al.}(2023){Gaia Collaboration}, {Drimmel},
  {Romero-G{\'o}mez}, {Chemin}, {Ramos}, {Poggio}, {Ripepi}, {Andrae},
  {Blomme}, {Cantat-Gaudin}, {Castro-Ginard}, {Clementini}, {Figueras},
  {Fouesneau}, {Fr{\'e}mat}, {Jardine}, {Khanna}, {Lobel}, {Marshall},
  {Muraveva}, {Brown}, {Vallenari}, {Prusti}, {de Bruijne}, {Arenou},
  {Babusiaux}, {Biermann}, {Creevey}, {Ducourant}, {Evans}, {Eyer}, {Guerra},
  {Hutton}, {Jordi}, {Klioner}, {Lammers}, {Lindegren}, {Luri}, {Mignard},
  {Panem}, {Pourbaix}, {Randich}, {Sartoretti}, {Soubiran}, {Tanga}, {Walton},
  {Bailer-Jones}, {Bastian}, {Jansen}, {Katz}, {Lattanzi}, {van Leeuwen},
  {Bakker}, {Cacciari}, {Casta{\~n}eda}, {De Angeli}, {Fabricius}, {Galluccio},
  {Guerrier}, {Heiter}, {Masana}, {Messineo}, {Mowlavi}, {Nicolas},
  {Nienartowicz}, {Pailler}, {Panuzzo}, {Riclet}, {Roux}, {Seabroke}, {Sordo},
  {Th{\'e}venin}, {Gracia-Abril}, {Portell}, {Teyssier}, {Altmann}, {Audard},
  {Bellas-Velidis}, {Benson}, {Berthier}, {Burgess}, {Busonero}, {Busso},
  {C{\'a}novas}, {Carry}, {Cellino}, {Cheek}, {Damerdji}, {Davidson}, {de
  Teodoro}, {Nu{\~n}ez Campos}, {Delchambre}, {Dell'Oro}, {Esquej},
  {Fern{\'a}ndez-Hern{\'a}ndez}, {Fraile}, {Garabato}, {Garc{\'\i}a-Lario},
  {Gosset}, {Haigron}, {Halbwachs}, {Hambly}, {Harrison}, {Hern{\'a}ndez},
  {Hestroffer}, {Hodgkin}, {Holl}, {Jan{\ss}en}, {Jevardat de Fombelle},
  {Jordan}, {Krone-Martins}, {Lanzafame}, {L{\"o}ffler}, {Marchal}, {Marrese},
  {Moitinho}, {Muinonen}, {Osborne}, {Pancino}, {Pauwels}, {Recio-Blanco},
  {Reyl{\'e}}, {Riello}, {Rimoldini}, {Roegiers}, {Rybizki}, {Sarro}, {Siopis},
  {Smith}, {Sozzetti}, {Utrilla}, {van Leeuwen}, {Abbas}, {{\'A}brah{\'a}m},
  {Abreu Aramburu}, {Aerts}, {Aguado}, {Ajaj}, {Aldea-Montero}, {Altavilla},
  {{\'A}lvarez}, {Alves}, {Anders}, {Anderson}, {Anglada Varela}, {Antoja},
  {Baines}, {Baker}, {Balaguer-N{\'u}{\~n}ez}, {Balbinot}, {Balog}, {Barache},
  {Barbato}, {Barros}, {Barstow}, {Bartolom{\'e}}, {Bassilana}, {Bauchet},
  {Becciani}, {Bellazzini}, {Berihuete}, {Bernet}, {Bertone}, {Bianchi},
  {Binnenfeld}, {Blanco-Cuaresma}, {Boch}, {Bombrun}, {Bossini}, {Bouquillon},
  {Bragaglia}, {Bramante}, {Breedt}, {Bressan}, {Brouillet}, {Brugaletta},
  {Bucciarelli}, {Burlacu}, {Butkevich}, {Buzzi}, {Caffau}, {Cancelliere},
  {Carballo}, {Carlucci}, {Carnerero}, {Carrasco}, {Casamiquela}, {Castellani},
  {Chaoul}, {Charlot}, {Chiaramida}, {Chiavassa}, {Chornay}, {Comoretto},
  {Contursi}, {Cooper}, {Cornez}, {Cowell}, {Crifo}, {Cropper}, {Crosta},
  {Crowley}, {Dafonte}, {Dapergolas}, {David}, {de Laverny}, {De Luise}, {De
  March}, {De Ridder}, {de Souza}, {de Torres}, {del Peloso}, {del Pozo},
  {Delbo}, {Delgado}, {Delisle}, {Demouchy}, {Dharmawardena}, {Di Matteo},
  {Diakite}, {Diener}, {Distefano}, {Dolding}, {Enke}, {Fabre}, {Fabrizio},
  {Faigler}, {Fedorets}, {Fernique}, {Fournier}, {Fouron}, {Fragkoudi}, {Gai},
  {Garcia-Gutierrez}, {Garcia-Reinaldos}, {Garc{\'\i}a-Torres}, {Garofalo},
  {Gavel}, {Gavras}, {Gerlach}, {Geyer}, {Giacobbe}, {Gilmore}, {Girona},
  {Giuffrida}, {Gomel}, {Gomez}, {Gonz{\'a}lez-N{\'u}{\~n}ez},
  {Gonz{\'a}lez-Santamar{\'\i}a}, {Gonz{\'a}lez-Vidal}, {Granvik}, {Guillout},
  {Guiraud}, {Guti{\'e}rrez-S{\'a}nchez}, {Guy}, {Hatzidimitriou}, {Hauser},
  {Haywood}, {Helmer}, {Helmi}, {Sarmiento}, {Hidalgo}, {H{\l}adczuk}, {Hobbs},
  {Holland}, {Huckle}, {Jasniewicz}, {Jean-Antoine Piccolo},
  {Jim{\'e}nez-Arranz}, {Juaristi Campillo}, {Julbe}, {Karbevska}, {Kervella},
  {Kordopatis}, {Korn}, {K{\'o}sp{\'a}l}, {Kostrzewa-Rutkowska},
  {Kruszy{\'n}ska}, {Kun}, {Laizeau}, {Lambert}, {Lanza}, {Lasne}, {Le
  Campion}, {Lebreton}, {Lebzelter}, {Leccia}, {Leclerc}, {Lecoeur-Taibi},
  {Liao}, {Licata}, {Lindstr{\o}m}, {Lister}, {Livanou}, {Lorca}, {Loup},
  {Madrero Pardo}, {Magdaleno Romeo}, {Managau}, {Mann}, {Manteiga},
  {Marchant}, {Marconi}, {Marcos}, {Marcos Santos}, {Mar{\'\i}n Pina},
  {Marinoni}, {Marocco}, {Martin Polo}, {Mart{\'\i}n-Fleitas}, {Marton},
  {Mary}, {Masip}, {Massari}, {Mastrobuono-Battisti}, {Mazeh}, {McMillan},
  {Messina}, {Michalik}, {Millar}, {Mints}, {Molina}, {Molinaro}, {Moln{\'a}r},
  {Monari}, {Mongui{\'o}}, {Montegriffo}, {Montero}, {Mor}, {Mora},
  {Morbidelli}, {Morel}, {Morris}, {Murphy}, {Musella}, {Nagy}, {Noval},
  {Oca{\~n}a}, {Ogden}, {Ordenovic}, {Osinde}, {Pagani}, {Pagano}, {Palaversa},
  {Palicio}, {Pallas-Quintela}, {Panahi}, {Payne-Wardenaar}, {Pe{\~n}alosa
  Esteller}, {Penttil{\"a}}, {Pichon}, {Piersimoni}, {Pineau}, {Plachy},
  {Plum}, {Pr{\v{s}}a}, {Pulone}, {Racero}, {Ragaini}, {Rainer}, {Raiteri},
  {Ramos-Lerate}, {Re Fiorentin}, {Regibo}, {Richards}, {Rios Diaz}, {Riva},
  {Rix}, {Rixon}, {Robichon}, {Robin}, {Robin}, {Roelens}, {Rogues},
  {Rohrbasser}, {Rowell}, {Royer}, {Ruz Mieres}, {Rybicki}, {Sadowski},
  {S{\'a}ez N{\'u}{\~n}ez}, {Sagrist{\`a} Sell{\'e}s}, {Sahlmann}, {Salguero},
  {Samaras}, {Sanchez Gimenez}, {Sanna}, {Santove{\~n}a}, {Sarasso},
  {Schultheis}, {Sciacca}, {Segol}, {Segovia}, {S{\'e}gransan}, {Semeux},
  {Shahaf}, {Siddiqui}, {Siebert}, {Siltala}, {Silvelo}, {Slezak}, {Slezak},
  {Smart}, {Snaith}, {Solano}, {Solitro}, {Souami}, {Souchay}, {Spagna},
  {Spina}, {Spoto}, {Steele}, {Steidelm{\"u}ller}, {Stephenson}, {S{\"u}veges},
  {Surdej}, {Szabados}, {Szegedi-Elek}, {Taris}, {Taylor}, {Teixeira},
  {Tolomei}, {Tonello}, {Torra}, {Torra}, {Torralba Elipe}, {Trabucchi},
  {Tsounis}, {Turon}, {Ulla}, {Unger}, {Vaillant}, {van Dillen}, {van Reeven},
  {Vanel}, {Vecchiato}, {Viala}, {Vicente}, {Voutsinas}, {Weiler}, {Wevers},
  {Wyrzykowski}, {Yoldas}, {Yvard}, {Zhao}, {Zorec}, {Zucker}, \&
  {Zwitter}}]{gaia2023}
{Gaia Collaboration}, {Drimmel}, R., {Romero-G{\'o}mez}, M., {et~al.} 2023,
  \aap, 674, A37

\bibitem[{{Heggie} \& {Hut}(2003)}]{heggie03}
{Heggie}, D. \& {Hut}, P. 2003, {The Gravitational Million-Body Problem: A
  Multidisciplinary Approach to Star Cluster Dynamics} (IOP Publishing Ltd)

\bibitem[{{Hopkins} \& {Quataert}(2011)}]{hopkins2011}
{Hopkins}, P.~F. \& {Quataert}, E. 2011, \mnras, 415, 1027

\bibitem[{{Julian} \& {Toomre}(1966)}]{JT1966}
{Julian}, W.~H. \& {Toomre}, A. 1966, \apj, 146, 810

\bibitem[{{Kalnajs}(1972)}]{kalnajs1972}
{Kalnajs}, A.~J. 1972, in Astrophysics and Space Science Library, Vol.~31, IAU
  Colloq. 10: Gravitational N-Body Problem, ed. M.~{Lecar}, 13

\bibitem[{{Kaur} \& {Sridhar}(2018)}]{kaur2018}
{Kaur}, K. \& {Sridhar}, S. 2018, \apj, 868, 134

\bibitem[{{Kaur} \& {Stone}(2022)}]{kaur2022}
{Kaur}, K. \& {Stone}, N.~C. 2022, \mnras, 515, 407

\bibitem[{{Kley} \& {Nelson}(2012)}]{kley2012}
{Kley}, W. \& {Nelson}, R.~P. 2012, \araa, 50, 211

\bibitem[{{Kocsis} \& {Tremaine}(2015)}]{kocsis2015}
{Kocsis}, B. \& {Tremaine}, S. 2015, \mnras, 448, 3265

\bibitem[{{Kormendy} \& {Ho}(2013)}]{kormendy2013}
{Kormendy}, J. \& {Ho}, L.~C. 2013, \araa, 51, 511

\bibitem[{{Labb{\'e}} {et~al.}(2023){Labb{\'e}}, {van Dokkum}, {Nelson},
  {Bezanson}, {Suess}, {Leja}, {Brammer}, {Whitaker}, {Mathews}, {Stefanon}, \&
  {Wang}}]{labbe2023}
{Labb{\'e}}, I., {van Dokkum}, P., {Nelson}, E., {et~al.} 2023, \nat, 616, 266

\bibitem[{{Lapiner} {et~al.}(2021){Lapiner}, {Dekel}, \&
  {Dubois}}]{lapiner2021}
{Lapiner}, S., {Dekel}, A., \& {Dubois}, Y. 2021, \mnras, 505, 172

\bibitem[{{Lau} \& {Binney}(2021)}]{lau2021}
{Lau}, J.~Y. \& {Binney}, J. 2021, \mnras, 507, 2241

\bibitem[{{Launhardt} {et~al.}(2002){Launhardt}, {Zylka}, \&
  {Mezger}}]{launhardt2002}
{Launhardt}, R., {Zylka}, R., \& {Mezger}, P.~G. 2002, \aap, 384, 112

\bibitem[{{Lescaudron} {et~al.}(2023){Lescaudron}, {Dubois}, {Beckmann}, \&
  {Volonteri}}]{lescaudron2023}
{Lescaudron}, S., {Dubois}, Y., {Beckmann}, R.~S., \& {Volonteri}, M. 2023,
  \aap, 674, A217

\bibitem[{{Lynden-Bell} \& {Kalnajs}(1972)}]{LBK1972}
{Lynden-Bell}, D. \& {Kalnajs}, A.~J. 1972, \mnras, 157, 1

\bibitem[{{Matthee} {et~al.}(2024){Matthee}, {Naidu}, {Brammer}, {Chisholm},
  {Eilers}, {Goulding}, {Greene}, {Kashino}, {Labbe}, {Lilly}, {Mackenzie},
  {Oesch}, {Weibel}, {Wuyts}, {Xiao}, {Bordoloi}, {Bouwens}, {van Dokkum},
  {Illingworth}, {Kramarenko}, {Maseda}, {Mason}, {Meyer}, {Nelson}, {Reddy},
  {Shivaei}, {Simcoe}, \& {Yue}}]{matthee2024}
{Matthee}, J., {Naidu}, R.~P., {Brammer}, G., {et~al.} 2024, \apj, 963, 129

\bibitem[{{Merritt}(2004)}]{merritt2004}
{Merritt}, D. 2004, \prl, 92, 201304

\bibitem[{{Merritt}(2013)}]{merritt2013}
{Merritt}, D. 2013, {Dynamics and Evolution of Galactic Nuclei},
  $1^{\mathrm{st}}$ edn. (Princeton Univ. Press, Princeton, NJ)

\bibitem[{{Merritt} {et~al.}(2007){Merritt}, {Berczik}, \&
  {Laun}}]{merritt2007}
{Merritt}, D., {Berczik}, P., \& {Laun}, F. 2007, \aj, 133, 553

\bibitem[{{Mo} {et~al.}(2010){Mo}, {van den Bosch}, \&
  {White}}]{galaxyformation2010}
{Mo}, H., {van den Bosch}, F.~C., \& {White}, S. 2010, {Galaxy Formation and
  Evolution}, $1^{\mathrm{st}}$ edn. (Princeton Univ. Press, Princeton, NJ)

\bibitem[{{Naidu} {et~al.}(2022){Naidu}, {Oesch}, {van Dokkum}, {Nelson},
  {Suess}, {Brammer}, {Whitaker}, {Illingworth}, {Bouwens}, {Tacchella},
  {Matthee}, {Allen}, {Bezanson}, {Conroy}, {Labbe}, {Leja}, {Leonova},
  {Magee}, {Price}, {Setton}, {Strait}, {Stefanon}, {Toft}, {Weaver}, \&
  {Weibel}}]{naidu2022}
{Naidu}, R.~P., {Oesch}, P.~A., {van Dokkum}, P., {et~al.} 2022, \apjl, 940,
  L14

\bibitem[{{Neumayer} {et~al.}(2020){Neumayer}, {Seth}, \&
  {B{\"o}ker}}]{neumayer2020}
{Neumayer}, N., {Seth}, A., \& {B{\"o}ker}, T. 2020, \aapr, 28, 4

\bibitem[{{Nieuwmunster} {et~al.}(2024){Nieuwmunster}, {Schultheis}, {Sormani},
  {Fragkoudi}, {Nogueras-Lara}, {Sch{\"o}del}, {McMillan}, {Smith}, \&
  {Sanders}}]{nieuwmunster2024}
{Nieuwmunster}, N., {Schultheis}, M., {Sormani}, M., {et~al.} 2024, \aap, 685,
  A93

\bibitem[{{Nogueras-Lara} {et~al.}(2020){Nogueras-Lara}, {Sch{\"o}del},
  {Gallego-Calvente}, {Gallego-Cano}, {Shahzamanian}, {Dong}, {Neumayer},
  {Hilker}, {Najarro}, {Nishiyama}, {Feldmeier-Krause}, {Girard}, \&
  {Cassisi}}]{nogueras2020}
{Nogueras-Lara}, F., {Sch{\"o}del}, R., {Gallego-Calvente}, A.~T., {et~al.}
  2020, Nature Astronomy, 4, 377

\bibitem[{{Ostriker}(1999)}]{ostriker1999}
{Ostriker}, E.~C. 1999, \apj, 513, 252

\bibitem[{{Panamarev} \& {Kocsis}(2022)}]{panamarev2022}
{Panamarev}, T. \& {Kocsis}, B. 2022, \mnras, 517, 6205

\bibitem[{{Papaloizou}(2021)}]{Papaloizou2021}
{Papaloizou}, J.~C.~B. 2021, Celestial Mechanics and Dynamical Astronomy, 133,
  30

\bibitem[{{Pelupessy} {et~al.}(2013){Pelupessy}, {van Elteren}, {de Vries},
  {McMillan}, {Drost}, \& {Portegies Zwart}}]{pelupessy2013}
{Pelupessy}, F.~I., {van Elteren}, A., {de Vries}, N., {et~al.} 2013, \aap,
  557, A84

\bibitem[{{Petts} {et~al.}(2015){Petts}, {Gualandris}, \& {Read}}]{petts2015}
{Petts}, J.~A., {Gualandris}, A., \& {Read}, J.~I. 2015, \mnras, 454, 3778

\bibitem[{{Pfister} {et~al.}(2021){Pfister}, {Dai}, {Volonteri}, {Auchettl},
  {Trebitsch}, \& {Ramirez-Ruiz}}]{bhgrowth2021}
{Pfister}, H., {Dai}, J.~L., {Volonteri}, M., {et~al.} 2021, \mnras, 500, 3944

\bibitem[{{Pfister} {et~al.}(2019){Pfister}, {Volonteri}, {Dubois}, {Dotti}, \&
  {Colpi}}]{pfister2019}
{Pfister}, H., {Volonteri}, M., {Dubois}, Y., {Dotti}, M., \& {Colpi}, M. 2019,
  \mnras, 486, 101

\bibitem[{{Portegies Zwart} \& {McMillan}(2018)}]{portegieszwart2018}
{Portegies Zwart}, S. \& {McMillan}, S. 2018, {Astrophysical Recipes; The art
  of AMUSE}, IOP Astronomy / AAS (IOP Publishing, Bristol, UK)

\bibitem[{{Portegies Zwart} {et~al.}(2013){Portegies Zwart}, {McMillan}, {van
  Elteren}, {Pelupessy}, \& {de Vries}}]{portegieszwart2013}
{Portegies Zwart}, S., {McMillan}, S.~L.~W., {van Elteren}, E., {Pelupessy},
  I., \& {de Vries}, N. 2013, Computer Physics Communications, 184, 456

\bibitem[{{Rauch} \& {Tremaine}(1996)}]{rauch1996}
{Rauch}, K.~P. \& {Tremaine}, S. 1996, \na, 1, 149

\bibitem[{{Read} {et~al.}(2006){Read}, {Goerdt}, {Moore}, {Pontzen}, {Stadel},
  \& {Lake}}]{read2006}
{Read}, J.~I., {Goerdt}, T., {Moore}, B., {et~al.} 2006, \mnras, 373, 1451

\bibitem[{{Reid} \& {Brunthaler}(2004)}]{reid2004}
{Reid}, M.~J. \& {Brunthaler}, A. 2004, \apj, 616, 872

\bibitem[{{Renaud} {et~al.}(2011){Renaud}, {Gieles}, \& {Boily}}]{renaud2011}
{Renaud}, F., {Gieles}, M., \& {Boily}, C.~M. 2011, \mnras, 418, 759

\bibitem[{{Renaud} {et~al.}(2021){Renaud}, {Romeo}, \& {Agertz}}]{renaud2021}
{Renaud}, F., {Romeo}, A.~B., \& {Agertz}, O. 2021, \mnras, 508, 352

\bibitem[{{Schawinski} {et~al.}(2015){Schawinski}, {Koss}, {Berney}, \&
  {Sartori}}]{schawinski2015}
{Schawinski}, K., {Koss}, M., {Berney}, S., \& {Sartori}, L.~F. 2015, \mnras,
  451, 2517

\bibitem[{{Spitzer}(1987)}]{spitzer1987}
{Spitzer}, L. 1987, {Dynamical evolution of globular clusters} (Princeton
  University Press)

\bibitem[{{Toomre}(1964)}]{toomre1964}
{Toomre}, A. 1964, \apj, 139, 1217

\bibitem[{{Tremaine} {et~al.}(1994){Tremaine}, {Richstone}, {Byun}, {Dressler},
  {Faber}, {Grillmair}, {Kormendy}, \& {Lauer}}]{tremaine1994}
{Tremaine}, S., {Richstone}, D.~O., {Byun}, Y.-I., {et~al.} 1994, \aj, 107, 634

\bibitem[{{Tremaine} \& {Weinberg}(1984)}]{TW1984}
{Tremaine}, S. \& {Weinberg}, M.~D. 1984, \mnras, 209, 729

\bibitem[{{{\"U}bler} {et~al.}(2024){{\"U}bler}, {Maiolino},
  {P{\'e}rez-Gonz{\'a}lez}, {D'Eugenio}, {Perna}, {Curti}, {Arribas}, {Bunker},
  {Carniani}, {Charlot}, {Rodr{\'\i}guez Del Pino}, {Baker}, {B{\"o}ker},
  {Cresci}, {Dunlop}, {Grogin}, {Jones}, {Kumari}, {Lamperti}, {Laporte},
  {Marshall}, {Mazzolari}, {Parlanti}, {Rawle}, {Scholtz}, {Venturi}, \&
  {Witstok}}]{ubler2024}
{{\"U}bler}, H., {Maiolino}, R., {P{\'e}rez-Gonz{\'a}lez}, P.~G., {et~al.}
  2024, \mnras, 531, 355

\bibitem[{{Vasiliev}(2019)}]{vasiliev2019}
{Vasiliev}, E. 2019, \mnras, 482, 1525

\bibitem[{{Volonteri}(2012)}]{volonteri2012}
{Volonteri}, M. 2012, Science, 337, 544

\bibitem[{{von Fellenberg} {et~al.}(2022){von Fellenberg}, {Gillessen},
  {Stadler}, {Baub{\"o}ck}, {Genzel}, {de Zeeuw}, {Pfuhl}, {Amaro Seoane},
  {Drescher}, {Eisenhauer}, {Habibi}, {Ott}, {Widmann}, \&
  {Young}}]{fellenberg2022}
{von Fellenberg}, S.~D., {Gillessen}, S., {Stadler}, J., {et~al.} 2022, \apjl,
  932, L6

\bibitem[{{{\v{S}}ubr} \& {Haas}(2014)}]{subr2014}
{{\v{S}}ubr}, L. \& {Haas}, J. 2014, \apj, 786, 121

\bibitem[{{Weinberg}(1989)}]{weinberg1989}
{Weinberg}, M.~D. 1989, \mnras, 239, 549

\bibitem[{{Weinberg}(1998)}]{weinberg1998}
{Weinberg}, M.~D. 1998, \mnras, 299, 499

\bibitem[{{Zoccali} {et~al.}(2024){Zoccali}, {Rojas-Arriagada}, {Valenti},
  {Contreras Ramos}, {Valenzuela-Navarro}, \& {Salvo-Guajardo}}]{zoccali2024}
{Zoccali}, M., {Rojas-Arriagada}, A., {Valenti}, E., {et~al.} 2024, \aap, 684,
  A214

\end{thebibliography}

%
\begin{appendix}
\section{Fokker-Planck approach}\label{sec:FP}
When a massive body 
moves through  space at velocity $\mathbf{v}$, multiple interactions with background stars 
lead to the diffusion  of kinetic energy by repeated small-angle scattering. This is put on a quantitative 
footing by computing the velocity diffusion coefficients as functions of the local stellar density $\rho_\star$  and velocity distribution 
function $\fv{}$. The scattering events are summed up by integrating over the  whole of space.  
When  stars are distributed isotropically about the orbital trajectory of the massive body, pair-wise deflections cancel out and 
the net velocity kicks orthogonal to the trajectory add up to zero (to first order). For that reason, 
it is useful to split the net rate of change\footnote{We follow the syntax of \citep{spitzer1987} and replace the time derivative operator  $\textrm{d} / \textrm{d} t $ with angled-brackets $\langle \dots \rangle$.} $\rate{v}{}$ in parallel and orthogonal components relative to the direction of motion of the body, 

\begin{equation}
\rate{v}{} = \rate{v}{\|} + \rate{v}{\bot}, 
\end{equation}
so that the velocity vector is incremented as \[\mathbf{v} \rightarrow \mathbf{v} + \Delta\mathbf{v} = \mathbf{v} + \rate{v}{\|}\,\delta t + \rate{v}{\bot}\,\delta t \] over the time interval $\delta t$. The kinetic energy per unit mass of the body, $E = \mathbf{v}\cdot\mathbf{v}/2$,  becomes 
\[ 
E^\prime = \frac{1}{2}\left( \mathbf{v} + \Delta\mathbf{v} \right)^2 = \frac{1}{2} \mathbf{v}\cdot\mathbf{v} +  \mathbf{v}\cdot\Delta\mathbf{v}  + \frac{1}{2} \left( \Delta\mathbf{v}_\| \right)^2 
+ \frac{1}{2} \left( \Delta\mathbf{v}_\bot \right)^2  \]
so that the rate of change $\rate{E}{} = \langle E^\prime - E\rangle $  may be expressed as  

\begin{equation}
\rate{E}{} = \mathbf{v}_\|\cdot \rate{v}{\|} + \frac{1}{2}\, \rates{v}{\|} + \frac{1}{2}\, \rates{v}{\bot}\, . \label{eq:Ediffusion}
\end{equation}
In this equation, the 1st-order orthogonal term $\mathbf{v}_\bot \cdot \rate{v}{\bot}$ cancels by an argument of symmetry. 
The diffusion coefficients in (\ref{eq:Ediffusion}) are computed from the (local) stellar velocity distribution function $\fv{\star}$ in integral form, 

 \begin{subeqnarray}
 \rate{v}{\|} & = & \Gamma\, \left(  1 + \frac{\Mbh}{m_\star}\right) \,\frac{\mathrm{d}}{\mathrm{d} v}\, h(\mathbf{v}\,|\,\fv{\star} )\\ 
 \frac{1}{2}\rates{v}{\|} & = & \frac{1}{2}\,\Gamma \, \frac{\mathrm{d}^2}{\mathrm{d} v^2}\, g(\mathbf{v}\,|\,\fv{\star}  )  \\
  \frac{1}{2}\rates{v}{\bot} &  = & \frac{1}{v}\,\Gamma \, \frac{\mathrm{d}}{\mathrm{d} v}\, g(\mathbf{v}\,|\,\fv{\star}  )  \label{eq:diffrates}
 \end{subeqnarray}
where $\Gamma = 4\pi G^2 m_\star^2 \ln \Lambda$ accounts for the Coulomb term ($\Lambda$) ; 
 integrals over the stellar distribution function $\fv{\star}$ define the {Rosenbluth potentials} $g, h$ as \citep[\S 7.4.4, Eqs. 7.83]{BT08}
\begin{equation}
\begin{array}{l}
 h(\mathbf{v}\,|\,\fv{\star} ) \equiv  \displaystyle{ \int \frac{\fv{\star}}{ \| \mathbf{v} - \mathbf{v}_\star \| }  \, \mathrm{\bf d}^3\mathbf{v}_\star} \\[8pt]
 g(\mathbf{v}\,|\,\fv{\star} ) \equiv  \displaystyle{ \int \fv{\star}\, \| \mathbf{v} - \mathbf{v}_\star \| \, \mathrm{\bf d}^3\mathbf{v}_\star}\ .
\end{array}   \label{eq:Rosenbluth}
\end{equation}
To ease the  evaluation of  the integrals  (\ref{eq:Rosenbluth}), define $\K(v_\phi, v_\bot) $ as 

\begin{equation}
        \K^2 \equiv  ( R\Omega - v_\phi)^2 + v_\bot^2 \label{eq:K}       
\end{equation}
and pick the sign of $\K$ to be that of $R\Omega - v_\phi$ (positive for a fast pattern speed / slow azimuthal velocity). From this function we obtain the derivative

\begin{equation} 
     \frac{\mathrm{d}\K}{\mathrm{d}v} = - \frac{1}{\K}\,\left(\, R\Omega\,\frac{v_\phi}{v} - v \right) \label{eq:dKdv}
\end{equation}
where all quantities are scalars, $v$ being the norm of the velocity, and the chain rule may be applied to the Rosenbluth potentials 
(\ref{eq:Rosenbluth}) to compute derivatives which appear in (\ref{eq:diffrates}), \ie through the operator 

\[  \frac{\mathrm{d}}{\mathrm{d}v} =  \frac{\mathrm{d}\K}{\mathrm{d}v}  \times \frac{\mathrm{d}}{\mathrm{d}\K}\, . \]
For instance, the second order derivative in (\ref{eq:diffrates}b) becomes 

\[ 
\frac{\mathrm{d}^2}{\mathrm{d}v^2}\, g = \frac{\mathrm{d}}{\mathrm{d}v}\,\left[ \frac{\mathrm{d}\K}{\mathrm{d}v}\, \frac{\mathrm{d}}{\mathrm{d}\K}\, \right]\, g = \frac{\mathrm{d}^2g}{\mathrm{d}\K^2}\left( \frac{\mathrm{d}\K}{\mathrm{d}v}\, \right)^2\, + \frac{\mathrm{d}g }{\mathrm{d}\K}\, \frac{\mathrm{d}^2\K}{\mathrm{d}v^2} \ . 
\]
The first- and second-order derivatives of the Rosenbluth potential \wrt $\K$ are given in Appendix~\ref{sec:AppendixRosenbluth}  for completeness. 

The Fokker-Planck equations consists in integrating the diffusion coefficients to determine the time-evolution of the distribution function in equilibrium. The application to a single massive body reduces the Fokker-Planck treatment to a response equation 
where the first-order term in (\ref{eq:Ediffusion}) dominates over the quadratic ones: this leads to the classic dynamical friction 
evolution with $dE / dt < 0$ due to a force term parallel but in the opposite sense to the motion of the body (see Fig.~\ref{fig:Cartoon}a and \eg \citealp{BT08}). The loss of kinetic energy and in-spiral of the massive body appears to be 
common to any distribution function giving rise to an isotropic velocity field. 

\subsection{Anisotropic configurations, dynamical traction}
The situation is different if we focus on orbits that are not isotropic in space. A second look at Fig.~\ref{fig:Cartoon} shows that the relative velocity with the background stars would remain the same if the massive body was at rest, and all the stars were flowing down the velocity vector, $-\vbh$. From that point of view, the dynamical friction becomes instead dynamical {traction}. 
 Dynamical traction would drag an BH in any direction defined by the 
 bulk background flow. Contrary to a dissipative frictional term, however, 
 the traction on a body increases its kinetic  energy and can be non-zero even if it is at rest in an inertial frame.
 For this to happen requires that the velocity field be anisotropic (as on Fig.~\ref{fig:Cartoon}[b]).   

This suggest that we embed an BH in a background potential made of a hot, isotropic component, and a second, cool 
component, made of stars drawn from a family of orbits carrying a net positive angular momentum. The situation is 
sketched on Fig.~\ref{fig:Cartoon}b. To simplify the 
discussion, we plot only stars rotating about the system barycentre and not the hot component. 
We suppose that the BH has come to rest at a distance $\simeq \delta\xbh$ given by (\ref{eq:dxBH})). The bulk motion is now in azimuth and we may rewrite (\ref{eq:Ediffusion}) in cylindrical coordinates as 

\begin{equation}
\rate{E}{} = \mathbf{v}_\phi\cdot \rate{v}{\phi} + \frac{1}{2}\, \rates{v}{\phi} + \frac{1}{2}\, \rates{v}{\bot}\, . \label{eq:Ediffusion2}
\end{equation}
The azimuthal acceleration $ \dot{\mathbf{v}}_\phi = \rate{v}{\phi} $ gives rise to a torque 

\[ \mathbf{T} = \mathbf{r}\times\, \Mbh \dot{\mathbf{v}} = \|\delta\xbh \rate{v}{\phi}\| \, \Mbh\, \hat{\mathbf{z}} \]
where we have combined the radial- and z-component of the velocity in the orthogonal component $\mathbf{v}_\bot$. 
This net torque will transfer  angular momentum, from the stars, to the BH. 
It is then a simple matter to estimate the magnitude of the angular momentum $L_z$ accrued by the BH as 

\begin{equation}
L_z \approx \|\mathbf{T} \| \, \delta t =  \|\delta\xbh \rate{v}{\phi}\| \, \Mbh \, \frac{2\pi}{\omega} \, . \label{eq:LzBH}
\end{equation}
In deriving  $L_z$ we have assumed the torque to be constant over one full period of the stars. They are bound together 
by their self-gravity and the potential generated by the hot component which we will take to be spherically symmetric 
 and uniform (no aggregates / fragment or radial profile). 

\section{Computing the diffusion coefficients}\label{sec:AppendixRosenbluth} 
For completeness we give here an outline of the integrals leading to (\ref{eq:diffrates}).  Starting with the distribution function (\ref{eq:df}), we compute the Rosenbluth potential $g(\mathbf{v} | \fv{\star} ) $ in (\ref{eq:Rosenbluth}) as  a function of the norm $v$ of the velocity of the massive body by expanding the vector dot products.  As a simplification, we extend the upper bound of integration in velocity space to infinity.  Inserting our choice of d.f. (\ref{eq:df}), the integral transforms to

\begin{eqnarray}
g[\mathbf{v}\,|\, \fv{\star} ] =  \displaystyle{\int} \, \mathrm{d}^3 \mathbf{v}_\star \fv{\star} \, \| \mathbf{v} - \mathbf{v}_\star \|  \nonumber\mbox{\hspace{23mm}} \\ 
 =  \displaystyle{\int} \pi\, \mathrm{d}v^2_{\star, \bot} \mathrm{d}v_{\star, \phi}\  f_\bot (v_{\star,\bot})\, \Dirac{R\Omega - v_{\star,\phi} } \hfill   \times 
      \nonumber \\   
  \sqrt{[\mathbf{v} - \mathbf{v}_\star ]\cdot [\mathbf{v} - \mathbf{v}_\star ] } \label{eq:RosenbluthInt}    \\ 
 =  \displaystyle{\int} 2\pi v_{\star,\bot} \mathrm{d}v_{\star, \bot}\,  f_\bot (v_{\star,\bot})\,  \, \left[ (R\Omega - v_\phi)^2 + v^2_\bot + v^2_{\star,\bot} \right]^{\frac{1}{2} } \nonumber
\end{eqnarray}
and the term under the radical defines the space-dependent parameter $\K(R,z) $ as 

\[ \K^2(R,z=0) = (R\Omega - v_\phi)^2 + v^2_\bot \ . \]
The integral (\ref{eq:RosenbluthInt}) is easily computed for the Gaussian form (\ref{eq:df_ortho}) taken for $f_\bot$. The simplification of a Dirac operator implies that the azimuthal motion of the background stars is largely dominated by the streaming flow and we neglect the energy invested in random motion for that degree of freedom. The same calculation leading to (\ref{eq:RosenbluthInt}) can be repeated to compute the first Rosenbluth potential $h(\mathbf{v}\,|\, \fv{\star}$. We find 

\begin{eqnarray}
h[\mathbf{v}\,|\, \fv{\star}] & = & \int_0^\infty  \frac{\pi\, \mathrm{d} v_{\star,\bot}^2\, f_\bot (v_{\star,\bot} ) }{\left[ \K^2 + v_{\star,\bot}^2 \right]^\frac{1}{2} }\nonumber \\ 
& = & - 2\pi \sqrt{\K^2} f_\bot (0) + \int_0^\infty \frac{\pi v_{\star,\bot}}{\sigma_\star^2}\, \mathrm{d} v_{\star,\bot} \, f_{\bot}(v_{\star,\bot}) 
\nonumber \\ 
& & \times  \sqrt{\K^2 + v_{\star,\bot}^2 }  
\end{eqnarray}
where the last integral is once more easily integrated by substituting (\ref{eq:df_ortho}). Note that $\sqrt{\K^2}$ yields the absolute value $|\K|$; the expression can be used in symbolic calculations to avoid conditional expressions. 

The task of computing the diffusion coefficients reduces to computing the first- and second-order derivatives of $h$ and $g$ \wrt $\K$, and then use the relation (\ref{eq:dKdv}).  After some straightforward algebra we find (dropping the dependence on d.f. $\fv{\star}$) 

\begin{equation*}
\frac{\mathrm{d}}{\mathrm{d}\K} h(v) = \mp  2\pi f_\bot (0) + \frac{\K}{\sqrt{2}\sigma} f_\bot (0) \left[ \sqrt{\pi} e^{\K^2/2\sigma_\star^2} \mathrm{erfc}\left( \frac{\K}{\sqrt{2}\sigma_\star} \right)\, \right]  
\end{equation*}
\begin{equation}
\frac{\mathrm{d}}{\mathrm{d}\K} g(v) = \frac{2}{3}\pi  f_\bot (0)  \, \left\{ - 3 \K^2+ 2\K e^{\K^2/2\sigma_\star^2} \, \sqrt{2\sigma^2\pi}\right\}  \label{eq:Rosenbluth_der}
\end{equation}
\begin{equation*}
\frac{\mathrm{d^2}}{\mathrm{d}\K^2} g(v) = \frac{2}{3}\pi f_\bot (0) \, \left\{ - 6 \K + 2\sqrt{2\sigma_\star^2\pi} \,  e^{\K^2/2\sigma_\star^2}   \, \left( 1 + \frac{\K^2}{\sigma^2} \right)  \right\}
\end{equation*}
and recall that $f_\bot(0) = \rho(R,z) / (2\pi m_\star\sigma_\star^2)$. The $\mp$ multiplicative factor   in the first of Eq.~(\ref{eq:Rosenbluth_der}) carries the sign of the derivative $\mathrm{d}\sqrt{\K^2}/\mathrm{d}\K$.  
 Note that \href{https://en.wikipedia.org/wiki/Error_function}{\textsc{erfc}} is the complementary error function. 
  
%

\section{Validation tests}\label{sec:validation}

\begin{center}
\begin{figure*}[t]
\begin{minipage}{\textwidth}
\includegraphics[scale=0.6123]{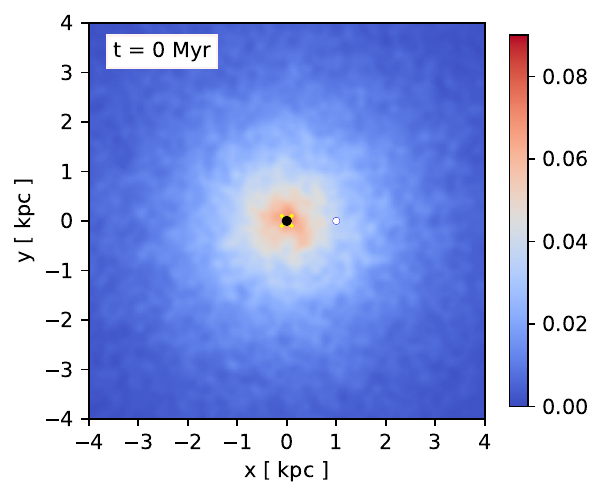 }
\includegraphics[scale=0.6123]{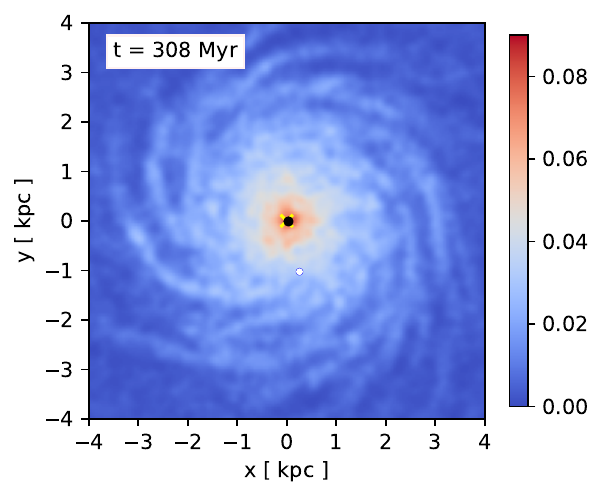}
\includegraphics[scale=0.6123]{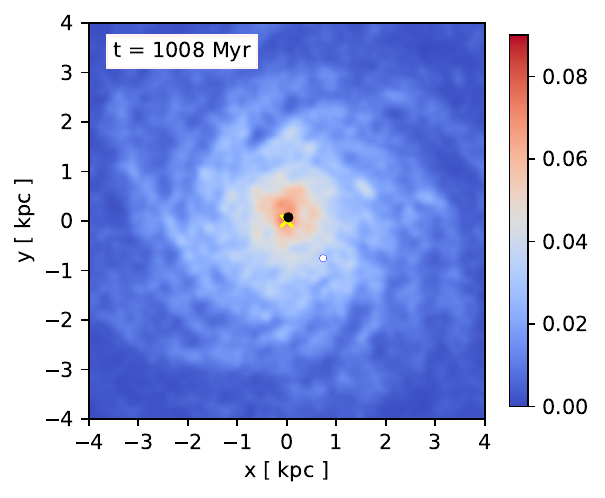}

\caption{Time evolution of a galactic disc with an BH at rest at the centre of coordinates. The black dot indicates the BH of mass  $= 1.25 \times 10^7 M_\odot$. A white dot indicates a test star which remains on a steady orbit. The time runs from  
 left to right as indicated. Notice how an over-density in the shape of a ring develops early on but is then wiped out by differential rotation. 
After $\simeq 1 \unit{Gyr}$ of integration time, the BH remains close to the barycentre with a velocity of $\sim 7\kms$ in amplitude. 
} 
\label{fig:BHatrest}
\end{minipage}
\end{figure*}
\end{center}

\begin{center}
\begin{figure*}[h]
\begin{minipage}{\textwidth}
\centering 
\begin{overpic}[abs, unit=1mm, scale=0.660]{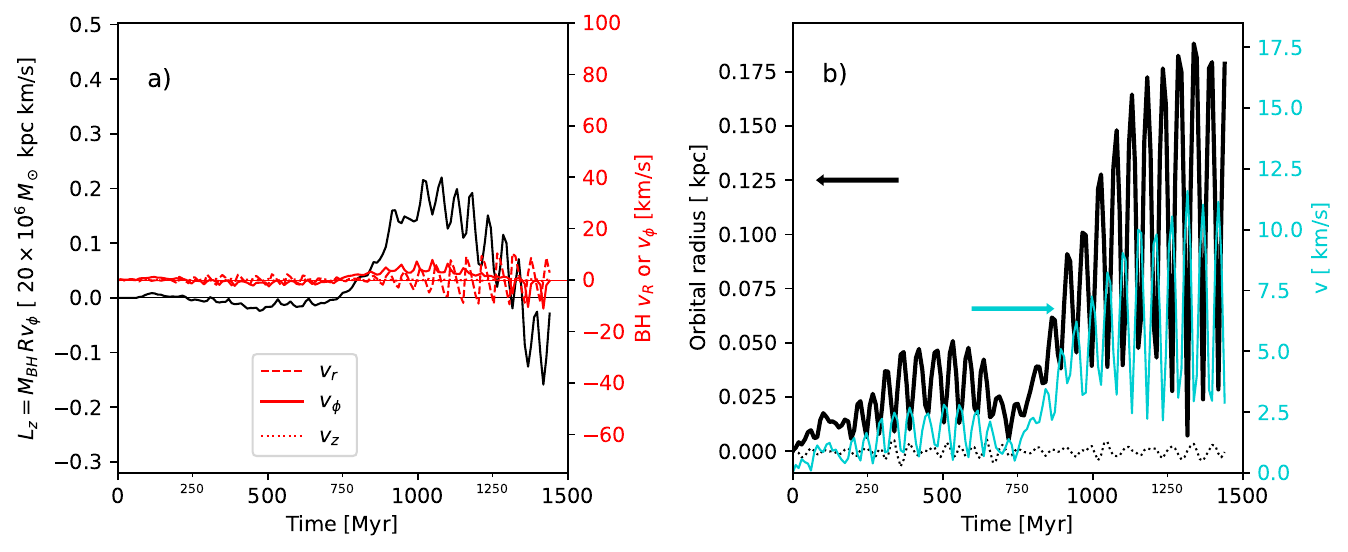}
\put(20,45){ { \begin{minipage}{2.75cm}  {\small \textit{BH at rest, cool disc} } \end{minipage} } } 
\end{overpic} 
\caption{Time evolution of the velocity, position and angular momentum of the BH as function of time. The BH initially sits at rest at the system's 
barycentre. a) The solid black curve graphs the angular momentum, which shows an irregular pattern of small amplitude; the scale's upper bound 
of $\simeq 0.5$ is the value expected for a radial orbit with $ r = 0.125 \kpc$ at velocity $v_\phi = 6.6 \kms$ ; the right-hand axis in red is the scale for individual velocity components ; b) The radius (in black, left-hand axis) and 
velocity $|\vbh|$  (in turquoise, right-hand axis) as function of time. The arrows indicate the spatial resolution $= 0.125 \kpc$ of the calculation, in black; and the expected velocity for an BH orbiting in the frozen background potential inside its influence radius, here $\simeq 0.180\kpc$, in turquoise. 
The black dotted line is the $z$ coordinate. Significant yet resolved motion develops only after a runtime $\sim 1.5 \unit{Gyr}$. } 
\label{fig:BHTimeEvol}
\end{minipage}
\end{figure*}
\end{center}

We ran tests with an axisymmetric configuration consisting of the  isochrone potential, to which we added the 
Miyamoto-Nagai stellar disc sampled with the Agama-generated self-consistent distribution functions. 
The BH was not included 
in the generation of the system's distribution function, but rather was put in afterwards, at rest,  at the centre of coordinates as an $m = 0$ perturbation. Our reasoning here 
is that if the numerical procedure failed, it would produce a {de facto} growing instability that would persist. 
With a stable numerical set-up, we expect a ring-like pattern to develop quickly before being  
washed out over a few dynamical times due to shearing and phase-mixing. 

The results are displayed on Figs.~\ref{fig:BHatrest} and \ref{fig:BHTimeEvol}. On the first figure, we graph the surface density 
of the stars from $t = 0$ up to $1008$ Myr of evolution, corresponding to roughly 10 orbital periods at $R = 1 \kpc$.  
Large-scale spiral patterns can be seen to develop and be present at all stages of evolution. The inner region runs differently, 
with a ring ($m  = 0$ ) mode soon appearing after one period (frame $t = 102 $ Myr) before vanishing at time $ t = 308 $ Myr 
and beyond. This is the expected evolution given the initial configuration. In the later stages of evolution, we find 
non-axisymmetric density fluctuations on a scale of $ \sim 1 \kpc$, but hardly any in the core region. 
The free motion of the BH  triggers fluctuations in the potential which remain on an amplitude coherent 
with the numerical resolution of the calculation. For example, the BH remains confined to its radius of influence ($\approx 180\,\pc$; see Fig. \ref{fig:BHTimeEvol}[b]) and essentially bound by the numerical resolution, indicated with a black arrow on the figure. For reference, 
we give the amplitude of the linear velocity of harmonic motion on a scale of the BH radius of influence as a red arrow : the residual 
motion of the BH which builds up over $\sim 1.5 \unit{Gyr}$ of evolution is therefore in agreement with analysis and sets a 
reference for the interpretation of our results.  

\subsection{Angular momentum conservation}\label{sec:conservation}
A crucial aspect for our study of angular momentum transfer is the conservation of angular momentum by the stars, and by the BH itself. The time evolution of the BH's angular momentum and velocity vector are displayed on Fig.~\ref{fig:BHTimeEvol}(a). The z-component of the angular momentum is shown in black (left-hand vertical axis). It appears to evolve somewhat 
stochastically, and crucially begins to change sign after $t = 750 $ Myr ; the same applies to the radial- and azimuthal velocity 
components, which oscillate about $0 \kms$ (the statistical expectancy). Overall the total system angular momentum was conserved to a relative error of $0.1 \%$ after $t = 1 000 $ Myr of evolution, and to $0.4 \% $ after  $t = 1 500 $ Myr: the numerical drift 
gives an important clue to what will come later when the BH is set in motion. 
This global behaviour is similar to Brownian motion due to the 
graininess of the potential \citep{merritt2007}. This is confirmed by monitoring the (3D) radius and velocity over time (Fig.~\ref{fig:BHTimeEvol}[b]). 
 We performed a second set of tests (no shown here) with the 
Barnes-Hut integrator replaced with the direct-summation Hermite code \textsc{Ph4}  to ensure that the trends are not enhanced / affected by the choice of integration method. The dispersion in the results obtained with different integrators for these reference configurations are
standards against which to identify a genuine physical trend in the calculations. 

\begin{figure}[h]

\centering
\includegraphics[scale=0.50]{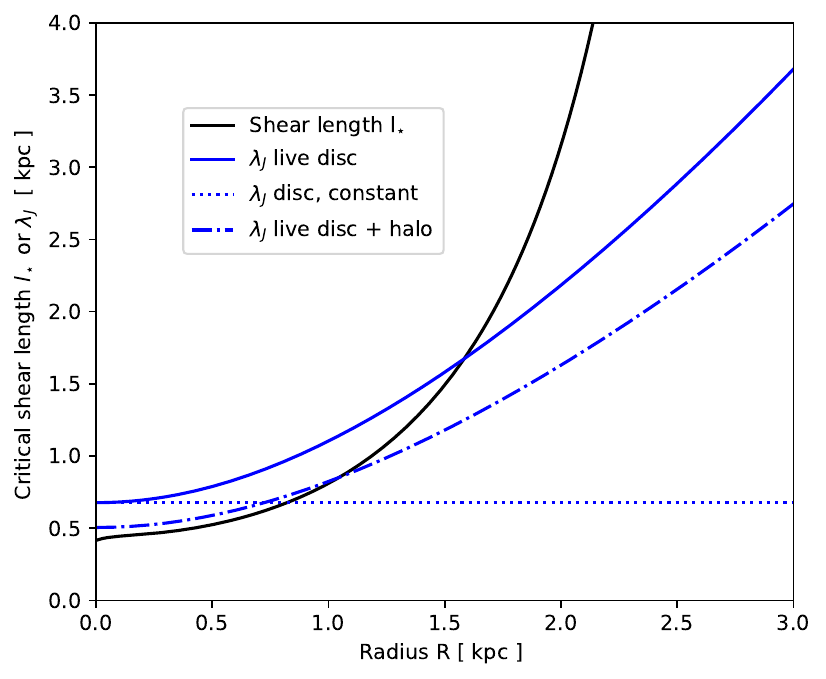}
\caption{The figure graphs the critical length $l_\star$ of Eq.~(\ref{eq:lstar3}) as a function of the cylindrical radius, $R$. The Jeans lengths $\lambda_J$ are shown in blue. The dotted line has $\lambda_J = $ constant, the central value. The solid blue line 
recomputes the Jeans length using the local density $\rho(R,0)$. The dot-dash blue line is the situation when both the disc and halo are live components. Notice how $\lstar \approx \lambda_J$ around the origin. 
 When $\lstar < \lambda_J$, the differential rotation contributes to erode a fragmentation mode.} 
\label{fig:Shear}

\end{figure}

\section{Stability analysis} \label{sec:stability}
We are interested in the stability of the stellar disc before a perturbation (the BH) is introduced. 
Details of the calculations and definitions can be found in \citet{BT08}. 

\subsection{Jeans modes, Toomre's Q}
The scale at which  self-gravity becomes significant is given by the Jeans length, 
\begin{equation}
\lambda_J = \sqrt{ \pi\, \sigma_\star^2 \slash  G\rho_0 }  \label{eq:lJeans}
\end{equation}
where $\sigma_\star$ is the three-dimensional stellar velocity dispersion, and $\rho_0$ the central density. We compute for $\sigma_\star \simeq 7.3 \kms$ 
a length $\lambda_J \approx 478 \pc$ from 
(\ref{eq:lJeans}), significantly larger than the vertical scale height $ b = 250 \pc$ used for the initial configuration. 
 Therefore we do not expect significant  evolution of the vertical disc structure. 

A stellar distribution is stable (in a local sense) against all axisymmetric radial modes of  fragmentation when  (see \S6.2 of \citealt{BT08}) 

\begin{equation}
 Q \equiv \frac{\sigma_R \kappa}{3.36\, G\Sigma} > 1.  \label{eq:ToomreQ}
\end{equation}
This \citet{toomre1964} Q parameter combines the radial epicyclic frequency $\kappa$, surface density $\Sigma$ as well as the radial velocity dispersion, $\sigma_R$. The epicyclic frequency is easily computed from (\ref{eq:MiyamotoNagai}) and (\ref{eq:Henon}) after summing over the potentials (\citealp[see][Eqs. 3.79]{BT08}). This leads to  a mean value 
 
 \[ \overline{Q} \simeq 0.87,  \] 
 with a trend of decreasing $Q$ with increasing radius ; in fact, $Q$ peaks at $Q \simeq 1.2$ near the origin, and comes down to $Q \approx 0.45$ when $R = 5 \kpc$. Therefore we expect over-dense patterns to form more easily outside the region $R = a$ while no or few Jeans-type  fragments should develop in the nearly-harmonic core. We recall  that $Q \simeq 1$ is a threshold value and hence  the self-gravity of large fluctuations may yet develop to some degree  (see \citealp{renaud2021} for a critical assessment  of $Q$). 

\subsection{Differential rotation and self-gravity}
The disc stars are not fully self-gravitating so the connection between velocity dispersion, angular velocity  and surface density  is not as in \citet{toomre1964}'s analysis leading to (\ref{eq:ToomreQ}). To quantify the importance of the external potential, we compare the scale $l_\star$ at which differential rotation is important, to the Jeans length $\lambda_J$. We fix $z = 0$ throughout.

Two points  separated radially  by a  length $\lstar$ have a relative azimuthal velocity 
to first-order in $\lstar/R$ given by 
\begin{equation}
\Delta v_\phi = \lstar\,  \left( \Omega + R\, \frac{\mathrm{d}\Omega}{\mathrm{d}R}\right)  \, . 
\end{equation}
We now focus on the situation where an over-dense clump of mass $M_J$ forms which spawns a volume of diameter $ = 2 \lambda_J$. 
A star approaching $M_J$ will not be bound to it if  $E = - GM_J/\lambda_J + \frac{1}{2} (\Delta v_\phi)^2 > 0$, or 

\begin{equation} 
\lstar > \sqrt{\frac{2\pi G\rho}{3\Delta^2}} \, \lambda_J \, ,   \label{eq:lstar3}
\end{equation} 
where we have made use of the  effective frequency
\footnote{Note that $\Delta = 0$ here does not have the usual meaning of a resonance since the ratio $\kappa :\Omega = \sqrt{2} : 1$ is not rational.}

\[ \Delta \equiv \left|   \frac{\kappa^2}{2\Omega} - \Omega \right|\, . \] 
Note that the density $\rho$ in (\ref{eq:lstar3}) is that of the live (disc) component alone; and that $\Delta = \Omega$ in the limit when $\mathrm{d}\Omega/\mathrm{d}R = 0 $ (solid body rotation). 
 When $\lstar \le \lambda_J$, the shear dissolves a Jeans-unstable fragment and stabilises the disc against local collapse ; otherwise when $\lstar > \lambda_J$,  the differential rotation can not stop the Jeans fragmentation mode from growing. 
 
 The transition from one 
 regime to the other is displayed on Fig.~\ref{fig:Shear} for the reference configuration (see \S\ref{sec:ICs}).  
 The black curve on Fig.~\ref{fig:Shear} is the solution for $\lstar$ from (\ref{eq:lstar3}). Inspection of the  figure shows that shearing takes place on a scale significantly smaller than the local Jeans length $\lambda_J = $ constant in the region inside radius $R = 0.82 \kpc$ 
 (dotted blue line). This  holds  true throughout the region $R = a = 1.5 \kpc$ if 
 we use the local value of $\lambda_J$  (solid blue curve; the cross-over to $\lstar > \lambda_J$ occurs at $R = 1.55(3) \kpc$).

In conclusion, we find that the Toomre criterion points to a marginally unstable core region with $\overline{Q} \sim 0.87 < 1$ 
on average. Because the stellar orbits are not fully self-gravitating, we argue that differential rotation helps curtail the growth of 
fragmentation modes in the central region: they should not form there unless triggered by a perturbation. 
By comparison, 
the tidal field of a background potential almost always contributes to slow down fragmentation modes; they have a limited impact when the density gradients are weak, which is the case in the core region 
(\citealp[see][\S5]{BT08} ; and \eg \citealp{renaud2011} for a general tensor treatment of tidal fields). On the larger scale, however, 
we computed a low value $Q \approx 0.45$ and so from the analysis we can anticipate  the formation of large spiral features outside the core length $a$. This is borne out by the outcome of numerical integrations shown on Fig.~\ref{fig:BHatrest}. 

\end{appendix} 

\end{document}